\documentclass[preprint,12pt]{elsarticle}
\usepackage[T1]{fontenc}
\usepackage{lmodern}
\usepackage{amsmath,amssymb,bm,mathrsfs}
\usepackage{booktabs,graphicx,float,tabularx}
\usepackage{pgfplots}
\usepgfplotslibrary{groupplots}
\pgfplotsset{compat=1.18}
\journal{Journal of Computational Physics}
\usepackage[ruled,vlined]{algorithm2e}
\usepackage{microtype,xurl}
\usepackage[section]{placeins}
\usepackage[hidelinks,pdfusetitle=false]{hyperref}
\hypersetup{
    pdftitle={An Implicit Discrete Adjoint Unified Gas-Kinetic Scheme for Sensitivity Analysis and Shape Optimization in Multiscale Gas Flows},
    pdfauthor={Yuze Zhu, Hangkong Wu, Kun Xu}
}
\usepackage{geometry}
\biboptions{numbers,sort&compress}

\newcommand{\dd}{\mathrm{d}}
\newcommand{\Kn}{\mathrm{Kn}}
\newcommand{\Pran}{\mathrm{Pr}}
\newcommand{\vect}[1]{\boldsymbol{#1}}

\makeatletter
\def\ps@pprintTitle{%
  \let\@oddhead\@empty
  \let\@evenhead\@empty
  \def\@oddfoot{\hfill\makebox[0pt][c]{\thepage}\hfill\llap{\footnotesize\itshape\today}}%
  \let\@evenfoot\@oddfoot
}
\makeatother
\begin{document}
\begin{frontmatter}
\title{An Implicit Discrete Adjoint Unified Gas-Kinetic Scheme for Sensitivity Analysis and Shape Optimization in Multiscale Gas Flows}
\author[a]{Yuze ZHU}
\author[a]{Hangkong WU}
\author[a,b,c]{Kun XU\corref{cor1}} 
\cortext[cor1]{Corresponding author. Email: makxu@ust.hk}

\affiliation[a]{
   organization= {Department of Mathematics, Hong Kong University of Science and Technology},
   city= {Clear Water Bay, Kowloon, Hong Kong},
   country= {China},
}

\affiliation[b]{
   organization= {Department of Mechanical and Aerospace Engineering, Hong Kong University of Science and Technology},
   city= {Clear Water Bay, Kowloon, Hong Kong},
   country= {China},
}

\affiliation[c]{
   organization= {Shenzhen Research Institute, Hong Kong University of Science and Technology},
   city= {Shenzhen},
   country= {China},
}

\begin{abstract}
Accurate and efficient gradient-based sensitivity evaluation remains challenging for performance optimization in multiscale flows with complex geometries. To address this challenge, this work develops a matrix-free discrete adjoint method for the steady implicit unified gas-kinetic scheme (UGKS) through the exact linearization and transposition of its converged fixed-point operator. The method employs a hybrid differentiation strategy that combines local automatic differentiation (AD) of complex nonlinear kinetic kernels with analytically derived tangent and transpose operators for the coupled macroscopic--microscopic implicit updates and mass-conservation projection. This formulation ensures exact discrete consistency across both flow and geometric dependencies while avoiding global Jacobian assembly and the memory overhead of end-to-end reverse-mode AD. Verification tests across continuum, transitional, and free-molecular regimes confirm machine-precision stepwise duality with the corresponding tangent-linear solver and close agreement with central finite differences. In a six-parameter shape-sensitivity analysis, a single adjoint solve achieves speedups of $7.54\times$ and $3.88\times$ over tangent-linear and central finite-difference evaluations, respectively. Finally, constrained shape optimizations of a periodic Knudsen pump and a Mach-2 NACA0012 airfoil converge within ten design updates, achieving up to $29.68\%$ improvement in normalized mass flow and $7.46\%$ reduction in aerodynamic drag across internal and external multiscale flows.
\end{abstract}


\end{frontmatter}

\section{Introduction}\label{sec:introduction}

Multiscale nonequilibrium gas flows arise in a wide range of applications, from high-altitude spacecraft re-entry to gas transport within microelectromechanical systems (MEMS) \cite{schouler2020survey,ho1998micro}. Accurate and efficient prediction of nonequilibrium effects is essential for evaluating aerothermodynamic loads, surface heat transfer, and microscale transport properties. Among the mesoscopic numerical methods developed for such problems, the unified gas-kinetic scheme (UGKS) provides a unified framework by coupling molecular transport and collisions through the time-dependent interface flux constructed from the integral solution of a kinetic model \cite{xu2010unified,huang2012unified,huang2013unified}. Benefiting from its asymptotic-preserving (AP) property, the scheme recovers the hydrodynamic continuum limit without resolving the molecular mean free path or collision time, while preserving genuine kinetic fidelity across transitional and free-molecular regimes. Subsequent advances in implicit strategies and multigrid acceleration have substantially enhanced its computational efficiency, broadening the applicability of UGKS toward demanding multiscale flow simulations \cite{zhu2016implicit,zhu2017unified,xu2022ugks}.

Beyond accurate flow simulation, system performance evaluation and gradient-based design optimization require the sensitivities of objective functions with respect to geometric and boundary-control parameters \cite{giles2000introduction}. Compared with macroscopic continuum models, kinetic schemes entail either discretizing the molecular velocity space or tracking a large number of particles, which incurs higher computational cost. Furthermore, detailed parameterizations of complex geometries and boundary profiles often introduce numerous design variables, further increasing the computational burden of sensitivity evaluation. Conventional approaches to sensitivity evaluation include finite-difference (FD), complex-step differentiation, and tangent-linear (TL) methods. However, the former two rely on additional nonlinear flow solves for each parameter direction, whereas the TL method requires a linearized solution for each design variable \cite{martins2013review}.  Consequently, evaluating directional sensitivities across high-dimensional parameter spaces becomes computationally prohibitive.

Adjoint methods overcome this computational bottleneck by evaluating the gradient of a targeted objective function with respect to an arbitrary number of design parameters via a single adjoint solve, rendering them particularly attractive for high-dimensional sensitivity analysis in computational fluid dynamics (CFD) \cite{kenway2019effective}. Depending on the sequence of differentiation and discretization, adjoint formulations are broadly categorized into continuous and discrete approaches. In the continuous adjoint framework, adjoint equations and boundary conditions are derived from the governing equations prior to numerical discretization \cite{pironneau1974optimum,jameson1988aerodynamic}. Conversely, the discrete adjoint is constructed directly by differentiating the fully discretized governing system and the numerical objective function \cite{giles2003algorithm}. Because differentiation and discretization do not generally commute, continuous-adjoint sensitivities may deviate from the exact derivatives of the discrete objective. In contrast, the discrete adjoint inherently guarantees exact consistency between the computed gradients and the discretized objective function, provided that all algorithmic and geometric dependencies are properly differentiated, making it particularly well suited for exact sensitivity analysis and gradient-based applications \cite{giles2003algorithm,martins2013review}.

Based on these advantages, adjoint-based sensitivity analysis has been extended to kinetic descriptions of multiscale gas flows through various numerical approaches. Sato et al.\ developed an adjoint-based topology optimization method for the BGK model \cite{sato2019topology}, while Guan et al.\ established adjoint formulations using information-preserving direct simulation Monte Carlo (IP-DSMC) and the discrete velocity method (DVM) \cite{guan2023topology,guan2024topology}. To cover all flow regimes, Yuan and Wu proposed kinetic adjoint methods combining continuous and discrete formulations for design optimization across continuum and rarefied flows \cite{yuan2024design,yuan2025adjoint}. Subsequently, Zhang et al.\ accelerated these kinetic adjoint calculations through the general synthetic iterative scheme (GSIS) \cite{zhang2026fast}. In the framework of gas-kinetic schemes (GKS), Wu et al.\ developed discrete adjoint methods based on automatic differentiation (AD) for aerodynamic shape optimization in continuum flows and subsequently extended the approach to implicit GKS calculations across different Mach number regimes \cite{wu2026discrete,wu2026implicit}. More recently, Zhang et al.\ analytically derived an asymptotic-preserving adjoint UGKS based on macroscopic-moment projection and microscopic lifting \cite{zhang2026asymptotic}.

Although these studies have established a foundation for kinetic adjoint methods, obtaining consistent sensitivities accurately and efficiently for steady implicit UGKS remains challenging when complex nonlinear operators are involved. In implicit UGKS, the converged steady state is defined by a nonlinear fixed-point operator that couples spatial reconstruction, kinetic fluxes, boundary conditions, macroscopic--microscopic implicit updates, and mesh geometry. Consistent sensitivity analysis therefore requires differentiating this entire dependency chain. However, manual analytical differentiation of the underlying numerical kernels is tedious, error-prone, and practically infeasible for certain nonlinear operators. On the other hand, applying end-to-end reverse-mode AD directly to the solver requires recording the full forward iteration history, resulting in excessive memory consumption. Consequently, a differentiation strategy is needed that preserves exact discrete consistency while maintaining computational and memory efficiency.

To this end, the present work develops a matrix-free discrete adjoint method for the steady implicit UGKS by exactly linearizing and transposing its converged fixed-point operator. Through a hybrid differentiation strategy, local AD is applied to complex nonlinear kinetic kernels, while analytically derived tangent and transpose operators handle the coupled macroscopic--microscopic implicit updates and mass-conservation projection. This formulation preserves the discrete algorithmic and geometric dependencies of the primal solver while avoiding global Jacobian assembly and the memory overhead of end-to-end reverse-mode AD. Sensitivity accuracy is verified through stepwise duality tests against the corresponding TL solver and comparisons with independent central FD benchmarks in cavity flows spanning continuum to free-molecular regimes. Computational efficiency is further assessed on a six-parameter thermal-pumping case against directional TL and central FD calculations. Finally, the method is applied to constrained shape optimization of a periodic thermally driven channel and a supersonic NACA0012 airfoil across different Knudsen numbers.

The remainder of this paper is organized as follows. Section~\ref{sec:primal} reviews the governing kinetic equations and the steady implicit UGKS solver. Section~\ref{sec:method} presents the discrete tangent-linear and adjoint formulations, along with geometric and boundary sensitivity evaluation and the constrained shape optimization procedure. Section~\ref{sec:results} verifies the sensitivity accuracy and computational efficiency, followed by constrained shape optimization studies for internal and external multiscale flows. Finally, Section~\ref{sec:conclusions} concludes the paper.

\section{Implicit unified gas-kinetic scheme}\label{sec:primal}

This section introduces the discrete fixed-point operator used for steady-state solutions in the implicit UGKS. First, the Shakhov kinetic model and spatial discretization are presented. Then, the kinetic boundary conditions used in this study are described, and finally, the coupled macroscopic--microscopic implicit update operator is formulated.

\subsection{Kinetic model and numerical discretization}
\label{sec:model}\label{sec:ugks}

A monatomic gas ($\gamma=5/3$) is considered in a two-dimensional physical domain. Let $\vect{x}=(x,y)$ denote the physical coordinates, $\vect{u}=(u,v)$ the in-plane molecular velocity, and $\xi$ the out-of-plane velocity component. All quantities are nondimensionalized by a reference length $L_0$, density $\rho_0$, temperature $T_0$, and characteristic molecular speed $c_0=\sqrt{2RT_0}$. The molecular distribution function $f(\vect{x},\vect{u},\xi,t)$ satisfies the Shakhov kinetic model~\cite{shakhov1968generalization},
\begin{equation}
    \frac{\partial f}{\partial t}
    +\vect{u}\cdot\nabla_{\vect{x}}f
    =\frac{f^S-f}{\tau},
    \label{eq:kinetic}
\end{equation}
where $f^S$ is the Shakhov collision target and $\tau$ is the relaxation time determined from the local pressure $p$ and dynamic viscosity $\mu$ through
\begin{equation}
    \tau=\frac{\mu}{p},
    \qquad
    \mu=\mu_{\mathrm{ref}}T^{\omega},
    \label{eq:relaxation_time}
\end{equation}
with viscosity exponent $\omega$. The reference viscosity is linked to the Knudsen number $\Kn$ through
\begin{equation}
    \mu_{\mathrm{ref}}
    =\frac{5(\alpha_{\mathrm{ref}}+1)
            (\alpha_{\mathrm{ref}}+2)\sqrt{\pi}}
           {4\alpha_{\mathrm{ref}}
            (5-2\omega_{\mathrm{ref}})
            (7-2\omega_{\mathrm{ref}})}\,\Kn,
    \label{eq:viscosity}
\end{equation}
where $\alpha_{\mathrm{ref}}$ and $\omega_{\mathrm{ref}}$ are molecular-model constants specified for each case.

To eliminate discretization along $\xi$, two reduced distribution functions are introduced:
\begin{equation}
    \begin{aligned}
        h(\vect{x},\vect{u},t)
        &=\int_{\mathbb R}f(\vect{x},\vect{u},\xi,t)\,\dd\xi,\\
        b(\vect{x},\vect{u},t)
        &=\int_{\mathbb R}\xi^2 f(\vect{x},\vect{u},\xi,t)\,\dd\xi.
    \end{aligned}
    \label{eq:reduction}
\end{equation}
With the peculiar velocity $\vect{c}=\vect{u}-\vect{U}$ and inverse temperature $\beta=1/T$, integrating the local Maxwellian and Shakhov distributions over $\xi$ yields the reduced Maxwellians
\begin{equation}
    H=\frac{\rho\beta}{\pi}e^{-\beta|\vect{c}|^2},
    \qquad
    B=\frac{H}{2\beta},
    \label{eq:maxwell}
\end{equation}
and the corresponding reduced Shakhov targets
\begin{equation}
    H^S=H\left[1+C_q(2\beta|\vect{c}|^2-4)\right],
    \qquad
    B^S=B\left[1+C_q(2\beta|\vect{c}|^2-2)\right],
    \label{eq:shakhov}
\end{equation}
where the heat-flux correction factor reads
\begin{equation}
    C_q=\frac{4(1-\Pran)\beta^2}{5\rho}
         (\vect{c}\cdot\vect{q}).
    \label{eq:shakhov_coefficient}
\end{equation}
Here, $\vect{q}$ denotes the macroscopic heat flux. For the monatomic Shakhov model, the Prandtl number is set to $\Pran=2/3$, whereas $\Pran=1$ recovers the BGK model. Hereafter, $z\in\{h,b\}$ denotes either reduced distribution and $z^S\in\{H^S,B^S\}$ its corresponding collision target.

The molecular velocity space is discretized by $N_v$ quadrature nodes $\vect{u}_k=(u_k,v_k)$ ($k=1,\ldots,N_v$) with weights $w_k$. The physical domain is divided into control volumes $\Omega_c$ of area $A_c$, where each face $f$ has centroid $\vect{x}_f$, length $L_f$, and outward unit normal $\vect{n}_{cf}$ relative to cell $c$. With $\vect{f}_c=(h_{c,k},b_{c,k})_{k=1}^{N_v}$ denoting the discrete reduced distributions in cell $c$, the conservative variables $\vect{W}_c$ and heat flux $\vect{q}_c$ are obtained by taking discrete velocity moments:
\begin{equation}
    \vect{W}_c=\mathcal M\vect{f}_c
    =\sum_{k=1}^{N_v}w_k
    \begin{pmatrix}
        h_{c,k}\\
        u_kh_{c,k}\\
        v_kh_{c,k}\\
        \tfrac12\bigl(|\vect{u}_k|^2h_{c,k}+b_{c,k}\bigr)
    \end{pmatrix},
    \label{eq:momentrecovery}
\end{equation}
\begin{equation}
    \vect{q}_c
    =\frac12\sum_{k=1}^{N_v}w_k(\vect{u}_k-\vect{U}_c)
      \left[|\vect{u}_k-\vect{U}_c|^2h_{c,k}+b_{c,k}\right].
    \label{eq:heat}
\end{equation}

Second-order spatial reconstruction is performed using least-squares gradients with the Venkatakrishnan limiter~\cite{venkatakrishnan1995convergence}. Following the standard UGKS formulation~\cite{xu2010unified,huang2012unified,huang2013unified}, the time-dependent interface distribution is given by the analytical integral solution along characteristics,
\begin{equation}
    \begin{aligned}
        z_f(t,\vect{u})={}&
        e^{-t/\tau_f}z_0(\vect{x}_f-\vect{u}t,\vect{u})\\
        &+\frac{1}{\tau_f}\int_0^t e^{-(t-t')/\tau_f}
          z^S\!\left(\vect{x}_f-\vect{u}(t-t'),\vect{u},t'\right)\,\dd t',
    \end{aligned}
    \label{eq:integral}
\end{equation}
where $z_0$ is the upwind-reconstructed initial distribution and $\tau_f$ is the local interface relaxation time held constant over the evolution. Averaging the microscopic flux over a local physical evolution interval $\vartheta_f$ yields
\begin{equation}
    \Phi^z_{cf,k}
    =\frac{L_f}{\vartheta_f}\int_0^{\vartheta_f}
     (\vect{u}_k\cdot\vect{n}_{cf})\,
     z_f(t,\vect{u}_k)\,\dd t,
    \qquad z\in\{h,b\},
    \label{eq:microflux}
\end{equation}
from which the steady microscopic residual in cell $c$ for velocity node $k$ reads
\begin{equation}
    \mathcal R^z_{c,k}
    =\frac{1}{A_c}\sum_{f\in\partial\Omega_c}\Phi^z_{cf,k}
     -\frac{z^S_{c,k}-z_{c,k}}{\tau_c}.
    \label{eq:microres}
\end{equation}
Applying the moment operator $\mathcal{M}$ to the same interface fluxes yields the macroscopic flux and steady residual,
\begin{equation}
    \vect\Phi^W_{cf}
    =\mathcal M(\vect\Phi^h_{cf},\vect\Phi^b_{cf}),
    \qquad
    \mathcal R^W_c
    =\frac{1}{A_c}\sum_{f\in\partial\Omega_c}\vect\Phi^W_{cf}.
    \label{eq:macrores}
\end{equation}

\subsection{Boundary conditions}\label{sec:boundaries}

Three types of boundary conditions are considered: diffuse reflection at solid walls, far-field boundaries, and translational periodic boundaries. At a solid wall with prescribed velocity $\vect{U}_w$, temperature $T_w$, and inward unit normal $\vect{n}_w$ (pointing into the gas domain), the wall-relative normal velocity is defined as
\begin{equation}
    u^w_{n,k}=(\vect{u}_k-\vect{U}_w)\cdot\vect{n}_w.
    \label{eq:wall_normal_velocity}
\end{equation}
The reduced unit-density wall Maxwellians are
\begin{equation}
    H^{(1)}_{w,k}
    =\frac{\beta_w}{\pi}
      e^{-\beta_w|\vect{u}_k-\vect{U}_w|^2},
    \qquad
    B^{(1)}_{w,k}=\frac{H^{(1)}_{w,k}}{2\beta_w},
    \label{eq:wallmaxwell}
\end{equation}
where $\beta_w=1/T_w$. The incident and reflected distributions satisfy
\begin{equation}
    (h_{w,k},b_{w,k})=
    \begin{cases}
        (h_{\mathrm{in},k},b_{\mathrm{in},k}),
            &u^w_{n,k}<0,\\[2pt]
        \rho_w\bigl(H^{(1)}_{w,k},B^{(1)}_{w,k}\bigr),
            &u^w_{n,k}>0.
    \end{cases}
    \label{eq:wall}
\end{equation}
The wall density $\rho_w$ follows from zero net normal mass flux,
\begin{equation}
    \rho_w=-\frac{J_{\mathrm{in}}}{J^{(1)}_{\mathrm{out}}},
    \label{eq:wallrho}
\end{equation}
where
\begin{equation}
    \begin{aligned}
        J_{\mathrm{in}}
        &=\sum_{u^w_{n,k}<0}w_ku^w_{n,k}h_{\mathrm{in},k},\\
        J^{(1)}_{\mathrm{out}}
        &=\sum_{u^w_{n,k}>0}w_ku^w_{n,k}H^{(1)}_{w,k}.
    \end{aligned}
    \label{eq:wallmassflux}
\end{equation}
At far-field boundaries, incoming distributions are prescribed by the free-stream reduced Maxwellians, whereas outgoing distributions are determined from the adjacent interior solution. For periodic boundaries, identical reduced distributions are enforced across paired faces using a translated reconstruction stencil, ensuring equal and opposite fluxes.

\subsection{Implicit steady-state iteration}\label{sec:implicit}

The implicit UGKS~\cite{zhu2016implicit} is adopted to obtain steady-state solutions. At iteration $n$, the discrete state is composed of the conservative variables $\vect{W}^n$ and reduced distributions $\vect{f}^n$:
\begin{equation}
    \vect{Q}^n=(\vect{W}^n,\vect{f}^n).
    \label{eq:primal_state}
\end{equation}
Each iteration proceeds through three stages: a macroscopic predictor, a microscopic correction, and conservative-moment recovery, together with a mass correction for closed or periodic systems. In this implicit formulation, the numerical pseudo-time step $\Delta\widetilde t_c$ is decoupled from the physical flux-evolution interval $\vartheta_f$ in Eq.~\eqref{eq:microflux}:
\begin{equation}
    \Delta\widetilde t_c=C_{\mathrm{CFL}}\,t_c^{\mathrm{spec}},
    \qquad
    \vartheta_f=C_{\mathrm{flux}}\min\!\left(t_{c_L}^{\mathrm{spec}},t_{c_R}^{\mathrm{spec}}\right),
    \label{eq:fluxtime}
\end{equation}
where $t_c^{\mathrm{spec}}$ is the local characteristic spectral time, $c_L$ and $c_R$ denote the cells adjacent to interface $f$, and $C_{\mathrm{CFL}}$ and $C_{\mathrm{flux}}$ are dimensionless scaling factors. Here, $\Delta\widetilde t_c$ enters only the implicit sweep coefficients, whereas $\vartheta_f$ enters the UGKS interface flux directly and depends on both the flow state and mesh geometry.

First, the macroscopic residual $\mathcal{R}_c^W$ is evaluated, and a macroscopic increment $\Delta\vect{W}$ is computed via forward--backward point-relaxation sweeps using a Lax--Friedrichs (LF) approximate Euler-flux operator. Across an interface shared by cell $c$ and its neighbor $j$, the neighbor contribution is approximated as
\begin{equation}
    \mathcal H_{cj}(\Delta\vect{W}_j)
    =\frac12\left[\vect{F}_E(\vect{W}_j^n+\Delta\vect{W}_j;\vect{n}_{cj})
      -\vect{F}_E(\vect{W}_j^n;\vect{n}_{cj})-r_{cj}\Delta\vect{W}_j\right],
    \label{eq:app_prflux}
\end{equation}
where $\vect{F}_E(\vect{W};\vect{n})$ is the Euler flux along unit normal $\vect{n}$ and $r_{cj}$ is the local spectral-radius dissipation coefficient. Initialized from zero increments, each local point-relaxation update reads
\begin{equation}
    \begin{gathered}
        \Delta\vect{W}_c^{\mathrm{new}}
        =\frac{1}{D_c^W}\left[\vect{b}_c^W-\sum_{j\in\mathcal N(c)}L_{cj}\mathcal H_{cj}\!\left(\Delta\vect{W}_j^{\mathrm{avail}}\right)\right],\\
        D_c^W=\frac{A_c}{\Delta\widetilde t_c}+\frac12\sum_{f\in\partial\Omega_c}L_f r_{cf},
        \qquad
        \vect{b}_c^W=-A_c\mathcal R_c^W,
    \end{gathered}
    \label{eq:app_prvisit}
\end{equation}
where $\mathcal N(c)$ comprises interior neighbors, periodic partners, and boundary ghost cells as applicable, and the superscript $\mathrm{avail}$ denotes the most recently updated neighbor increment at the current stage of the sweep. Denoting the resulting macroscopic prediction operator by $\mathcal{P}_B$, the predicted conservative state is obtained as
\begin{equation}
    \widetilde{\vect{W}}=\vect{W}^n+\theta\,\mathcal P_B(\vect{b}^W),
    \label{eq:macroprediction}
\end{equation}
where $\vect{b}^W$ collects the cellwise area-weighted residuals $\vect{b}_c^W$ and $\theta$ is a relaxation factor.

Next, the predicted state $\widetilde{\vect{W}}$ is used to update the collision source in the microscopic correction stage. With $S^z(\vect{W},\vect{f})=(z^S-z)/\tau$ ($z\in\{h,b\}$) denoting the collision source, the microscopic defect is formed as
\begin{equation}
    r^z=-\mathcal R^z-S^z(\vect{W}^n,\vect{f}^n)
        +S^z(\widetilde{\vect{W}},\vect{f}^n).
    \label{eq:microdefect}
\end{equation}
Collecting $r^h$ and $r^b$ into $\vect{r}$, the provisional reduced distributions are obtained via
\begin{equation}
    \widehat{\vect{f}}
    =\vect{f}^n+\mathcal P_{\mathrm{kin}}\vect{r},
    \label{eq:microprovisional}
\end{equation}
where $\mathcal P_{\mathrm{kin}}$ is the microscopic implicit correction operator, implemented via frozen-coefficient LU-SGS sweeps. For interior and periodic couplings, the forward and backward triangular sweeps and the resulting linear operator take the form
\begin{equation}
    \begin{gathered}
        (\mathsf D+\mathsf L)\vect{y}=\vect{r},
        \qquad
        (\mathsf D+\mathsf U)\Delta\vect{f}=\mathsf D\vect{y},\\
        \mathcal P_{\mathrm{kin}}=\theta(\mathsf D+\mathsf U)^{-1}\mathsf D(\mathsf D+\mathsf L)^{-1},
    \end{gathered}
    \label{eq:app_kinmatrix}
\end{equation}
where $\mathsf D$, $\mathsf L$, and $\mathsf U$ represent the diagonal, strictly lower-triangular, and strictly upper-triangular parts of the frozen upwind microscopic operator. In practice, $\mathcal P_{\mathrm{kin}}$ is applied matrix-free through successive triangular sweeps without forming inverse matrices; the explicit sweep coefficients and diffuse-wall increment coupling are detailed in \ref{app:sweeps}.

Finally, the accepted reduced distributions $\vect f^+$ are determined prior to conservative-moment recovery. For open external flows, the provisional distributions are accepted directly ($\vect f^+=\widehat{\vect f}$). For closed and periodic systems, the prescribed total gas mass $M_\star$ is strictly enforced:
\begin{equation}
    \sum_c A_c\rho_c=M_\star.
    \label{eq:massconstraint}
\end{equation}
With the mass row operator $\vect{m}^T$ defined by
\begin{equation}
    \vect{m}^T\vect{f}
    =\sum_c\sum_{k=1}^{N_v}A_cw_kh_{c,k},
    \label{eq:massrow}
\end{equation}
the linear kinetic response to a normalized density source $\vect s$ (defined in \ref{app:sweeps}) and its total mass moment are evaluated as
\begin{equation}
    \vect z_M=\mathcal P_{\mathrm{kin}}\vect s,
    \qquad
    \sigma=\vect m^T\vect z_M.
    \label{eq:massresponse}
\end{equation}
For $\sigma\neq0$, the mass-projected distributions are computed via
\begin{equation}
    \begin{gathered}
        \chi=\frac{\vect m^T\widehat{\vect f}-M_\star}{\sigma},\\
        \vect f^+=\widehat{\vect f}-\vect z_M\chi.
    \end{gathered}
    \label{eq:massschur}
\end{equation}
In all cases, the updated conservative variables are then recovered as $\vect W^+=\mathcal M\vect f^+$. Further details of the scalar mass correction and its linearization are provided in \ref{app:sweeps} and \ref{app:transpose}.

Combining the macroscopic predictor, microscopic correction, and moment recovery (with mass projection where applicable) yields the single-step implicit update operator $G$:
\begin{equation}
    \vect Q^{n+1}
    =G(\vect Q^n,\vect X,\vect{\mathcal B})
    =(\vect W^+,\vect f^+),
    \label{eq:fixedpointmap}
\end{equation}
where $\vect X$ denotes the mesh-node coordinates and $\vect{\mathcal B}$ the prescribed boundary data. At steady-state convergence, the discrete flow state satisfies the fixed-point relation
\begin{equation}
    \vect Q=G(\vect Q,\vect X,\vect{\mathcal B}).
    \label{eq:primal_fixedpoint}
\end{equation}

\section{Discrete adjoint formulation for the implicit UGKS}\label{sec:method}

This section formulates the matrix-free discrete adjoint method for the steady implicit UGKS and the associated shape optimization framework. First, the fixed-point tangent-linear and adjoint equations and their stepwise duality properties are established. The hybrid differentiation strategy is then presented in two complementary parts: local automatic differentiation (AD) for nonlinear kinetic and geometric kernels, and analytical tangent and transpose operators for the coupled implicit updates. Finally, the constrained shape optimization procedure is outlined.

\subsection{Fixed-point adjoint formulation and discrete duality}\label{sec:discretefixedpoint}

For a converged steady state satisfying Eq.~\eqref{eq:primal_fixedpoint}, a scalar objective function $I(\vect{Q},\vect{X},\vect{\mathcal B})$ is considered. Let $\vect{a}$ and $\vect{\eta}$ denote the geometric and boundary design parameters, respectively. The mesh coordinates and prescribed boundary data, together with their first-order variations along arbitrary perturbation directions $\vect{a}_d$ and $\vect{\eta}_d$, are written as
\begin{equation}
    \begin{gathered}
        \vect{X}=P(\vect{a}),
        \qquad
        \vect{\mathcal B}=V(\vect{\eta}),\\
        \vect{X}_d=P_a\vect{a}_d,
        \qquad
        \vect{\mathcal B}_d=V_{\eta}\vect{\eta}_d,
    \end{gathered}
    \label{eq:controldirections}
\end{equation}
where $P$ and $V$ denote the parameterization functions, with Jacobians $P_a=\partial\vect{X}/\partial\vect{a}$ and $V_{\eta}=\partial\vect{\mathcal B}/\partial\vect{\eta}$.

\begin{figure}[h]
    \centering
    \includegraphics[width=0.85\textwidth]{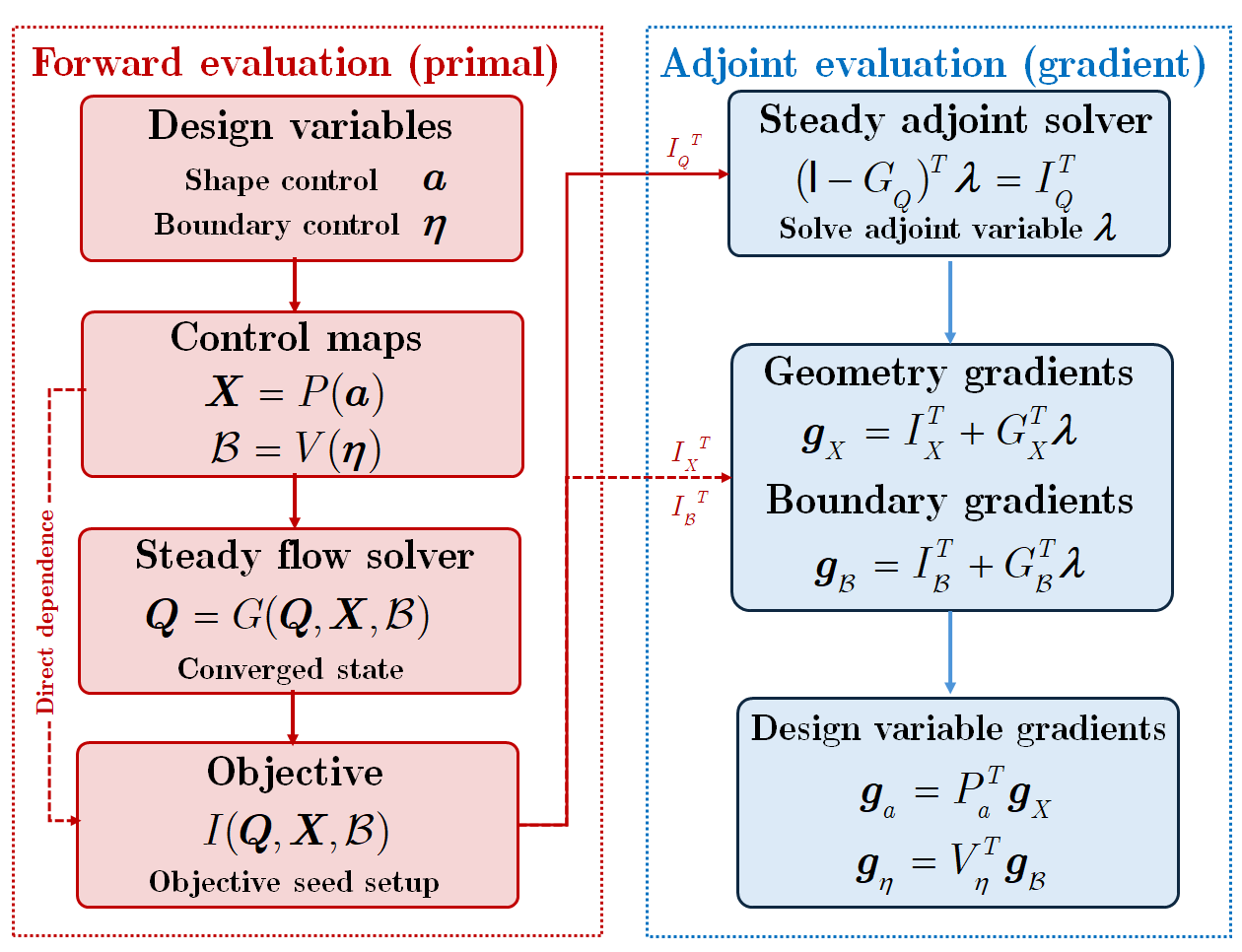}
    \caption{Schematic of the forward primal evaluation and discrete adjoint gradient-assembly workflow for shape and boundary parameters.}
    \label{fig:flowchart}
\end{figure}

Linearizing Eq.~\eqref{eq:primal_fixedpoint} and the objective function yields the discrete tangent-linear system and total objective variation:
\begin{equation}
    \begin{gathered}
        (\mathsf I-G_Q)\vect{Q}_d
        =G_X\vect{X}_d+G_{\mathcal B}\vect{\mathcal B}_d,\\
        I_d=I_Q\vect{Q}_d+I_X\vect{X}_d+I_{\mathcal B}\vect{\mathcal B}_d,
    \end{gathered}
    \label{eq:fixedpointtangent}
\end{equation}
where $G_Q$, $G_X$, and $G_{\mathcal B}$ denote the partial Jacobians of the update operator $G$ with respect to the discrete state, mesh coordinates, and boundary data, respectively. To eliminate the explicit dependence on the state variation $\vect{Q}_d$, the discrete adjoint state $\vect{\lambda}$ is defined by the transposed linear system
\begin{equation}
    (\mathsf I-G_Q)^T\vect{\lambda}=I_Q^T.
    \label{eq:fixedpointadjoint}
\end{equation}
Substituting Eq.~\eqref{eq:fixedpointadjoint} into Eq.~\eqref{eq:fixedpointtangent} yields the sensitivity gradients with respect to the mesh coordinates and boundary data,
\begin{equation}
    \vect{g}_X=I_X^T+G_X^T\vect{\lambda},
    \qquad
    \vect{g}_{\mathcal B}=I_{\mathcal B}^T+G_{\mathcal B}^T\vect{\lambda}.
    \label{eq:physicalgradients}
\end{equation}
Projecting these sensitivities onto the design spaces through the transposed parameterization Jacobians gives the design gradients and directional derivative:
\begin{equation}
    \begin{gathered}
        \vect{g}_a=P_a^T\vect{g}_X,
        \qquad
        \vect{g}_{\eta}=V_{\eta}^T\vect{g}_{\mathcal B},\\
        I_d=\vect{g}_a^T\vect{a}_d+\vect{g}_{\eta}^T\vect{\eta}_d.
    \end{gathered}
    \label{eq:controlgradients}
\end{equation}
Figure~\ref{fig:flowchart} summarizes this forward evaluation and discrete adjoint gradient-assembly workflow. Whereas the tangent-linear system in Eq.~\eqref{eq:fixedpointtangent} must be solved once per perturbation direction, a single solution of Eq.~\eqref{eq:fixedpointadjoint} provides the complete gradient vector simultaneously.

To solve Eqs.~\eqref{eq:fixedpointtangent} and~\eqref{eq:fixedpointadjoint} iteratively without assembling global Jacobians, the direct variations from mesh coordinates and boundary data are collected as
\begin{equation}
    \vect b=G_X\vect X_d+G_{\mathcal B}\vect{\mathcal B}_d,
    \qquad
    I_{\mathrm{dir}}=I_X\vect X_d+I_{\mathcal B}\vect{\mathcal B}_d.
    \label{eq:direct_variations}
\end{equation}
At the converged steady state, the stationary tangent-linear and discrete adjoint iterations read
\begin{equation}
    \vect Q_d^{\,m+1}=G_Q\vect Q_d^{\,m}+\vect b,
    \qquad
    \vect\lambda^{\,m+1}=G_Q^T\vect\lambda^{\,m}+I_Q^T.
    \label{eq:dual_iterations}
\end{equation}
When $G_Q^T$ is constructed as the exact discrete transpose of $G_Q$, both linear iterations share identical spectral radii and inherit the multiscale convergence behavior of the implicit UGKS solver. Moreover, starting from zero initial vectors ($\vect Q_d^{\,0}=\vect 0$ and $\vect\lambda^{\,0}=\vect 0$), the cumulative directional sensitivities after $m$ iterations satisfy
\begin{equation}
    I_d^{\mathrm{TL},m}
    =I_{\mathrm{dir}}+I_Q\sum_{\ell=0}^{m-1}G_Q^\ell\vect b
    =I_{\mathrm{dir}}
       +\left(\sum_{\ell=0}^{m-1}(G_Q^T)^\ell I_Q^T\right)^T\vect b
    =I_d^{\mathrm{Adj},m}.
    \label{eq:stepwise_duality}
\end{equation}
Thus, the tangent-linear and adjoint sensitivities are algebraically identical at every matched iteration step $m$. Verifying this stepwise equivalence against the tangent-linear solver serves as a key validation step for discrete adjoint implementations~\cite{giles2003algorithm}, confirming discrete duality across all flow regimes.

\subsection{Local automatic differentiation for nonlinear kinetic and geometric kernels}
\label{sec:local_ad}\label{sec:shape_boundary_derivatives}

Evaluating the matrix-free actions of $G_Q$, $G_X$, $G_{\mathcal B}$ and their transposes requires balancing discrete consistency against memory cost. To achieve this balance, a hybrid differentiation strategy is adopted by partitioning the single-step operator $G$ into two layers: local AD is applied to cell- and face-level nonlinear kinetic and geometric kernels in this subsection, whereas analytical tangent and transpose operators handle the coupled macroscopic--microscopic implicit updates in Section~\ref{sec:implicit_derivatives}.

Within the local AD layer, each cell- or face-level nonlinear operation, including mesh-metric evaluation, limited spatial reconstruction, UGKS interface flux integration, and kinetic boundary reflection, is represented by a local kernel $\mathcal{K}_\ell$:
\begin{equation}
    \vect{\phi}_\ell
    =\mathcal{K}_\ell\!\left(\vect{Q}_\ell,\vect{\mathcal{G}}_\ell,\vect{\mathcal{B}}_\ell\right),
    \label{eq:local_ad_kernel}
\end{equation}
where $\vect{\phi}_\ell$ denotes the evaluated local quantity, such as a mesh metric, reconstructed value, interface flux, or boundary distribution. The inputs $\vect{Q}_\ell$, $\vect{\mathcal{G}}_\ell$, and $\vect{\mathcal{B}}_\ell$ collect the corresponding local flow states, geometric data, and prescribed boundary data, respectively. Along the active branch at the converged state, forward-mode local AD evaluates the directional variation from local input perturbations:
\begin{equation}
    \vect{\phi}_{\ell,d}
    =\mathcal{K}_{\ell,Q}\vect{Q}_{\ell,d}
     +\mathcal{K}_{\ell,\mathcal{G}}\vect{\mathcal{G}}_{\ell,d}
     +\mathcal{K}_{\ell,\mathcal{B}}\vect{\mathcal{B}}_{\ell,d},
    \label{eq:local_ad_jvp}
\end{equation}
where $\mathcal{K}_{\ell,Q}$, $\mathcal{K}_{\ell,\mathcal{G}}$, and $\mathcal{K}_{\ell,\mathcal{B}}$ denote the local partial Jacobians of $\mathcal{K}_\ell$ with respect to $\vect{Q}_\ell$, $\vect{\mathcal{G}}_\ell$, and $\vect{\mathcal{B}}_\ell$, respectively. In reverse mode, given an output adjoint seed $\overline{\vect{\phi}}_\ell$, a single reverse AD pass over the same local kernel computes the vector--Jacobian products:
\begin{equation}
    \overline{\vect{Q}}_\ell
    =\mathcal{K}_{\ell,Q}^T\overline{\vect{\phi}}_\ell,
    \qquad
    \overline{\vect{\mathcal{G}}}_\ell
    =\mathcal{K}_{\ell,\mathcal{G}}^T\overline{\vect{\phi}}_\ell,
    \qquad
    \overline{\vect{\mathcal{B}}}_\ell
    =\mathcal{K}_{\ell,\mathcal{B}}^T\overline{\vect{\phi}}_\ell.
    \label{eq:local_ad_vjp}
\end{equation}
These local adjoint seeds are immediately accumulated into the corresponding global state, geometric, and boundary arrays, after which the local computational tape is released. Because no iteration-level tape is retained, the memory requirement does not grow with the number of nonlinear iterations.

A key feature of UGKS is the multiscale interface flux in Eq.~\eqref{eq:microflux}, which couples free molecular transport and collisional relaxation over the local physical evolution interval $\vartheta_f$. Because $\vartheta_f$ in Eq.~\eqref{eq:fluxtime} depends directly on both the flow state and mesh geometry, differentiating this evolution interval is essential for accurate sensitivities. Expressing the collected discrete residuals as $\vect{\mathcal R}=\mathscr R(\vect{Q},\vect{X},\vect{\mathcal B},\vect{\vartheta})$ with $\vect{\vartheta}=\mathcal T(\vect{Q},\vect{X})$, the chain rule yields
\begin{equation}
    \begin{gathered}
        \vect{\mathcal R}_d=\mathscr R_Q\vect{Q}_d+\mathscr R_X\vect{X}_d+\mathscr R_{\mathcal B}\vect{\mathcal B}_d+\mathscr R_{\vartheta}\vect{\vartheta}_d,\\
        \vect{\vartheta}_d=\mathcal T_Q\vect{Q}_d+\mathcal T_X\vect{X}_d,
    \end{gathered}
    \label{eq:residual_time_chain}
\end{equation}
where the partial derivatives of $\mathscr R$ are evaluated at fixed $\vect{\vartheta}$. In reverse mode, given a residual adjoint seed $\overline{\vect{\mathcal R}}$, the corresponding contributions back-propagated through the local AD kernels are given by
\begin{equation}
    \begin{gathered}
        \overline{\vect{\vartheta}}=\mathscr R_{\vartheta}^T\overline{\vect{\mathcal R}},
        \qquad
        \overline{\vect{\mathcal B}}^{\mathcal R}=\mathscr R_{\mathcal B}^T\overline{\vect{\mathcal R}},\\
        \overline{\vect{Q}}^{\mathcal R}=\mathscr R_Q^T\overline{\vect{\mathcal R}}+\mathcal T_Q^T\overline{\vect{\vartheta}},
        \qquad
        \overline{\vect{X}}^{\mathcal R}=\mathscr R_X^T\overline{\vect{\mathcal R}}+\mathcal T_X^T\overline{\vect{\vartheta}}.
    \end{gathered}
    \label{eq:residual_time_adjoint}
\end{equation}

The remaining nonlinear operator, including mesh metrics, limited spatial reconstruction, and kinetic boundary conditions, are differentiated uniformly through the local AD kernels in Eqs.~\eqref{eq:local_ad_jvp}--\eqref{eq:local_ad_vjp}. In particular, at diffuse-reflection solid walls, perturbations in the incident distributions, wall unit normal $\vect{n}_w$, and prescribed wall temperature $T_w$ are propagated consistently through the zero-net-mass-flux normalization in Eq.~\eqref{eq:wallrho}. At far-field and periodic boundaries, incoming and outgoing velocity half-ranges as well as periodic translation vectors and paired reconstruction stencils are differentiated directly without introducing auxiliary boundary variables. The explicit derivative expressions for mesh metrics, spatial reconstruction, $\vartheta_f$, kinetic boundaries, and objective normalizations are documented in \ref{app:geometry}.

\subsection{Analytical tangent and transpose of the implicit updates}\label{sec:implicit_derivatives}

While local AD differentiates the nonlinear residual kernels, the coupled macroscopic--microscopic implicit updates are differentiated analytically. Because the implicit sweeps act only as a preconditioner without affecting the converged steady state, derivatives of the frozen sweep coefficients multiply vanishing increments and drop out identically. Consequently, sensitivities within the implicit sweeps enter exclusively through the physical residuals and collision sources, while closed and periodic domains retain an additional direct cell-area contribution from the mass-conservation projection.

For the macroscopic predictor in Eqs.~\eqref{eq:app_prflux}--\eqref{eq:macroprediction}, linearizing the Lax--Friedrichs neighbor flux difference at zero increment yields
\begin{equation}
    \mathsf H_{cj}=\frac{L_{cj}}{2}\left[\mathsf A_{E,j}(\vect{n}_{cj})-r_{cj}\mathsf I\right],
    \label{eq:pr_tangent_flux}
\end{equation}
where $\mathsf A_{E,j}(\vect{n}_{cj})$ denotes the Jacobian of the Euler flux $\vect{F}_E(\vect{W}_j;\vect{n}_{cj})$ with respect to $\vect{W}_j$. The linearized local point-relaxation update and predicted macroscopic variation read
\begin{equation}
    \begin{gathered}
        \Delta\vect{W}_{c,d}^{\mathrm{new}}=\frac{1}{D_c^W}\left[-A_c(\mathcal R_c^W)_d-\sum_{j\in\mathcal N(c)}\mathsf H_{cj}\Delta\vect{W}_{j,d}^{\mathrm{avail}}\right],\\
        \widetilde{\vect{W}}_d=\vect{W}_d+\theta\Delta\vect{W}_d,
    \end{gathered}
    \label{eq:pr_tangent_visit}
\end{equation}
where $(\mathcal R_c^W)_d$ is the directional derivative of the UGKS macroscopic residual and the sweeps follow the exact forward--backward ordering of the primal solver. In reverse mode, the direct predictor term passes $\overline{\widetilde{\vect{W}}}$ to the macroscopic state seed, while $\theta\overline{\widetilde{\vect{W}}}$ initializes the transposed point-relaxation sweeps. Traversing the cells in strictly reverse sweep order, each local adjoint update contributes to the residual and neighbor seeds via
\begin{equation}
    \left(\overline{\mathcal R}_c^W\right)_{\mathrm{visit}}
    =-\frac{A_c}{D_c^W}\overline{\Delta\vect{W}}_c^{\mathrm{new}},
    \qquad
    \left(\overline{\Delta\vect{W}}_j^{\mathrm{avail}}\right)_{\mathrm{visit}}
    =-\frac{1}{D_c^W}\mathsf H_{cj}^T\overline{\Delta\vect{W}}_c^{\mathrm{new}}.
    \label{eq:pr_transpose_visit}
\end{equation}

For the microscopic correction, let $S^z$ and $\widetilde S^z$ denote the collision sources evaluated at $(\vect{W},\vect{f})$ and $(\widetilde{\vect{W}},\vect{f})$, respectively. Differentiating Eqs.~\eqref{eq:microdefect}--\eqref{eq:microprovisional} yields the linearized defect, collision source, and provisional kinetic variation:
\begin{equation}
    \begin{gathered}
        (r^z)_d=-(\mathcal R^z)_d-(S^z)_d+(\widetilde S^z)_d,
        \qquad
        (S^z)_d=\frac{(z^S)_d-z_d}{\tau}-\frac{z^S-z}{\tau^2}\tau_d,\\
        \widehat{\vect{f}}_d=\vect{f}_d+\mathcal P_{\mathrm{kin}}\vect{r}_d,
    \end{gathered}
    \label{eq:micro_tangent_update}
\end{equation}
where $z\in\{h,b\}$ and the variations of $z^S$ and $\tau$ are evaluated from the corresponding macroscopic state. In reverse mode, an adjoint seed $\overline{\widehat{\vect{f}}}$ on the provisional distribution is passed directly to the microscopic state seed and propagated to the defect seed through the transposed LU-SGS operator in Eq.~\eqref{eq:app_kinmatrix}:
\begin{equation}
    \begin{gathered}
        \overline{\vect{r}}=\mathcal P_{\mathrm{kin}}^T\overline{\widehat{\vect{f}}},\\
        (\mathsf D+\mathsf U)^T\vect{\zeta}=\overline{\widehat{\vect{f}}},
        \qquad
        (\mathsf D+\mathsf L)^T\overline{\vect{r}}=\theta\mathsf D^T\vect{\zeta},
    \end{gathered}
    \label{eq:micro_transpose_sweeps}
\end{equation}
where $\vect{\zeta}$ denotes an intermediate adjoint vector and diffuse-wall increment coupling is incorporated within the reverse sweeps. The defect seed $\overline{\vect{r}}$ is then back-propagated to the microscopic residual seed $-\overline{\vect{r}}$ and through the collision sources $S^z$ and $\widetilde S^z$ to supply $\overline{\widetilde{\vect{W}}}$ for the transposed macroscopic predictor.

For closed and periodic systems enforcing the mass constraint in Eq.~\eqref{eq:massconstraint}, the converged scalar mass correction maintains zero total-mass variation under state and mesh perturbations. Combining the accepted-state adjoint seeds through conservative-moment recovery yields the transposed mass projection:
\begin{equation}
    \begin{gathered}
        \vect{p}=\overline{\vect{f}}^{+}+\mathcal M^T\overline{\vect{W}}^{+},
        \qquad
        \nu=\frac{\vect{z}_M^T\vect{p}}{\sigma},\\
        \overline{\widehat{\vect{f}}}=\vect{p}-\vect{m}\nu,
        \qquad
        \overline{A}_c^{\mathrm{mass}}=-\rho_c\nu,
    \end{gathered}
    \label{eq:massborderadjoint}
\end{equation}
where $\overline{A}_c^{\mathrm{mass}}$ accounts for nonuniform local cell-area variations under mesh deformation and is accumulated with the geometric sensitivities prior to nodal back-propagation. Complete tangent and transpose derivations for the implicit sweeps, diffuse-wall coupling, and mass projection are provided in \ref{app:transpose}.

\subsection{Constrained shape optimization}\label{sec:optimization_procedure}

For gradient-based shape optimization, a scaled objective function and its design gradient are defined as
\begin{equation}
    F(\vect a)=s_I\frac{I(\vect a)}{I_0},
    \qquad
    \nabla_{\vect a}F=\frac{s_I}{I_0}\vect g_a,
    \label{eq:optimization_objective}
\end{equation}
where $I_0$ denotes the baseline objective value, $s_I=-1$ corresponds to mass-flow maximization, and $s_I=1$ corresponds to aerodynamic drag minimization. The constrained optimization problem is solved using sequential least-squares quadratic programming (SLSQP)~\cite{kraft1988software}.

Throughout the optimization, the shape parameterization deforms the physical mesh according to Eq.~\eqref{eq:controldirections}, while the boundary states, molecular velocity quadrature, and mesh topology are held fixed. At each trial design $\vect a$, the steady UGKS state is converged, the discrete adjoint system is solved via the hybrid matrix-free formulation in Sections~\ref{sec:local_ad} and~\ref{sec:implicit_derivatives}, and the accumulated nodal sensitivities are projected onto the design space through Eq.~\eqref{eq:controlgradients}. The optimization procedure is summarized in Algorithm~\ref{alg:optimization}.

\begin{algorithm}[!htbp]
\small
\LinesNumbered
\DontPrintSemicolon
\KwIn{Reference mesh, shape parameterization $P(\vect{a})$, flow conditions, and geometric constraints.}
\KwOut{Optimized shape parameters $\vect{a}^*$ and corresponding objective value $I(\vect{a}^*)$.}
Initialize the mesh-deformation operator, baseline objective $I_0$, and SLSQP optimizer\;
\While{SLSQP convergence criteria are not satisfied}{
    Obtain trial shape parameters $\vect{a}$ from SLSQP\;
    Update $\vect{X}=P(\vect{a})$ and evaluate geometric constraints and Jacobians\;
    Converge the steady UGKS state $\vect{Q}$ [Eq.~\eqref{eq:primal_fixedpoint}]\;
    Evaluate $I(\vect{Q},\vect{X})$ and its partial derivatives $I_Q$ and $I_X$\;
    Solve the discrete adjoint system $(\mathsf{I}-G_Q)^T\vect{\lambda}=I_Q^T$ [Eqs.~\eqref{eq:fixedpointadjoint} and~\eqref{eq:dual_iterations}]\;
    Assemble $\vect{g}_X=I_X^T+G_X^T\vect{\lambda}$ and project $\vect{g}_a=P_a^T\vect{g}_X$ [Eqs.~\eqref{eq:physicalgradients}--\eqref{eq:controlgradients}]\;
    Supply $F(\vect{a})$, $\nabla_{\vect{a}}F$ [Eq.~\eqref{eq:optimization_objective}], and constraints to SLSQP\;
}
\caption{Discrete-adjoint-based constrained shape optimization.}
\label{alg:optimization}
\end{algorithm}

In addition to explicit parameter bounds, problem-specific geometric constraints are imposed on the total domain area and local cross-sectional dimensions (channel width or airfoil thickness) at prescribed reference stations. For closed and periodic systems, the geometric area-preservation constraint confines the admissible design manifold and remains distinct from the physical mass constraint in Eq.~\eqref{eq:massconstraint}, which enforces total gas mass conservation in the flow equations.
\section{Numerical results}\label{sec:results}

The numerical assessment first verifies discrete sensitivity accuracy and exact duality in cavity flows spanning the continuum to free-molecular regimes, and subsequently demonstrates constrained shape optimization for a thermally driven Knudsen pump and a supersonic airfoil. In the verification studies, tangent-linear (TL) and discrete adjoint sensitivities are compared at matched linear iteration counts to confirm stepwise discrete duality [Eq.~\eqref{eq:stepwise_duality}]. Independent central finite-difference (FD) benchmarks along a perturbation direction $\vect d$ are evaluated as
\begin{equation}
    I_d^{\mathrm{FD}}=
    \frac{I(\vect a+\epsilon\vect d)-I(\vect a-\epsilon\vect d)}
         {2\epsilon},
    \label{eq:central_fd}
\end{equation}
where $\epsilon$ denotes the perturbation step size and each perturbed configuration is converged independently to steady state; perturbations to the boundary-temperature parameters $\vect{\eta}$ are evaluated analogously. The relative discrepancy between the converged adjoint derivative and the finite-difference benchmark is quantified by $E_{\mathrm{Adj},FD}=|I_d^{\mathrm{Adj}}-I_d^{\mathrm{FD}}|/|I_d^{\mathrm{FD}}|$.

\subsection{Cavity-flow verification}\label{sec:cavity_verification}

A unit square cavity of side length $L$ is driven by an upper lid moving at a constant speed $U_w=0.15$, while the remaining three walls are stationary. All four walls are maintained at a reference temperature $T_w=1$ with fully diffuse reflection imposed. The physical domain is discretized using a uniform $60\times60$ quadrilateral mesh, and the viscosity follows the variable hard sphere (VHS) model with $\omega=0.81$ and $\alpha_{\mathrm{ref}}=1$. Three flow conditions spanning the continuum to free-molecular regimes are considered, characterized by the Reynolds number ($\mathrm{Re}$, based on $U_w$ and $L$) and the Knudsen number ($\Kn$): $\mathrm{Re}=1000$, $\Kn=0.075$, and $\Kn=10$. The case-specific flow parameters, reference-viscosity exponents, and velocity-space quadratures are summarized in Table~\ref{tab:cavity_cases}. Throughout all simulations, the kinetic boundary fluxes and limited reconstructions are differentiated consistently across the primal, tangent-linear, and adjoint solvers.

\begin{table}[htbp]
    \centering
    \footnotesize
    \caption{Simulation parameters for the three lid-driven cavity test cases.}
    \label{tab:cavity_cases}
    \setlength{\tabcolsep}{6pt}
    \begin{tabular}{@{}lccc@{}}
        \toprule
        Condition & $\Kn$ & $\omega_{\mathrm{ref}}$ & Velocity-space quadrature\\
        \midrule
        $\mathrm{Re}=1000$ & $2.05189\times10^{-4}$ & $0.81$ & $28\times28$ Gauss--Hermite\\
        $\Kn=0.075$ & $0.075$ & $0.5$ & $28\times28$ Gauss--Hermite\\
        $\Kn=10$ & $10$ & $0.81$ & $61\times61$ uniform on $[-4,4]^2$ (Newton--Cotes)\\
        \bottomrule
    \end{tabular}
\end{table}

\begin{figure}[htbp]
    \centering
    \begin{minipage}[t]{0.49\linewidth}
        \centering
        \includegraphics[width=\linewidth]{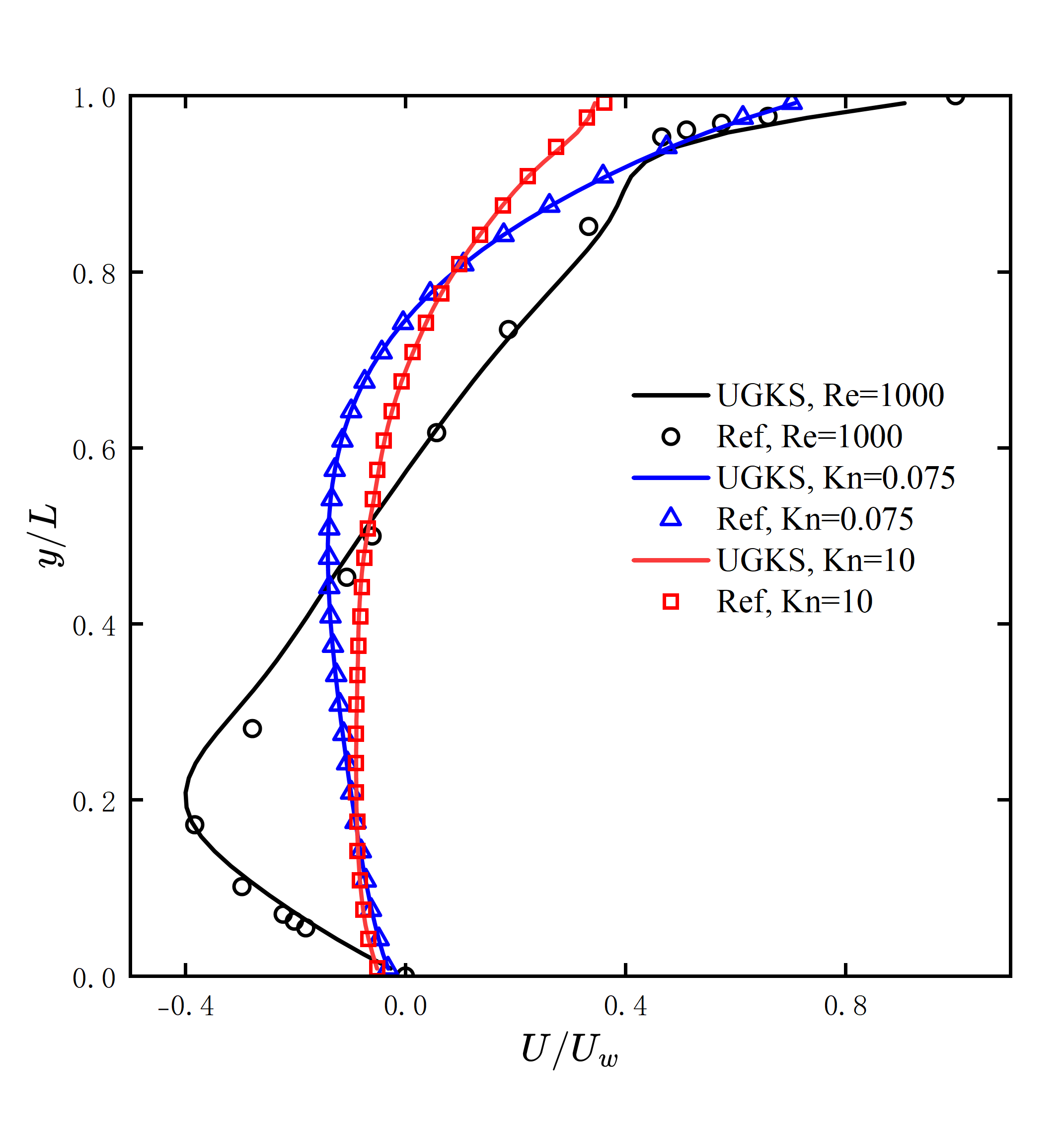}
        \par\smallskip{\small (a) $U/U_w$ at $x/L=0.5$}
    \end{minipage}\hfill
    \begin{minipage}[t]{0.49\linewidth}
        \centering
        \includegraphics[width=\linewidth]{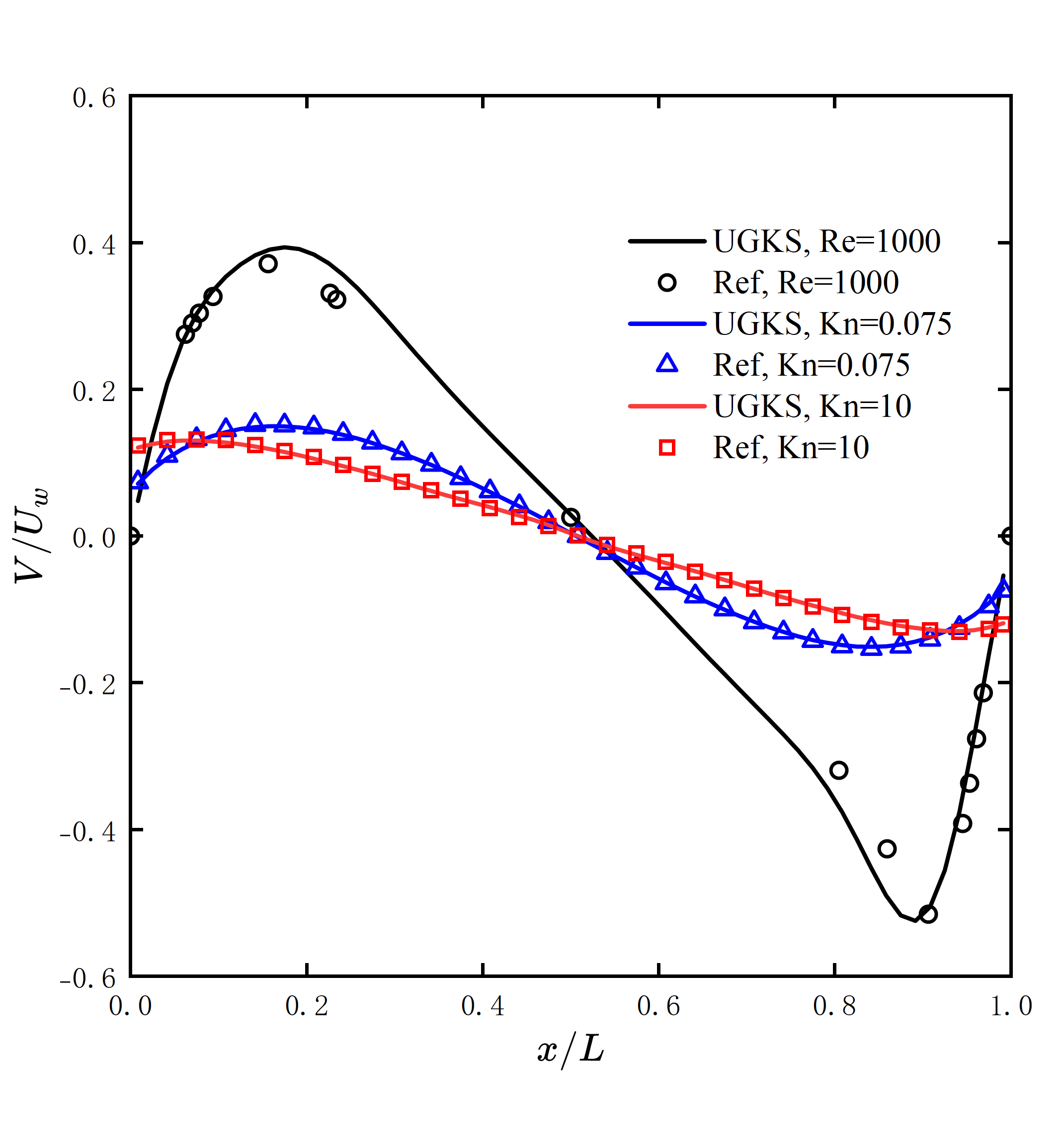}
        \par\smallskip{\small (b) $V/U_w$ at $y/L=0.5$}
    \end{minipage}
    \caption{Cavity centerline velocity profiles: present UGKS solutions (solid lines) compared with benchmark data (symbols)~\cite{ghia1982high,john2011effects,huang2012unified}.}
    \label{fig:cavity_centerline_profiles}
\end{figure}

Figure~\ref{fig:cavity_centerline_profiles} compares the normalized centerline velocity profiles against published benchmark datasets~\cite{ghia1982high,john2011effects,huang2012unified}. Across all three regimes, the present UGKS solutions reproduce the primary flow structures and agree closely with the reference data, exhibiting only minor local discrepancies near the velocity extrema.

For sensitivity verification, the design variable is taken as the uniform temperature of the west wall, $T_{\mathrm{west}}$, and the quantity of interest is the domain-integrated heat-flux magnitude,
\begin{equation}
    I_{\mathrm{cav}}=\sum_c A_c|\vect q_c|,
    \qquad
    I_d=\frac{\dd I_{\mathrm{cav}}}{\dd T_{\mathrm{west}}},
    \label{eq:cavity_objective}
\end{equation}
while the cavity geometry and lid velocity remain fixed. This perturbation direction directly tests the differentiation of the diffuse-wall reflection distribution and its wall-density normalization, followed by the coupled kinetic transport response across the domain. In all three cases, the prescribed total gas inventory is held fixed across the primal, tangent-linear, and adjoint calculations as well as the finite-difference benchmarks, with the mass constraint and its derivatives treated as described in Section~\ref{sec:implicit_derivatives} and \ref{app:transpose}.

\begin{figure}[htbp]
    \centering
    \begin{minipage}[t]{0.325\linewidth}
        \centering
        \includegraphics[width=\linewidth]{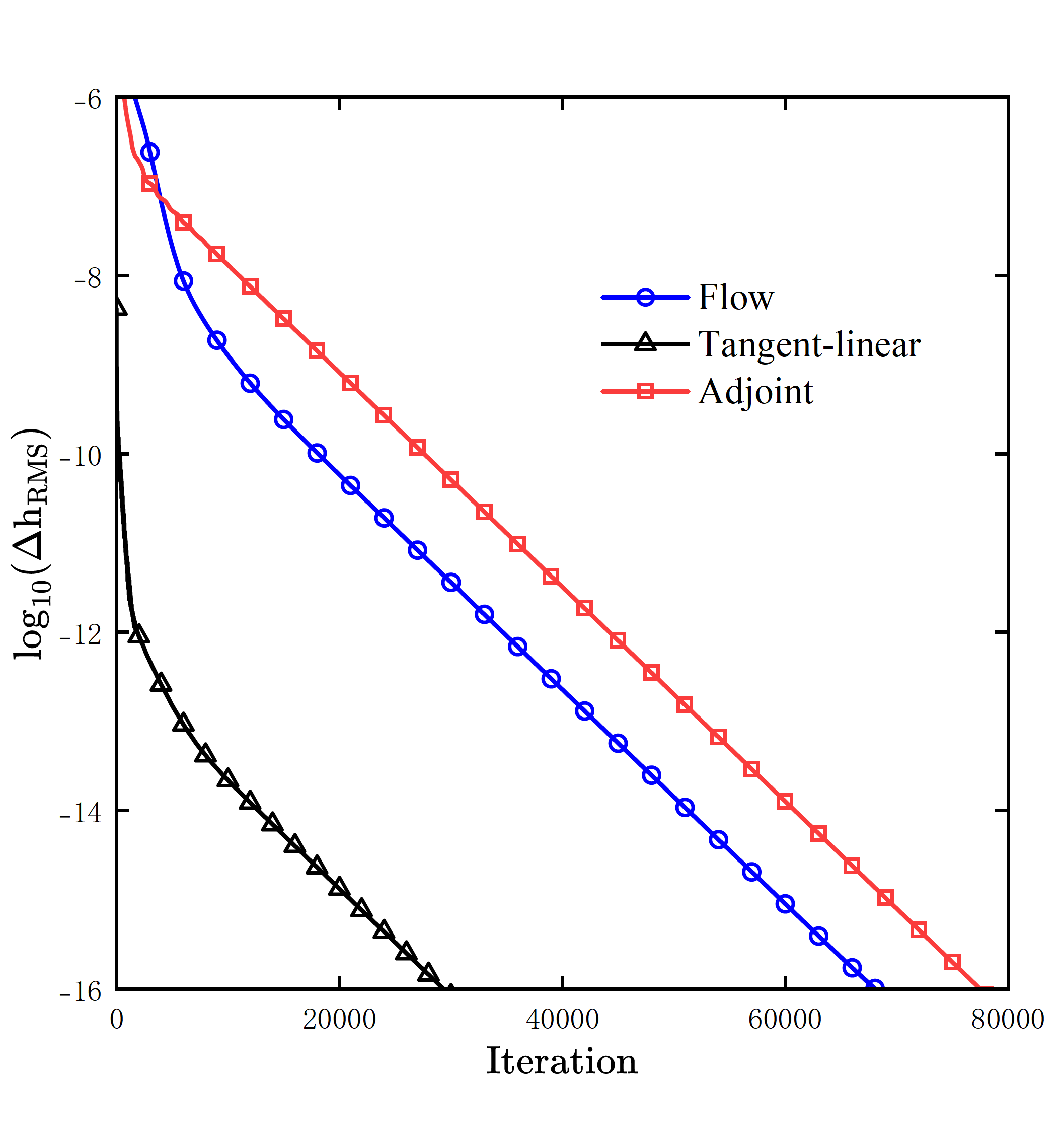}
        \par\smallskip{\small (a) $\mathrm{Re}=1000$}
    \end{minipage}\hfill
    \begin{minipage}[t]{0.325\linewidth}
        \centering
        \includegraphics[width=\linewidth]{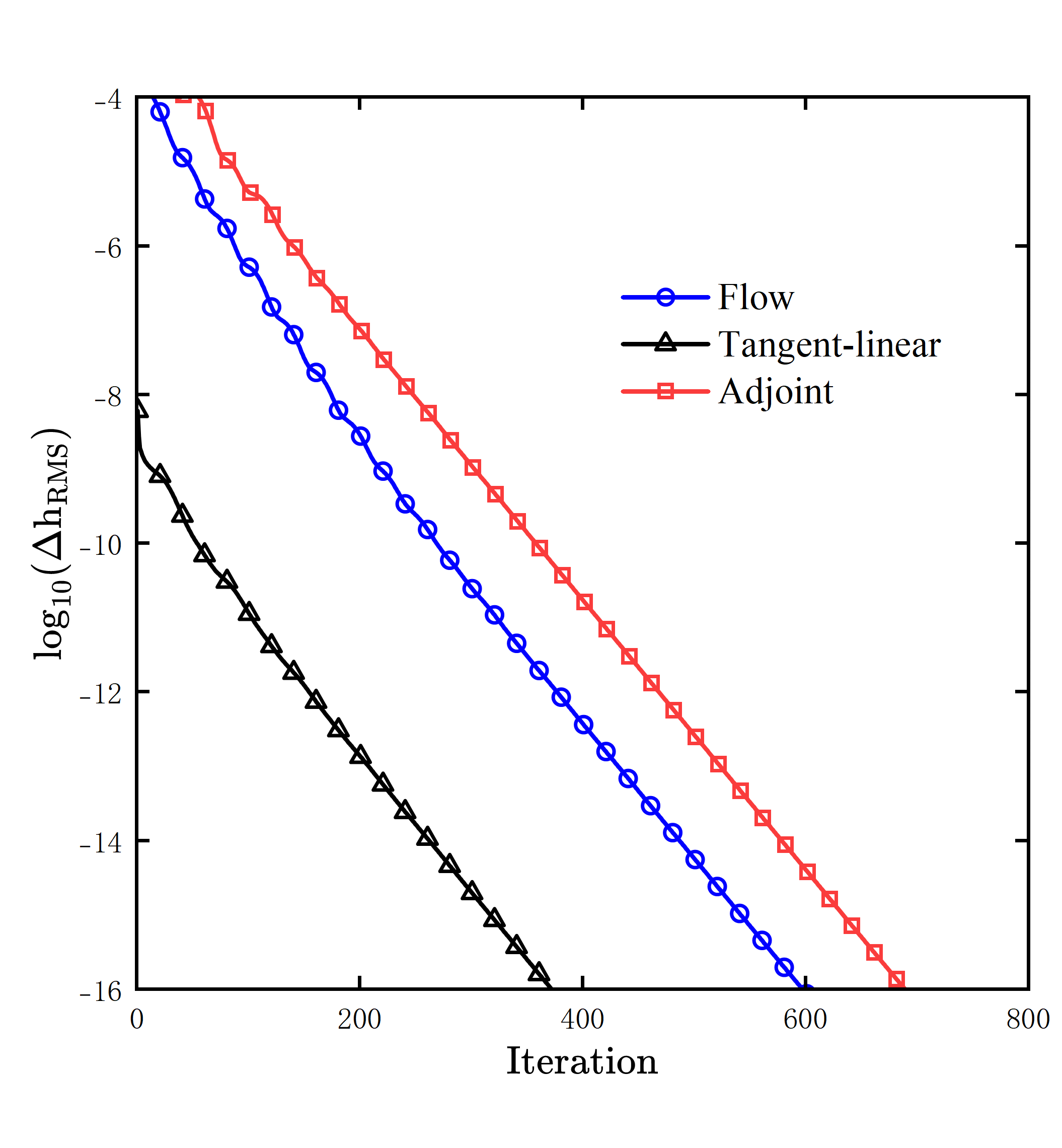}
        \par\smallskip{\small (b) $\Kn=0.075$}
    \end{minipage}\hfill
    \begin{minipage}[t]{0.325\linewidth}
        \centering
        \includegraphics[width=\linewidth]{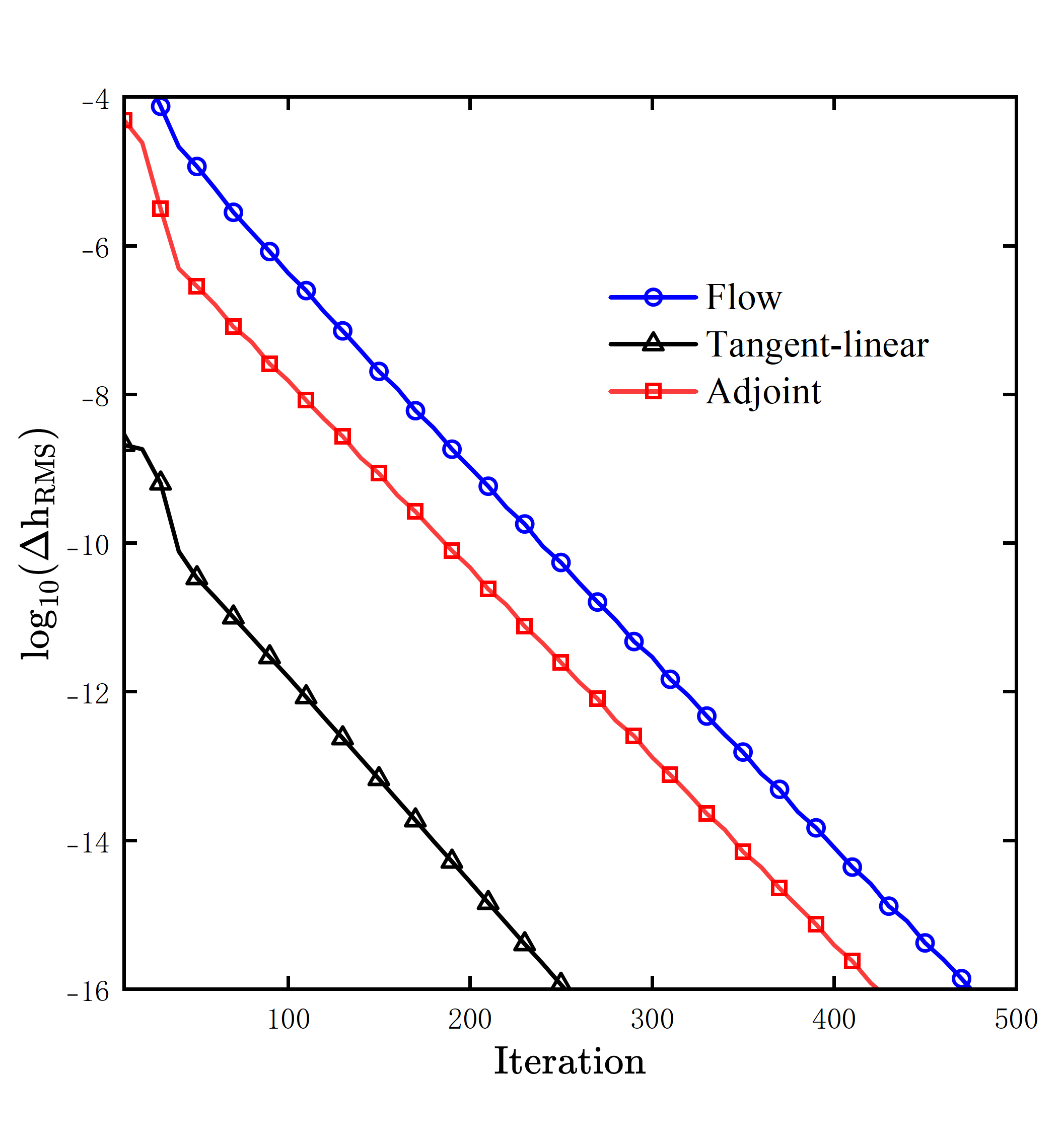}
        \par\smallskip{\small (c) $\Kn=10$}
    \end{minipage}
    \caption{Cavity-flow convergence histories: RMS updates of the microscopic distribution and its tangent and adjoint variables.}
    \label{fig:cavity_residuals}
\end{figure}

Figure~\ref{fig:cavity_residuals} compares the iterative convergence histories of the reduced distribution $h$, its tangent perturbation, and the corresponding adjoint component. For a discrete vector $z$ of length $N_z$, the root-mean-square (RMS) update norm is defined as
\begin{equation}
    \Delta z_{\mathrm{RMS}}^{\,m}
    =\left[\frac{1}{N_z}\sum_{j=1}^{N_z}
       \left(z_j^{m+1}-z_j^m\right)^2\right]^{1/2}.
    \label{eq:update_rms}
\end{equation}
Consistent with exact discrete-adjoint duality~\cite{giles2003algorithm}, all three solvers exhibit matching asymptotic convergence rates with sustained monotonic decay before reaching their respective round-off plateaus.

\begin{figure}[htbp]
    \centering
    \begin{minipage}[t]{0.325\linewidth}
        \centering
        \includegraphics[width=\linewidth]{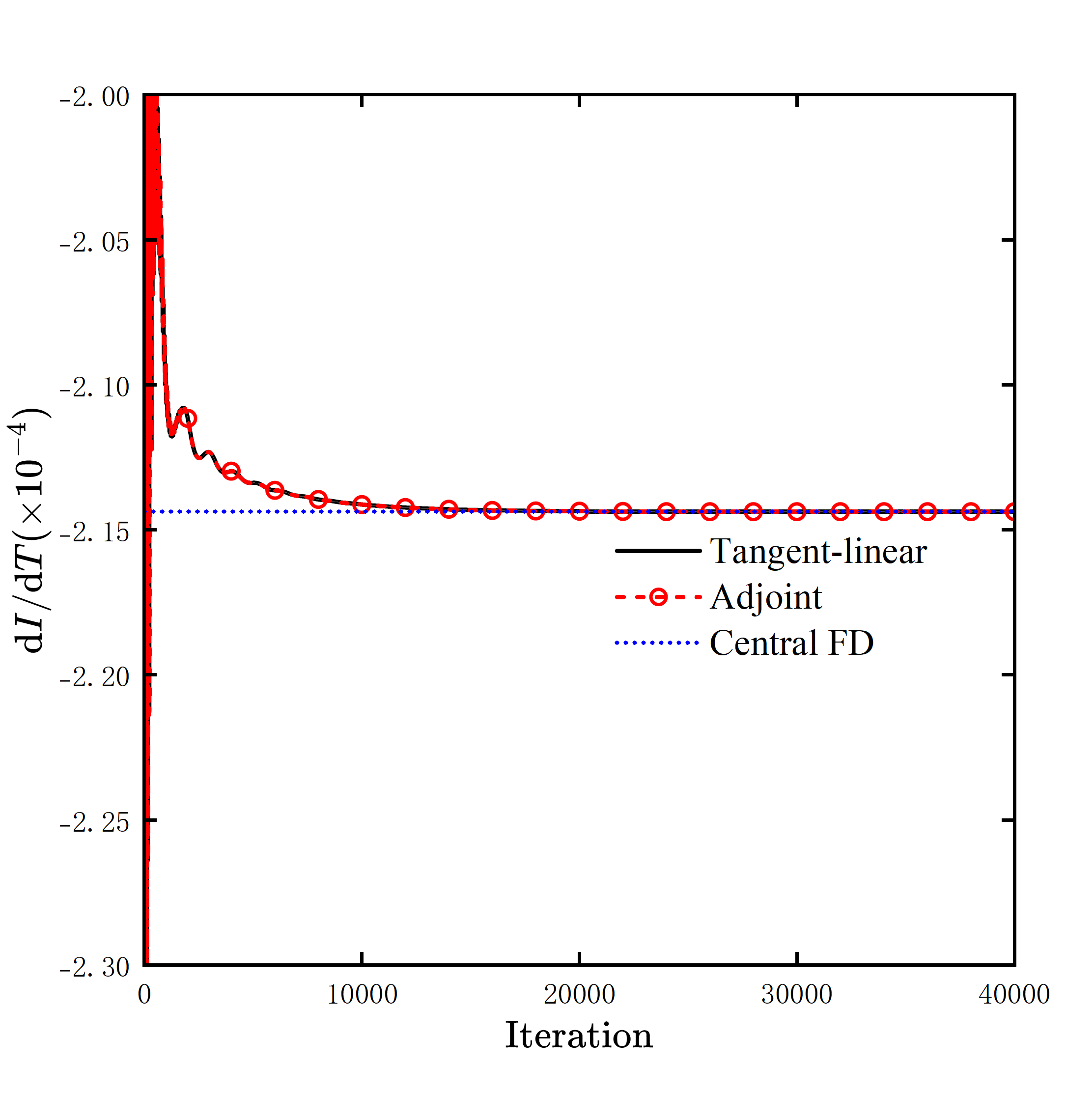}
        \par\smallskip{\small (a) $\mathrm{Re}=1000$}
    \end{minipage}\hfill
    \begin{minipage}[t]{0.325\linewidth}
        \centering
        \includegraphics[width=\linewidth]{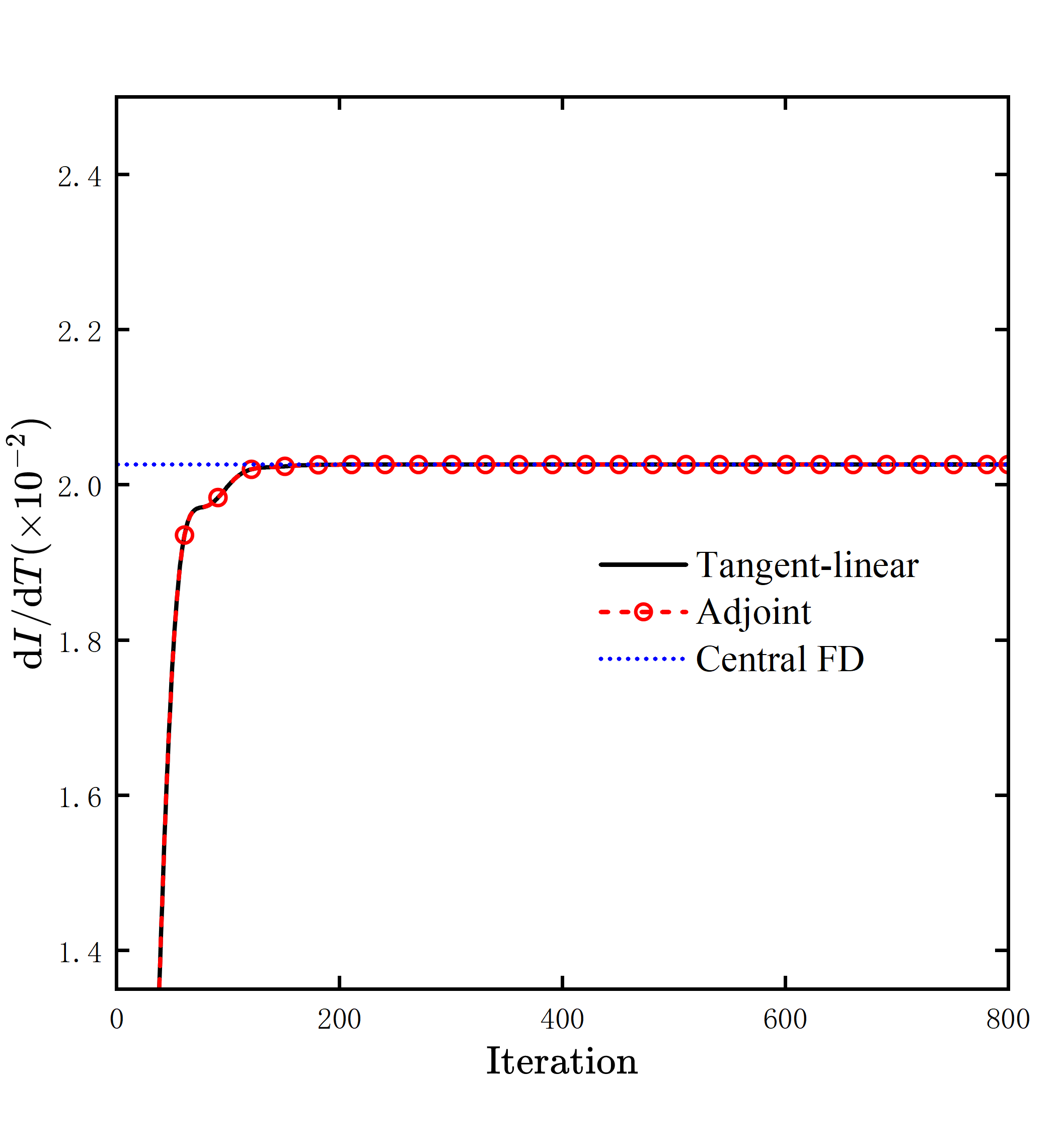}
        \par\smallskip{\small (b) $\Kn=0.075$}
    \end{minipage}\hfill
    \begin{minipage}[t]{0.325\linewidth}
        \centering
        \includegraphics[width=\linewidth]{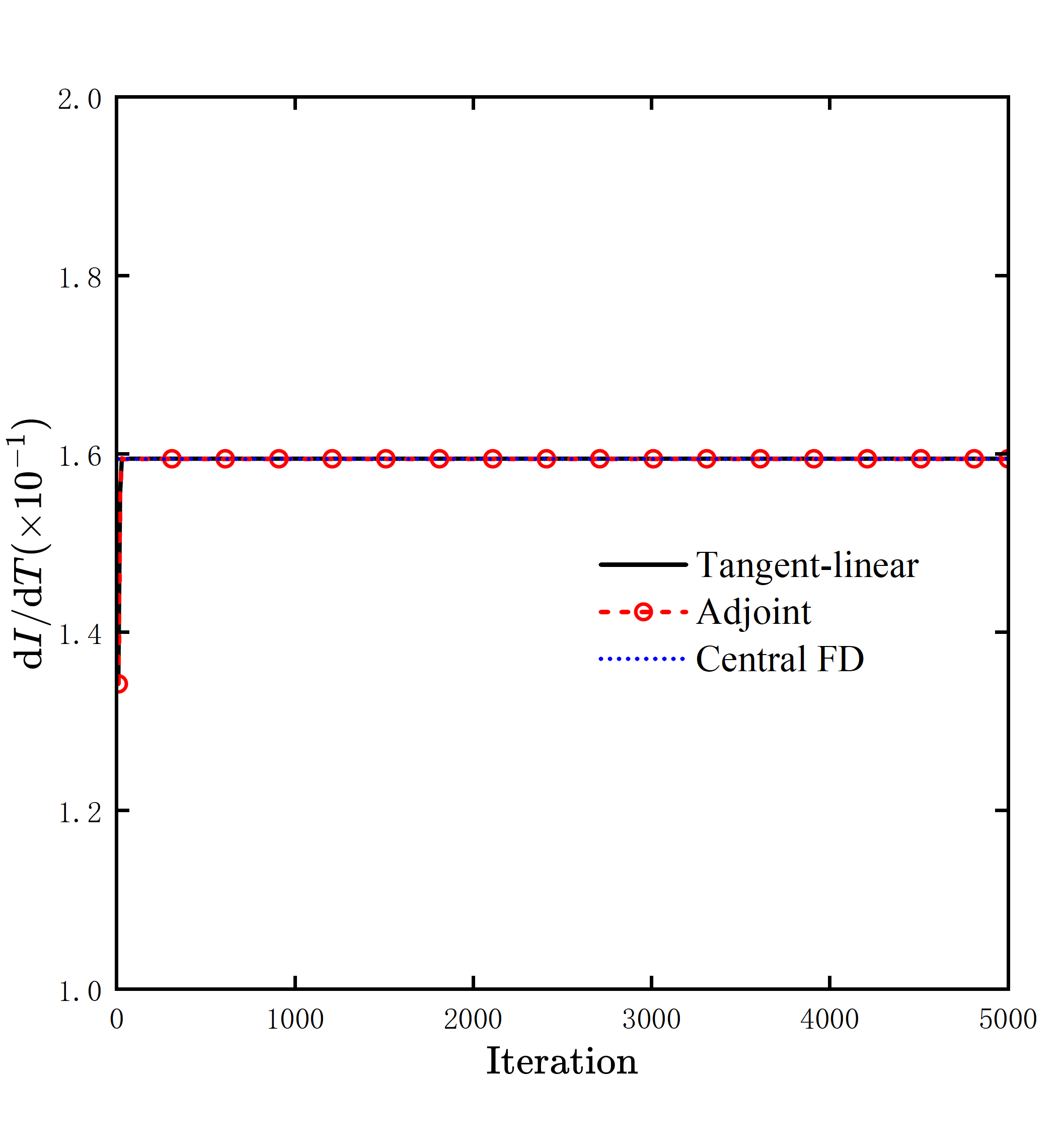}
        \par\smallskip{\small (c) $\Kn=10$}
    \end{minipage}
    \caption{Iteration histories of $\dd I_{\mathrm{cav}}/\dd T_{\mathrm{west}}$ obtained from the tangent-linear and adjoint solvers.}
    \label{fig:cavity_sensitivities}
\end{figure}

The sensitivity histories in Fig.~\ref{fig:cavity_sensitivities} directly verify stepwise discrete duality [Eq.~\eqref{eq:stepwise_duality}]. Across all three regimes, the TL and adjoint trajectories overlap throughout the iterative transient and converge to identical asymptotic values that match the central finite-difference benchmarks. Notably, $\dd I_{\mathrm{cav}}/\dd T_{\mathrm{west}}$ is small and negative at $\mathrm{Re}=1000$, whereas it becomes positive and orders of magnitude larger at $\Kn=0.075$ and $10$.

Table~\ref{tab:cavity_sensitivity} summarizes the converged sensitivity values and relative errors. The maximum absolute discrepancy between the final TL and adjoint sensitivities across all three cases is $6.4\times10^{-16}$, confirming exact discrete duality to machine precision. The relative discrepancies between the adjoint derivatives and central finite-difference benchmarks are $1.035\times10^{-5}$, $5.858\times10^{-8}$, and $3.085\times10^{-10}$ for $\mathrm{Re}=1000$, $\Kn=0.075$, and $\Kn=10$, respectively.

\begin{table}[htbp]
    \centering
    \footnotesize
    \caption{Cavity sensitivity verification for a uniform west-wall temperature perturbation. The finite-difference step is $\epsilon=10^{-6}$; relative errors are dimensionless.}
    \label{tab:cavity_sensitivity}
    \setlength{\tabcolsep}{4pt}
    \begin{tabular}{@{}lrrrr@{}}
        \toprule
        Condition & TL & Adjoint & FD & $E_{\mathrm{Adj},FD}$\\
        \midrule
        $\mathrm{Re}=1000$ &
        $-2.1437905112\times10^{-4}$ & $-2.1437905112\times10^{-4}$ &
        $-2.1437683279\times10^{-4}$ & $1.035\times10^{-5}$\\
        $\Kn=0.075$ &
        $2.0261885998\times10^{-2}$ & $2.0261885998\times10^{-2}$ &
        $2.0261887185\times10^{-2}$ & $5.858\times10^{-8}$\\
        $\Kn=10$ &
        $1.5942949466\times10^{-1}$ & $1.5942949466\times10^{-1}$ &
        $1.5942949461\times10^{-1}$ & $3.085\times10^{-10}$\\
        \bottomrule
    \end{tabular}
\end{table}
\FloatBarrier

\subsection{Shape optimization of a periodic Knudsen pump}\label{sec:pump}

\begin{figure}[htbp]
    \centering
    \includegraphics[width=0.5\linewidth]{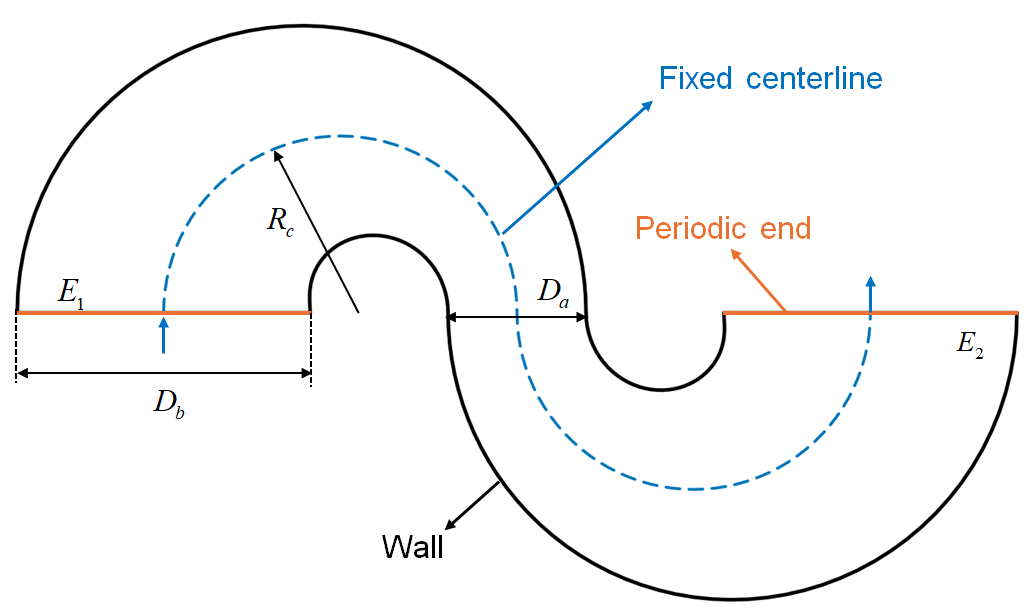}
    \caption{Schematic of the periodic S-shaped thermal micropump showing the fixed centerline, reference widths $D_a$ and $D_b$, arc radius $R_c$, and periodic end sections $E_1$ and $E_2$.}
    \label{fig:pump_schematic}
\end{figure}

A periodic S-shaped channel is considered, following the curved-channel thermal micropump configurations of Bond et al.~\cite{bond2014numerical,bond2016numerical}. As shown in Fig.~\ref{fig:pump_schematic}, the domain consists of two tangent semicircular bends bounded by stationary, fully diffuse walls and connected across the periodic end sections $E_1$ and $E_2$ by a translation of $(4R_c,0)$ with zero imposed pressure jump.

Parameterized by the angular coordinate $q\in[0,2\pi]$, the fixed reference centerline is defined as
\begin{equation}
    \vect x_{c,0}(q)=
    \begin{cases}
        R_c(-\cos q,\ \sin q),&0\leq q\leq\pi,\\
        R_c(2-\cos(q-\pi),\ -\sin(q-\pi)),&\pi\leq q\leq2\pi,
    \end{cases}
    \label{eq:centerline}
\end{equation}
and the baseline channel width and wall coordinates are given by
\begin{equation}
    w_0(q)=D_a+(D_b-D_a)\cos^2(q/2),
    \qquad
    \vect x_{\pm,0}(q)=\vect x_{c,0}(q)
       \pm\frac{w_0(q)}{2}\vect n_{\mathrm{ref}}(q),
    \label{eq:referencewidth}
\end{equation}
where $\vect n_{\mathrm{ref}}$ denotes the reference centerline unit normal.

The wall temperature profile is fixed and piecewise linear in the normalized reference arclength $s=q/(2\pi)$, while only the channel geometry varies during optimization with the total gas inventory held fixed at its baseline value $M_\star$. Two operating regimes at $\Kn=0.1$ and $10$ (based on $D_a$) are examined; all geometric, physical, and discretization parameters are summarized in Table~\ref{tab:pump_setup}.

\begin{table}[htbp]
    \centering
    \footnotesize
    \caption{Geometric, operating, and discretization parameters for the periodic S-shaped micropump.}
    \label{tab:pump_setup}
    \setlength{\tabcolsep}{6pt}
    \begin{tabular}{@{}lll@{}}
        \toprule
        Category & Parameter & Value / specification\\
        \midrule
        Geometry & Reference widths & $D_a=1$, \quad $D_b/D_a=\sqrt{17}-2$\\
                 & Centerline arc radius & $R_c/D_a=(1+\sqrt{17})/4$\\
                 & Periodic translation & $(4R_c,0)$ with zero pressure jump\\
        \midrule
        Flow \& boundary & Knudsen number ($\Kn$, based on $D_a$) & $0.1$, $10$\\
                         & Wall temperature $T_w(s)$ & Piecewise linear through $(0,2)$, $(1/4,1)$, $(3/4,3)$, $(1,2)$\\
                         & Viscosity model (VHS) & $\omega=\omega_{\mathrm{ref}}=0.81$, \quad $\alpha_{\mathrm{ref}}=1$\\
        \midrule
        Discretization & Spatial mesh & $100\times24$ and $200\times48$ quadrilateral cells\\
                       & Velocity-space quadrature & $121\times121$ uniform on $[-8,8]^2$ (Newton--Cotes)\\
        \bottomrule
    \end{tabular}
\end{table}

The objective functional is the density-normalized mass flow rate averaged over all streamwise cross-sections,
\begin{equation}
    \mathcal F=\frac{1}{N_x}\sum_{i=0}^{N_x-1}
                    \sum_{j=0}^{N_y-1}\Phi^\rho_{i,j},
    \qquad
    I=\frac{\mathcal F}{\bar\rho D_a},
    \qquad
    \bar\rho=\frac{\sum_c A_c\rho_c}{A},
    \quad A=\sum_c A_c,
    \label{eq:objective}
\end{equation}
where fluxes are oriented positively along increasing centerline arclength and $D_a$ is fixed. Although $\bar\rho=M_\star/A$ holds at converged states satisfying the mass constraint $\sum_c A_c\rho_c=M_\star$, the objective is differentiated via the general discrete density expression in Eq.~\eqref{eq:objective}---retaining both state and cell-area dependencies---as detailed in \ref{app:geometry}.

\begin{figure}[h]
    \centering
    \begin{minipage}[t]{0.48\linewidth}
        \centering
        \includegraphics[width=\linewidth]{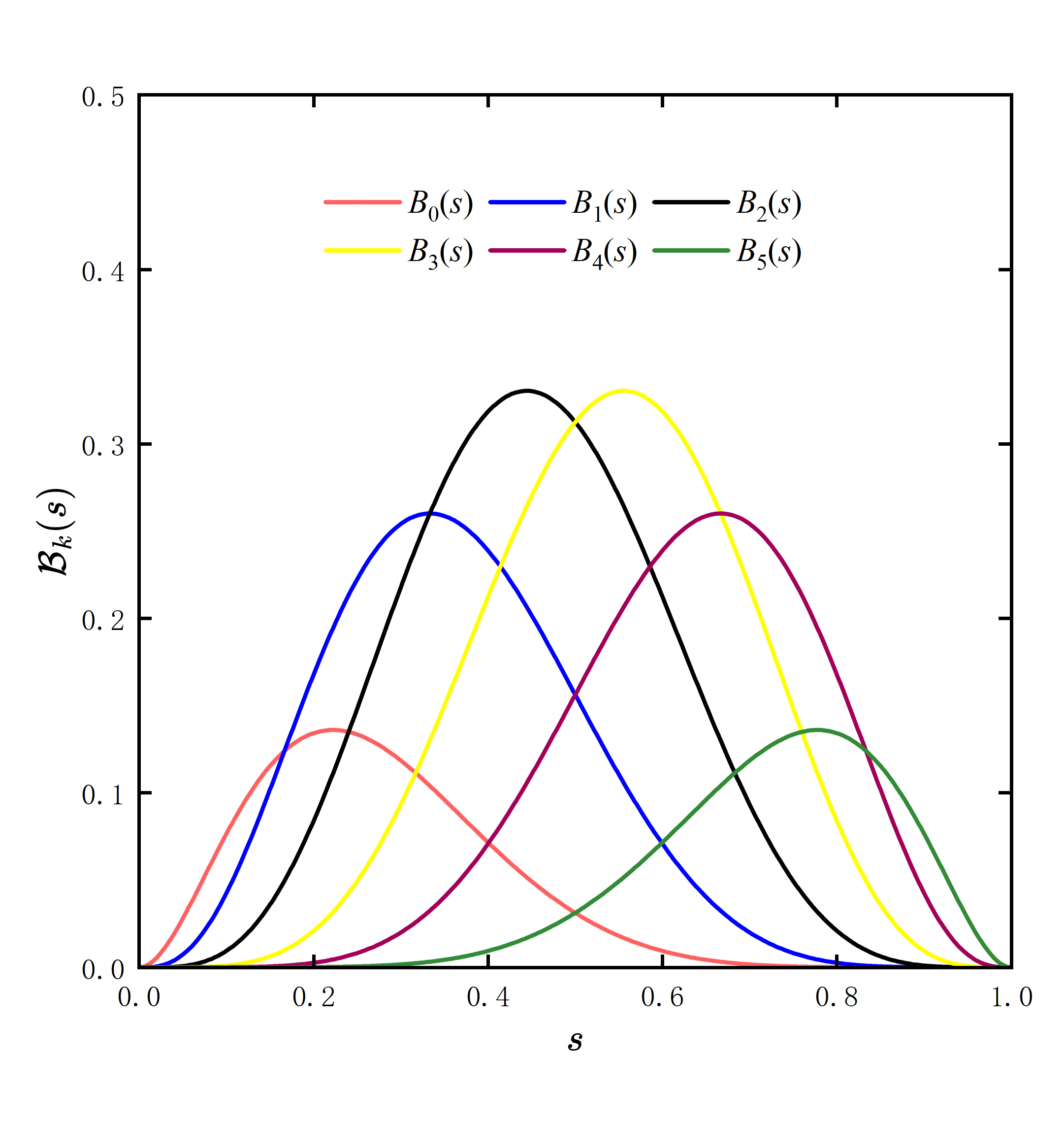}
        \par\smallskip{\small (a) Width basis functions}
    \end{minipage}\hfill
    \begin{minipage}[t]{0.48\linewidth}
        \centering
        \includegraphics[width=\linewidth]{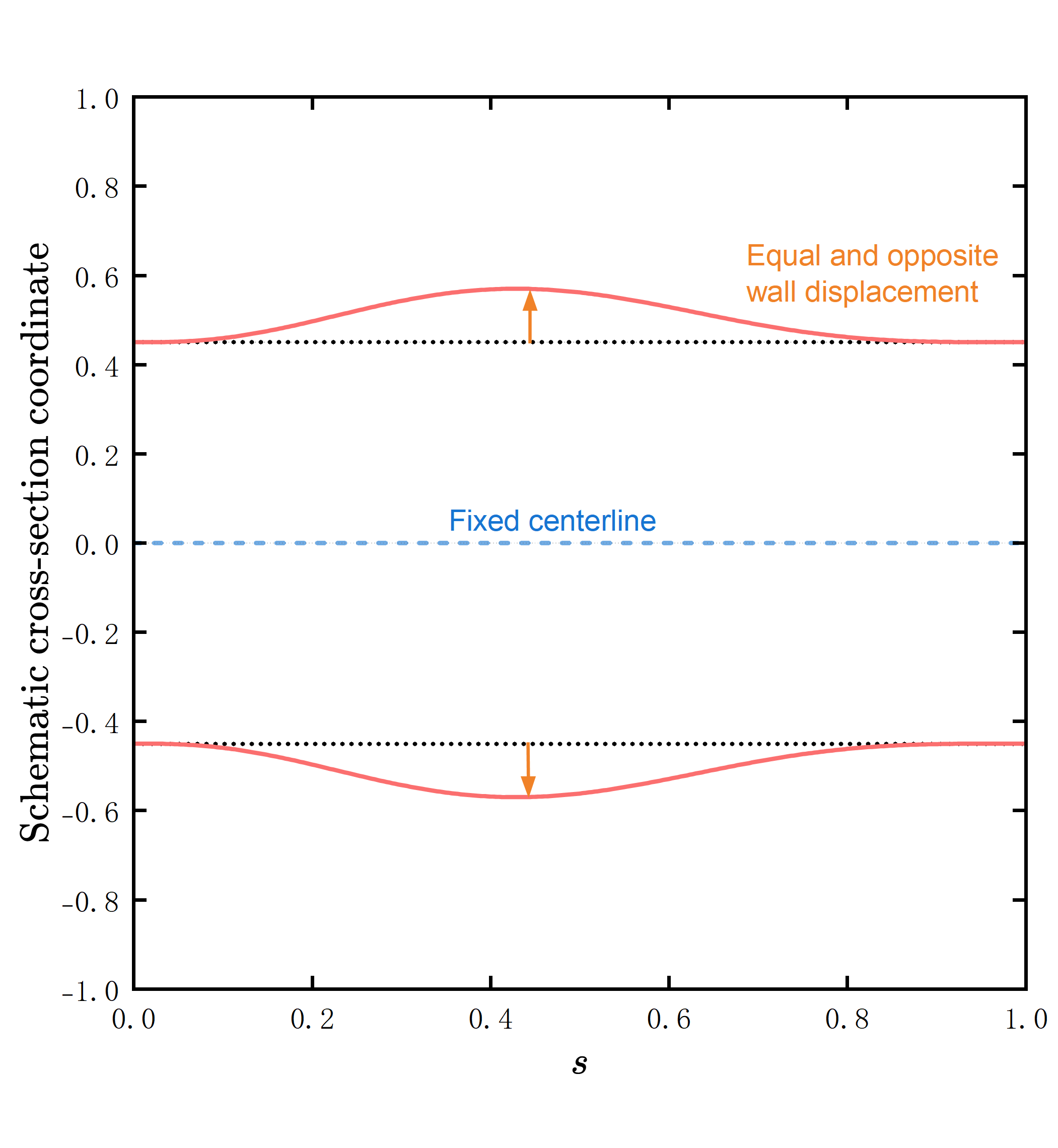}
        \par\smallskip{\small (b) Symmetric wall displacement}
    \end{minipage}
    \caption{Pump width parameterization with fixed endpoint positions and wall tangents.}
    \label{fig:pump_parameterization}
\end{figure}

Six design parameters $\vect a=(a_0,\dots,a_5)^T$ govern the channel-width perturbation through a squared-envelope Bernstein basis~\cite{farin2007curves},
\begin{equation}
    \begin{gathered}
        \Delta w(s,\vect a)=\sum_{k=0}^{5}a_k\mathcal B_k(s),\\
        \mathcal B_k(s)=16s^2(1-s)^2B_k^5(s),\qquad
        B_k^5(s)=\binom{5}{k}s^k(1-s)^{5-k}.
    \end{gathered}
    \label{eq:basis}
\end{equation}
Because $\mathcal B_k(0)=\mathcal B_k(1)=0$ and $\mathcal B_k'(0)=\mathcal B_k'(1)=0$, endpoint positions and wall tangents are preserved at the periodic boundaries. The width perturbation displaces the opposing walls symmetrically about the fixed centerline,
\begin{equation}
    \vect x_\pm(s)=\vect x_{\pm,0}(s)
        \pm\frac{\Delta w(s)}2\vect n_{\mathrm{ref}}(s),
    \label{eq:wallmotion}
\end{equation}
and the wall displacement is propagated into the interior mesh via linear elasticity on the reference grid (Fig.~\ref{fig:pump_parameterization}).

\subsubsection{Shape sensitivity and computational cost}\label{sec:pump_sensitivity_verification}

The sensitivity verification and six-parameter efficiency benchmark employ the squared-envelope Bernstein parameterization in Eq.~\eqref{eq:basis}. Sensitivity verification is performed with respect to the second width coefficient ($a_1$ under zero-based indexing) at fixed gas inventory and wall-temperature profile, without imposing a domain-area constraint.

\begin{figure}[!htbp]
    \centering
    \begin{minipage}[t]{0.48\linewidth}
        \centering
        \includegraphics[width=\linewidth]{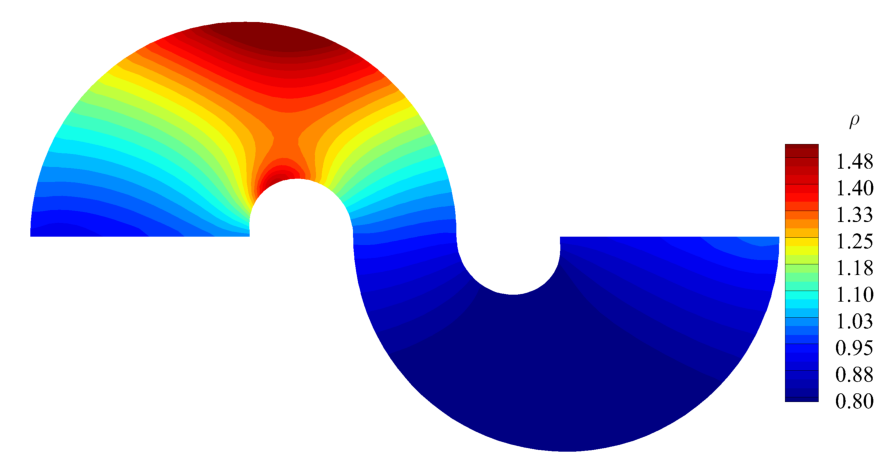}
        \par\smallskip{\small (a) Primal density $\rho$}
    \end{minipage}\hfill
    \begin{minipage}[t]{0.48\linewidth}
        \centering
        \includegraphics[width=\linewidth]{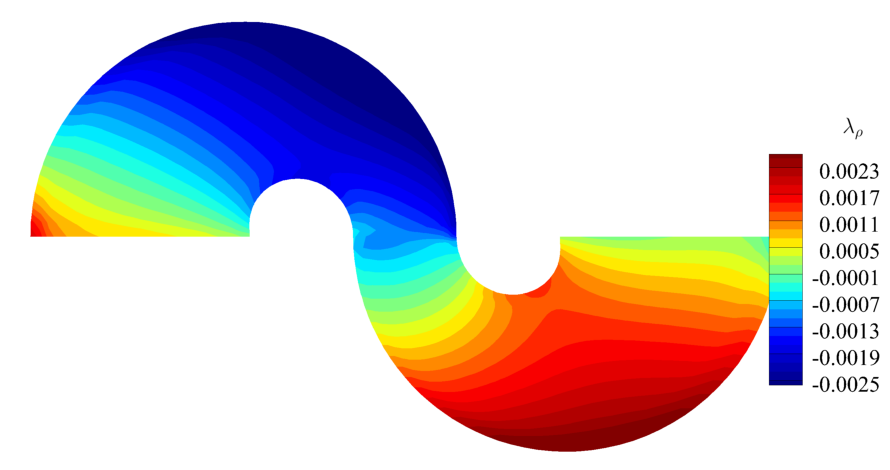}
        \par\smallskip{\small (b) Adjoint $\lambda_\rho$}
    \end{minipage}

    \medskip
    \begin{minipage}[t]{0.48\linewidth}
        \centering
        \includegraphics[width=\linewidth]{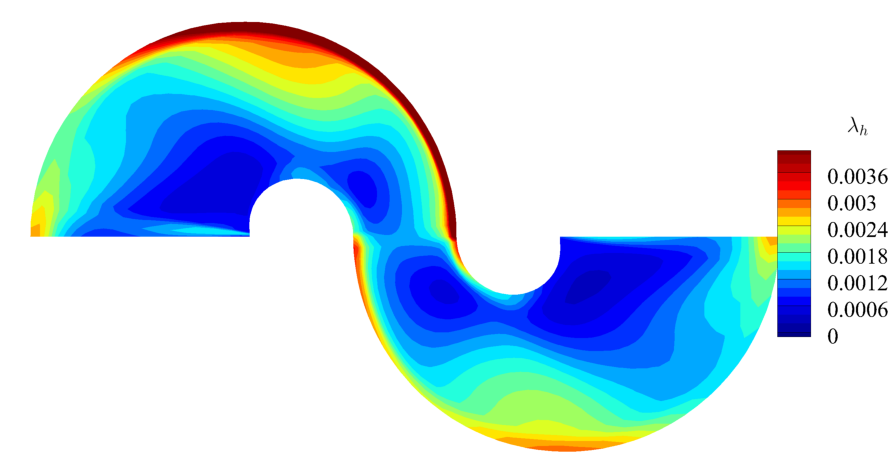}
        \par\smallskip{\small (c) Adjoint $\lambda_h$}
    \end{minipage}
    \caption{Representative converged primal and adjoint fields of the baseline periodic Knudsen pump at $\Kn=0.1$: (a) density $\rho$, (b) macroscopic density-adjoint component $\lambda_\rho$, and (c) microscopic adjoint component $\lambda_h$.}
    \label{fig:pump_adjoint_fields}
\end{figure}

Figure~\ref{fig:pump_adjoint_fields} presents the converged density and representative adjoint fields for the baseline periodic pump at $\Kn=0.1$. The density varies across the curved channel, while the two adjoint components exhibit distinct spatial distributions. The plotted $\lambda_h$ field is more pronounced near the curved walls, whereas $\lambda_\rho$ changes sign between the two bends. These contours illustrate the spatial structure of the adjoint solution associated with the mass-flow objective.

\FloatBarrier

\begin{figure}[h!]
    \centering
    \begin{minipage}[t]{0.48\linewidth}
        \centering
        \includegraphics[width=\linewidth]{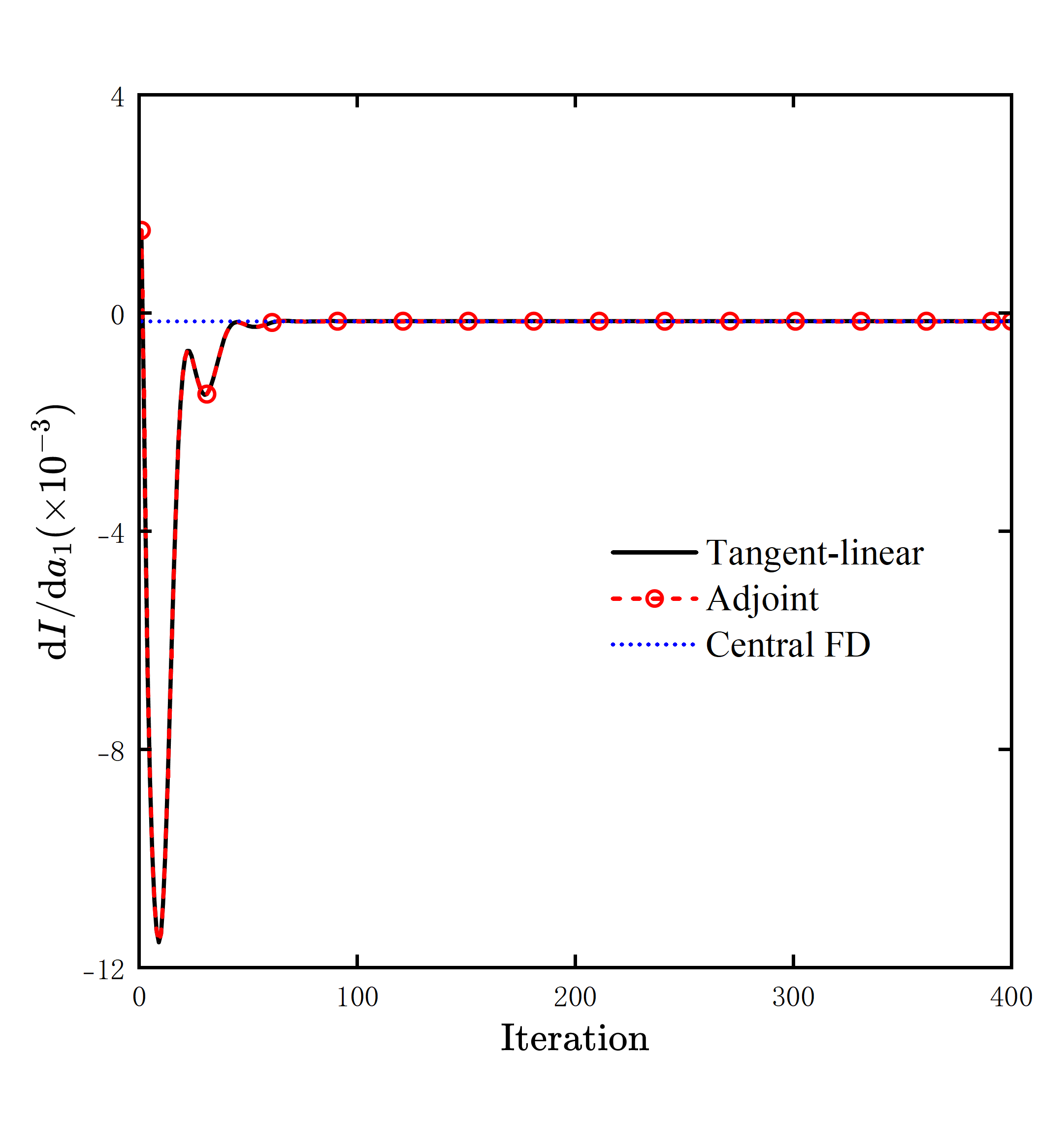}
        \par\smallskip{\small (a) $\Kn=0.1$}
    \end{minipage}\hfill
    \begin{minipage}[t]{0.48\linewidth}
        \centering
        \includegraphics[width=\linewidth]{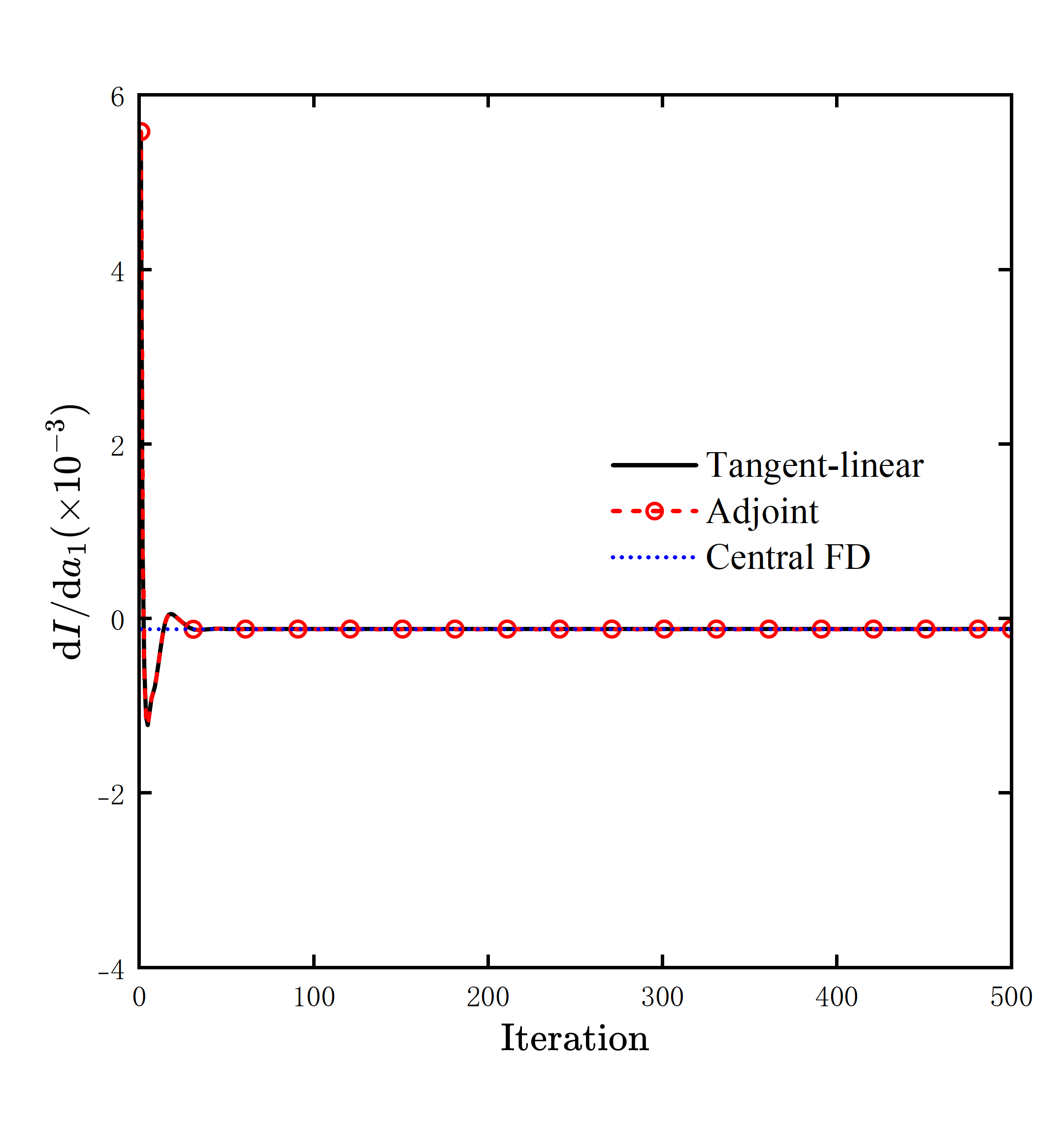}
        \par\smallskip{\small (b) $\Kn=10$}
    \end{minipage}
    \caption{Pump shape-sensitivity histories for the second width coefficient $a_1$, compared against central finite-difference benchmarks ($\epsilon=10^{-4}$).}
    \label{fig:pump_sensitivity_history}
\end{figure}

Figure~\ref{fig:pump_sensitivity_history} compares the iterative histories of $\dd I/\dd a_1$ obtained from the TL and adjoint solvers at $\Kn=0.1$ and $10$. 
Both formulations agree at every iteration step and converge to identical asymptotic values, with maximum stepwise differences of $1.34\times10^{-14}$ and $2.26\times10^{-15}$, respectively. The converged derivatives are $-1.48311\times10^{-4}$ at $\Kn=0.1$ and $-1.21625\times10^{-4}$ at $\Kn=10$, differing from the central finite-difference benchmarks ($\epsilon=10^{-4}$) by $0.180\%$ and $0.320\%$, respectively.

Computational efficiency is assessed on the full six-parameter pump problem at $\Kn=1$ using the same spatial and velocity grids, executed with 56 OpenMP threads on a single compute node. A single adjoint solve yields all six gradient components simultaneously, whereas TL requires six directional solves and central FD requires twelve perturbed flow solves.

\begin{figure}[htbp]
    \centering
    \includegraphics[width=0.48\linewidth]{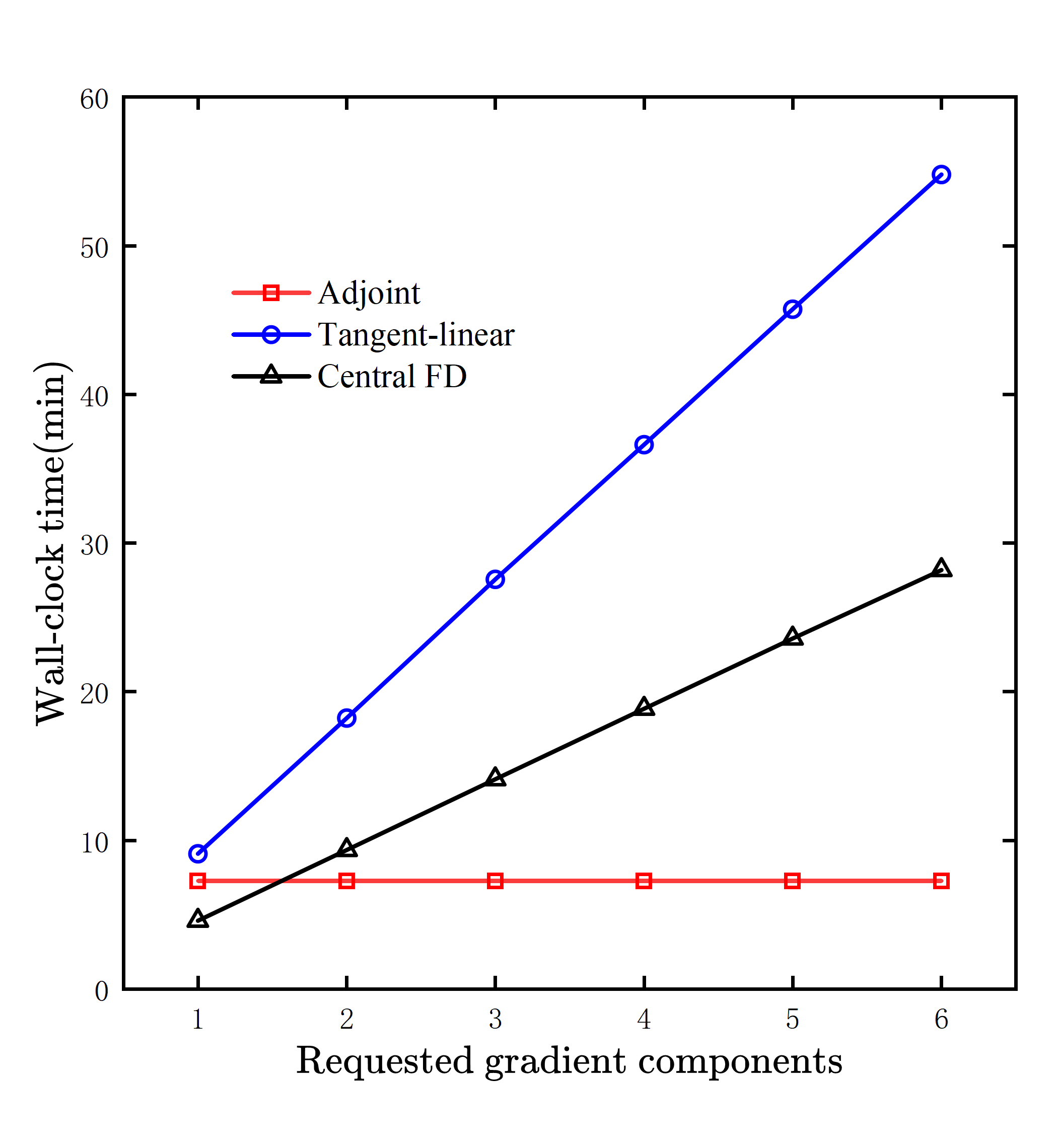}
    \caption{Wall-clock time versus number of design parameters for the six-parameter pump gradient evaluation at $\Kn=1$.}
    \label{fig:pump_efficiency_comparison}
\end{figure}

For the complete six-parameter gradient, the adjoint solver requires $435.84~\mathrm{s}$, compared with $3287.41~\mathrm{s}$ for TL and $1690.76~\mathrm{s}$ for FD, representing speedups of $7.54\times$ and $3.88\times$, respectively. As shown in Fig.~\ref{fig:pump_efficiency_comparison}, the adjoint wall-clock cost remains invariant as the number of design parameters increases, whereas the TL and FD costs scale linearly. Quantified by the relative $\ell_2$ gradient discrepancy $\|\vect g^{\mathrm{Adj}}-\vect g^{\mathrm{ref}}\|_2/\|\vect g^{\mathrm{ref}}\|_2$, the converged adjoint gradient differs from the TL and FD gradients by $5.12\times10^{-11}$ and $7.3178\times10^{-4}$ ($0.0732\%$), respectively.

\FloatBarrier

\subsubsection{Constrained shape optimization}\label{sec:pump_shape_results}

Shape optimization is performed using the parameterization in Eq.~\eqref{eq:basis}, which preserves the periodic endpoint positions and tangents. Under symmetric wall displacement, the enclosed channel area is affine in the design coefficients (see \ref{app:geometry}),
\begin{equation}
    A(\vect a)=A_0+\vect c_A^T\vect a.
    \label{eq:affinearea}
\end{equation}
The constrained optimization problem is formulated as
\begin{equation}
    \begin{aligned}
        \underset{\vect a}{\operatorname{maximize}}\quad &I(\vect Q(\vect a),P(\vect a)),\\
        \text{subject to}\quad
        &A(\vect a)=A_0,\\
        &w(s_j,\vect a)\geq0.5w_0(s_j),\\
        &-0.6\leq a_k\leq0.6,\qquad k=0,\ldots,5.
    \end{aligned}
    \label{eq:optimization}
\end{equation}
The minimum local width ratio across the discrete streamwise stations $s_j$ is denoted by $r_w=\min_j[w(s_j)/w_0(s_j)]$. At each trial design, the primal and adjoint systems are converged following Algorithm~\ref{alg:optimization}, and the scaled objective, design gradient, and geometric constraints are supplied to SLSQP.

\begin{figure}[htbp]
    \centering
    \begin{minipage}[t]{0.48\linewidth}
        \centering
        \includegraphics[width=\linewidth]{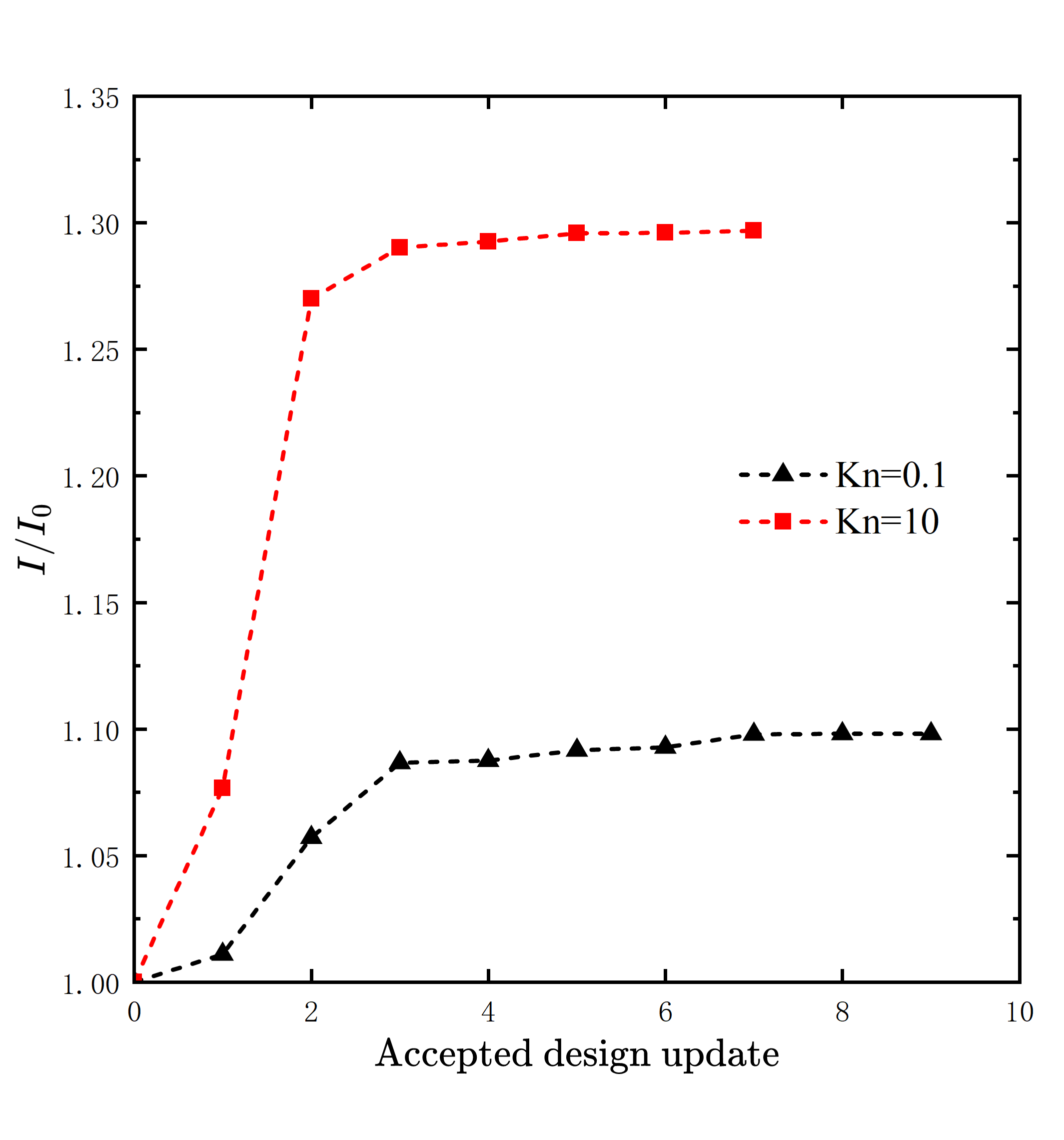}
        \par\smallskip{\small (a) Normalized mass flow}
    \end{minipage}\hfill
    \begin{minipage}[t]{0.48\linewidth}
        \centering
        \includegraphics[width=\linewidth]{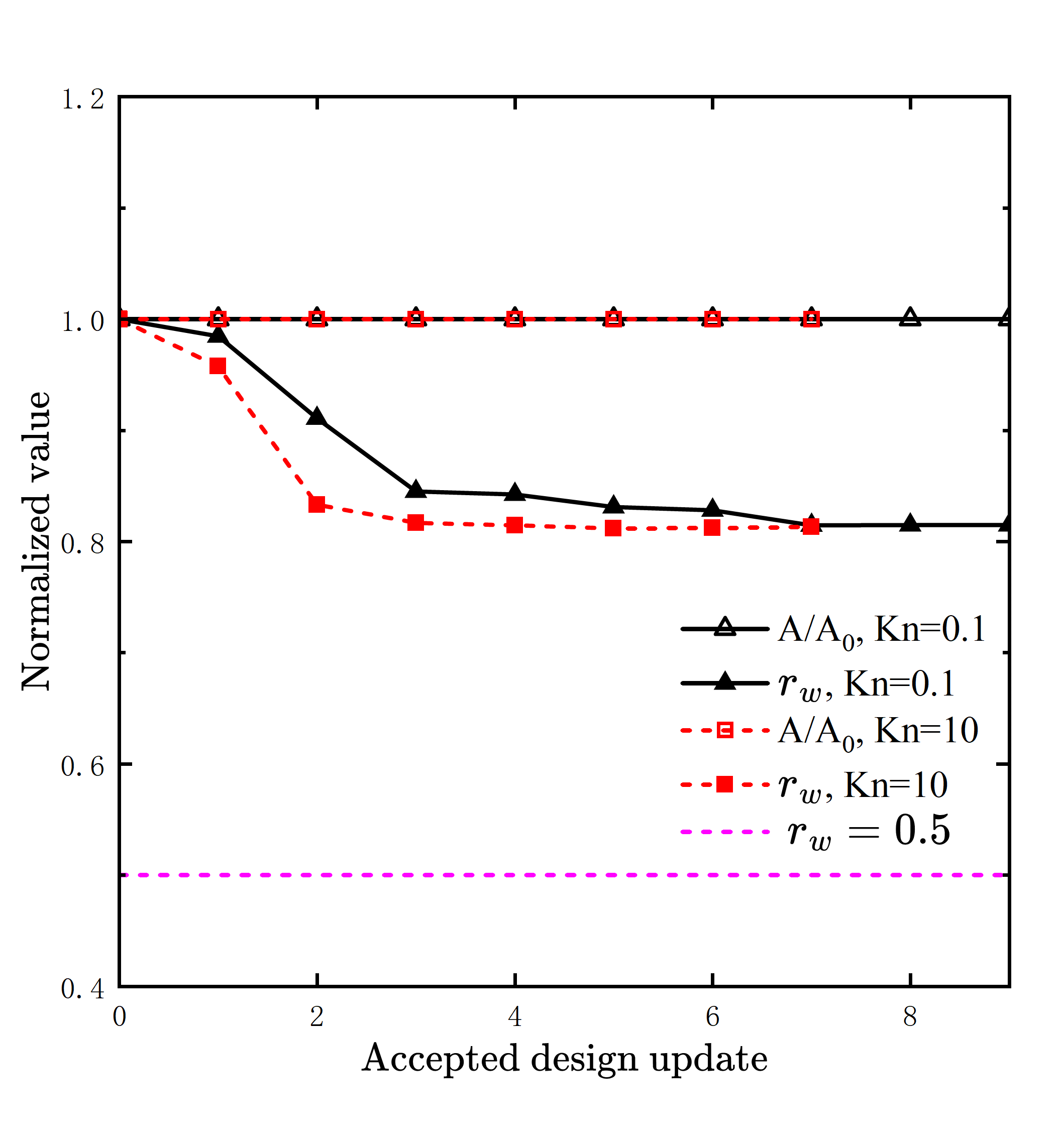}
        \par\smallskip{\small (b) Area and local width constraints}
    \end{minipage}
    \caption{Pump optimization histories for the two Knudsen numbers. The dashed lower-bound line in (b) is $r_w=0.5$.}
    \label{fig:pump_optimization_history}
\end{figure}

Figure~\ref{fig:pump_optimization_history} shows the convergence histories of the normalized objective and geometric constraints. 
The normalized mass flow $I/I_0$ increases to $1.09812$ at $\Kn=0.1$ and $1.29683$ at $\Kn=10$, representing net gains of $9.81\%$ and $29.68\%$ after nine and seven accepted design updates, respectively. Most of the performance gain occurs in the early iterations before the objective plateaus over the final updates. Throughout the optimization, the total domain area remains strictly equal to $A_0$, and the final minimum width ratios reach $0.8147$ and $0.8132$---well above the prescribed lower bound of $0.5$---confirming that the improvement is not obtained by driving a local constriction to the geometric admissibility limit.

\begin{figure}[htbp]
    \centering
    \begin{minipage}[t]{0.48\linewidth}
        \centering
        \includegraphics[width=\linewidth]{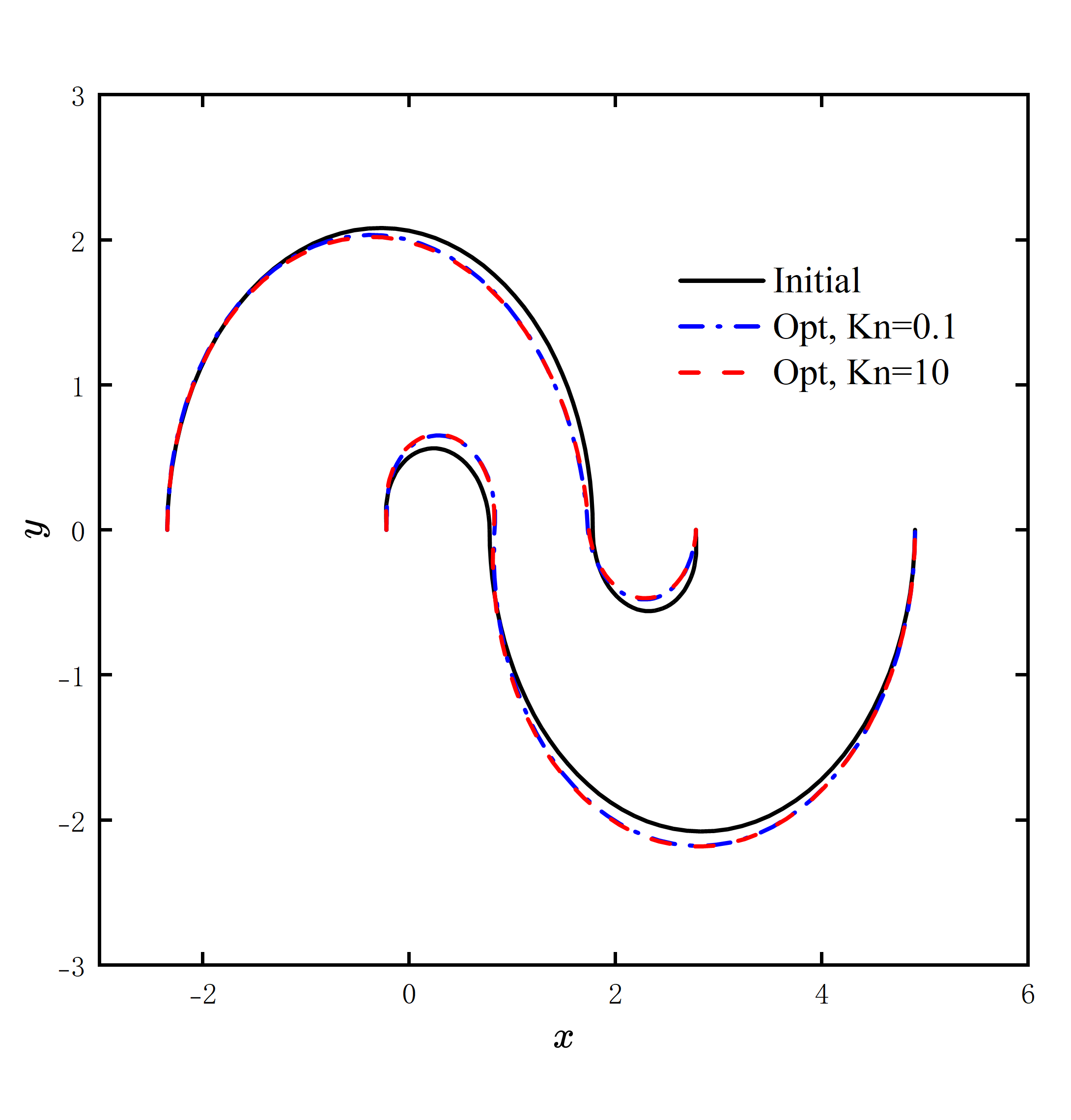}
        \par\smallskip{\small (a) Channel walls}
    \end{minipage}\hfill
    \begin{minipage}[t]{0.48\linewidth}
        \centering
        \includegraphics[width=\linewidth]{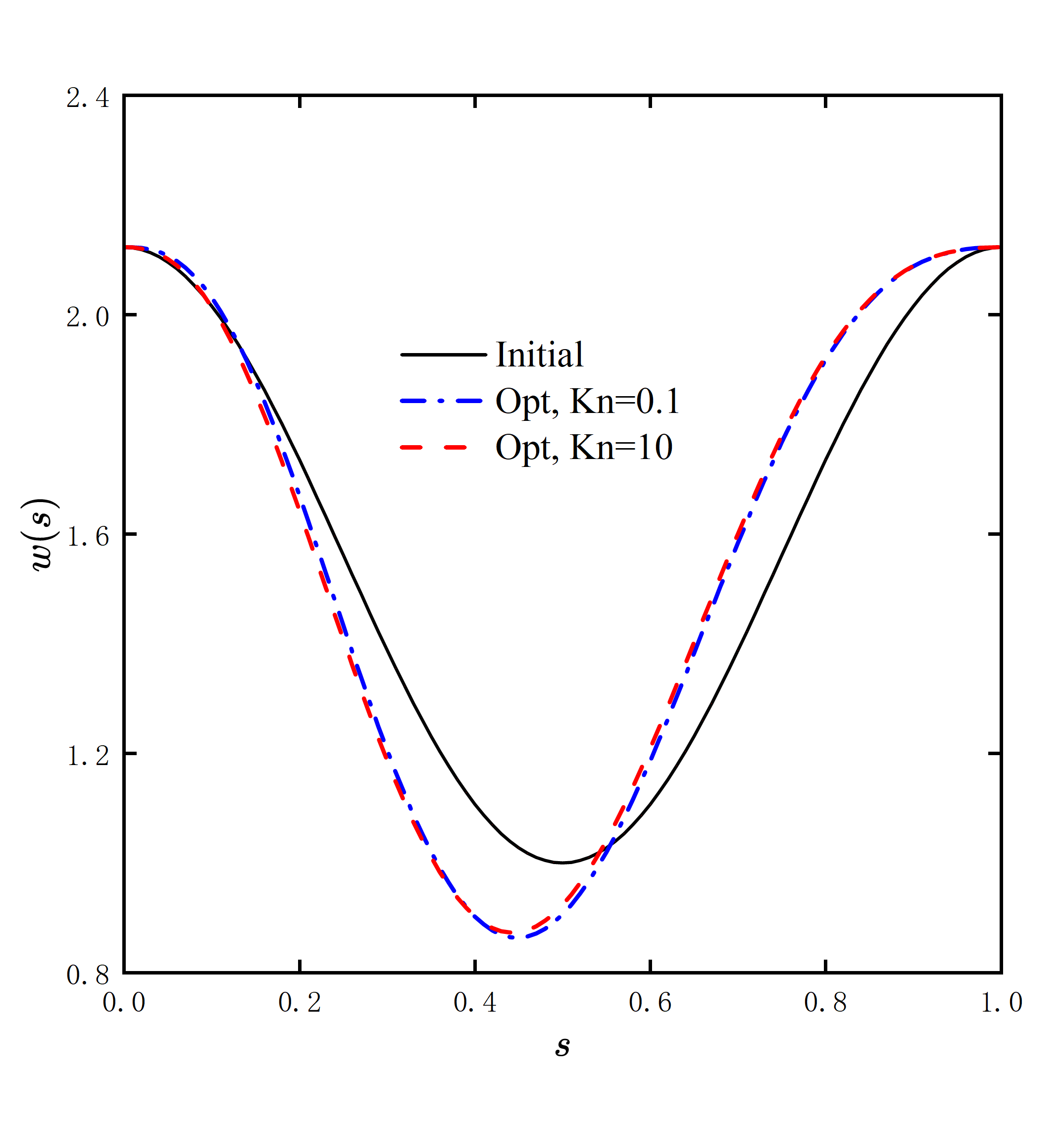}
        \par\smallskip{\small (b) Width distributions}
    \end{minipage}
    \caption{Initial and optimized pump geometries. The width lower bound is $0.5w_0(s)$, and the endpoint widths and tangents remain fixed.}
    \label{fig:pump_shape_comparison}
\end{figure}

As shown in Fig.~\ref{fig:pump_shape_comparison}, the optimized geometries redistribute the available cross-sectional area along the channel: the passage narrows upstream of the throat and widens downstream, while the centerline, periodic end sections, and endpoint tangents remain invariant. Although the local width constraint remains inactive, several shape coefficients reach their prescribed parameter bounds of $\pm0.6$.

\begin{figure}[h!]
    \centering
    \begin{tabular}{@{}cc@{}}
        \includegraphics[width=0.48\linewidth]{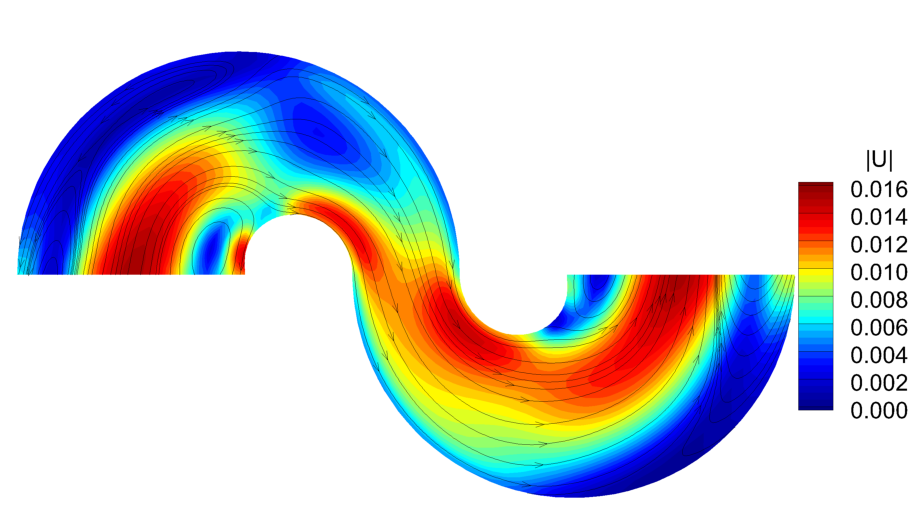}&
        \includegraphics[width=0.48\linewidth]{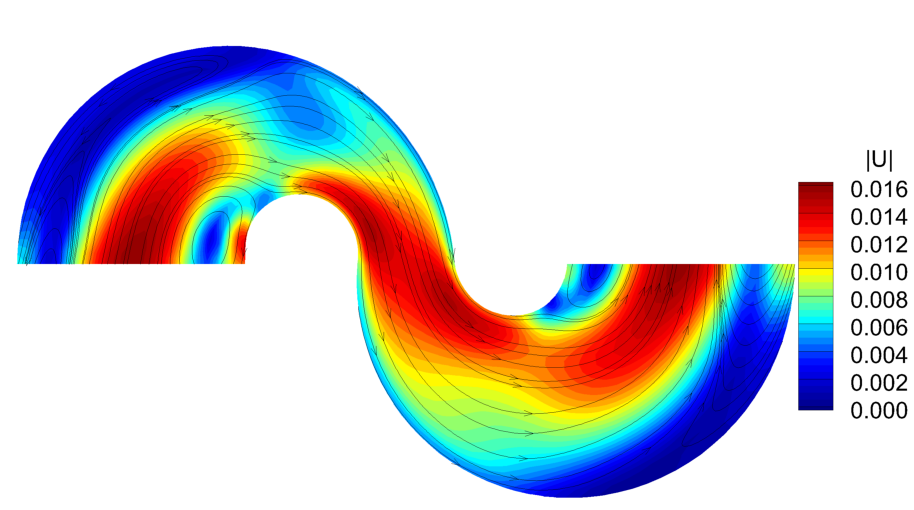}\\
        {\small (a) Initial, $\Kn=0.1$}&{\small (b) Optimized, $\Kn=0.1$}\\[5pt]
        \includegraphics[width=0.48\linewidth]{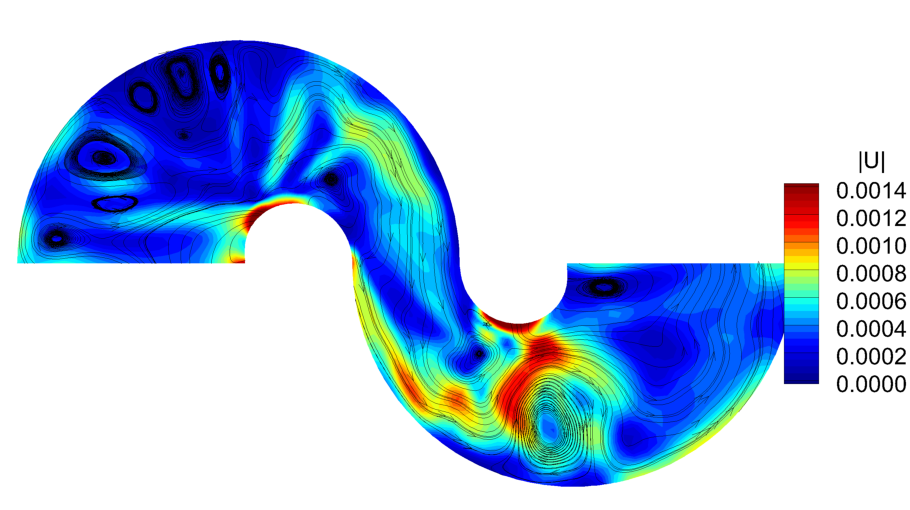}&
        \includegraphics[width=0.48\linewidth]{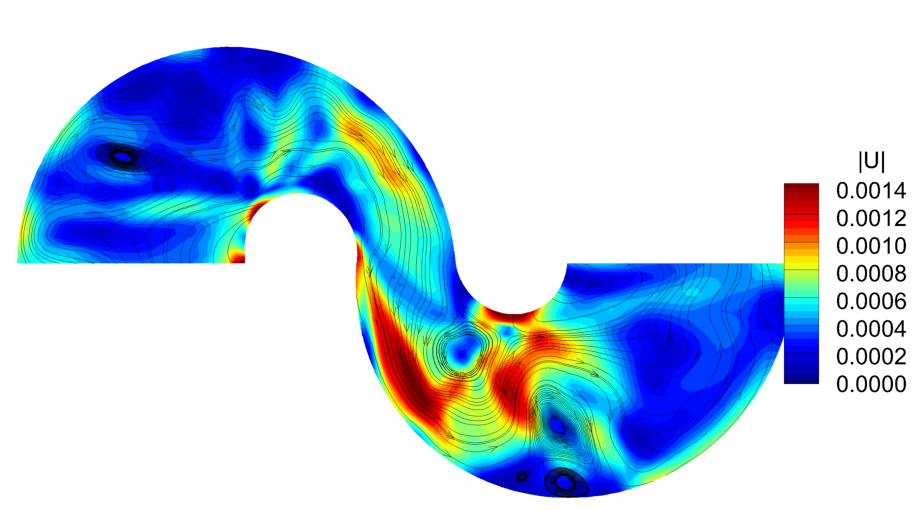}\\
        {\small (c) Initial, $\Kn=10$}&{\small (d) Optimized, $\Kn=10$}
    \end{tabular}
    \caption{Nondimensional velocity-magnitude contours and streamlines in the initial and optimized Knudsen pumps. The same color scale is used within each operating-condition pair.}
    \label{fig:pump_velocity_magnitude}
\end{figure}

Figure~\ref{fig:pump_velocity_magnitude} compares the velocity-magnitude contours and streamlines before and after optimization under identical wall-temperature distributions. 
At $\Kn=0.1$, the optimization connects and broadens the high-speed thermal-creep core across the central S-bend throat while compressing the low-speed recirculation zones along the outer bend walls, establishing a smoother through-flow pathway. At $\Kn=10$, the baseline flow exhibits pronounced cellular recirculation vortices in the upper outer bend and fragmented high-speed pockets near the inner corners; shape optimization suppresses these parasitic vortices and expands the coherent streamwise transport regions across the lower bend and throat. Consequently, although the final normalized mass-flow objective at $\Kn=10$ ($0.000522656$) remains smaller in absolute magnitude than at $\Kn=0.1$ ($0.0107515$), suppressing flow-blocking recirculation enables the substantially larger relative gain ($29.68\%$ versus $9.81\%$) reported in Fig.~\ref{fig:pump_optimization_history}. Furthermore, re-evaluating the baseline and optimized geometries on a finer $200\times48$ mesh confirms spatial grid independence: the initial normalized mass flow agrees with the $100\times24$ prediction within $0.32\%$, and the relative optimization gain differs by less than $0.5\%$.
\FloatBarrier

\subsection{Shape optimization of a supersonic NACA0012 airfoil}\label{sec:naca}

\subsubsection{Flow conditions, discretization, and design variables}\label{sec:naca_setup}

External supersonic flow past a NACA0012 airfoil of chord $c$ is considered at free-stream Mach number $2$, zero angle of attack, and chord-based Knudsen numbers $\Kn=0.01$ and $0.5$, extending the sensitivity framework to compressible open-domain aerodynamic design~\cite{yuan2025adjoint}. The gas is a monatomic hard-sphere gas ($\gamma=5/3$, $\Pran=1$, $\omega=\omega_{\mathrm{ref}}=0.5$, and $\alpha_{\mathrm{ref}}=1$). The airfoil surface is stationary and fully diffuse at the free-stream temperature ($T_w=T_\infty$), while incoming far-field distributions are prescribed from the free-stream Maxwellian.

\begin{figure}[h]
    \centering
    \begin{minipage}[t]{0.48\linewidth}
        \centering
        \includegraphics[width=\linewidth]{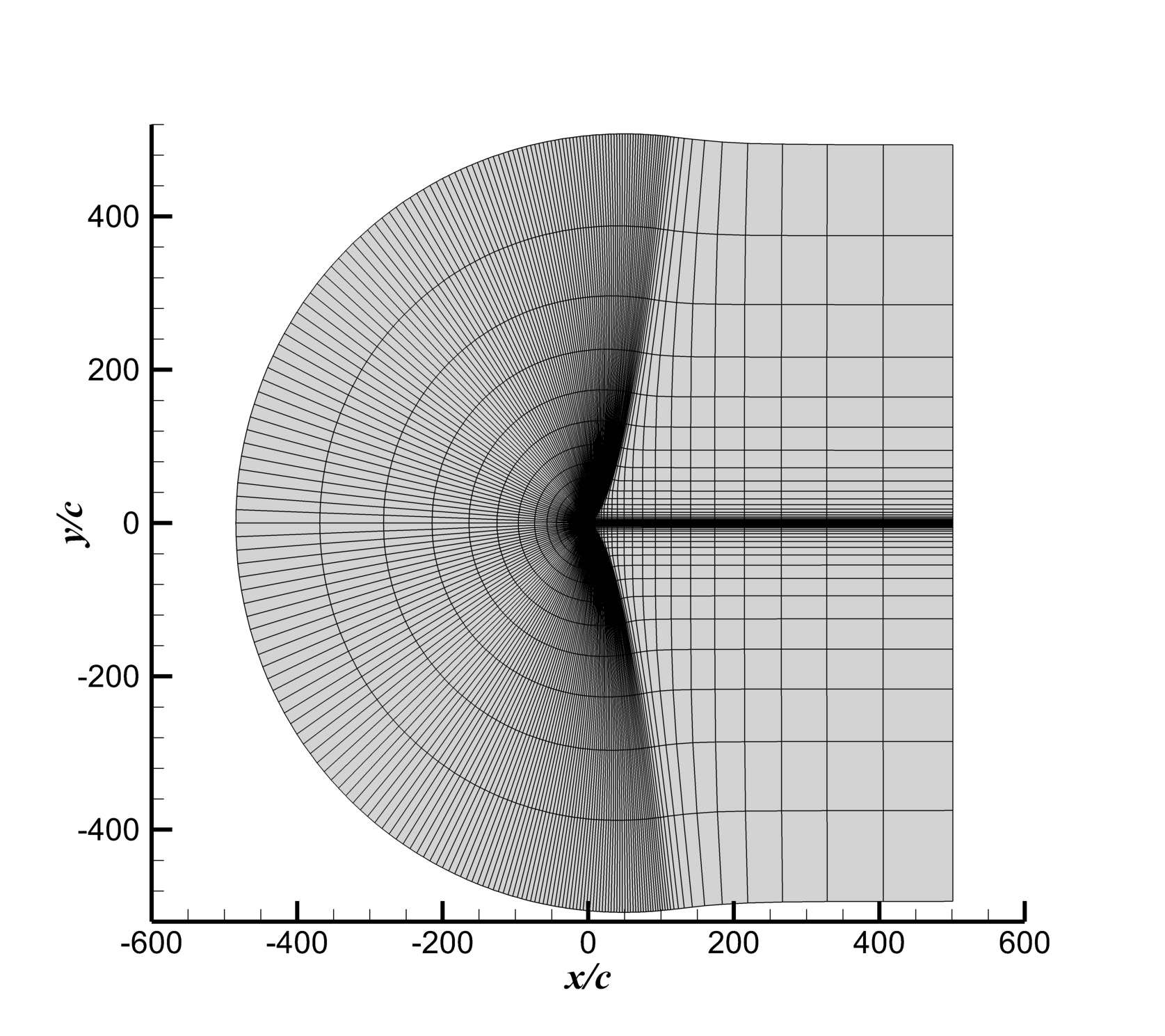}
        \par\smallskip{\small (a) Computational domain}
    \end{minipage}\hfill
    \begin{minipage}[t]{0.48\linewidth}
        \centering
        \includegraphics[width=\linewidth]{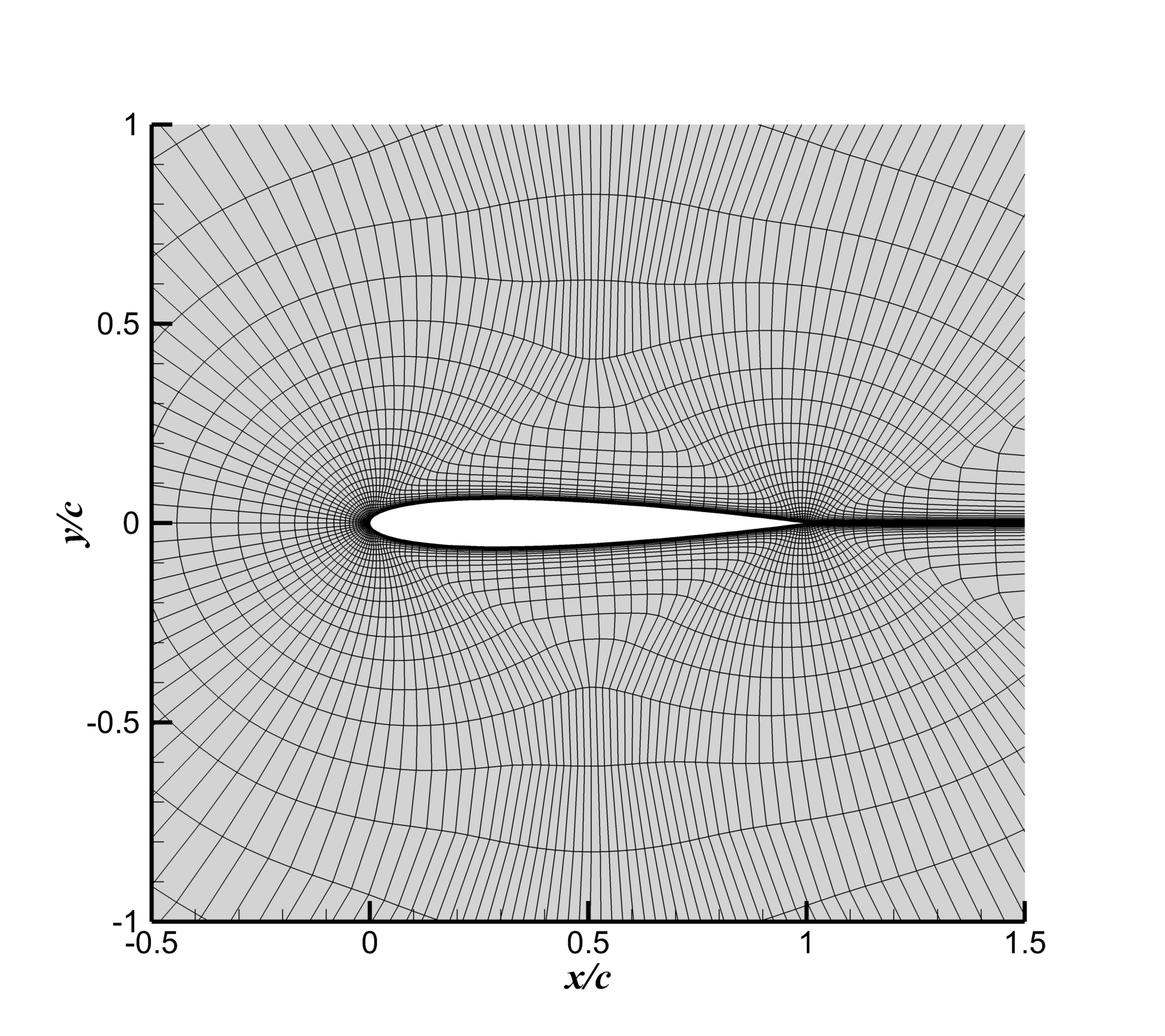}
        \par\smallskip{\small (b) Near-airfoil mesh}
    \end{minipage}
    \caption{Physical-space mesh for the NACA0012 calculations ($224\times64$ cells).}
    \label{fig:naca_mesh}
\end{figure}

\begin{figure}[tbp]
    \centering
    \includegraphics[width=0.58\linewidth]{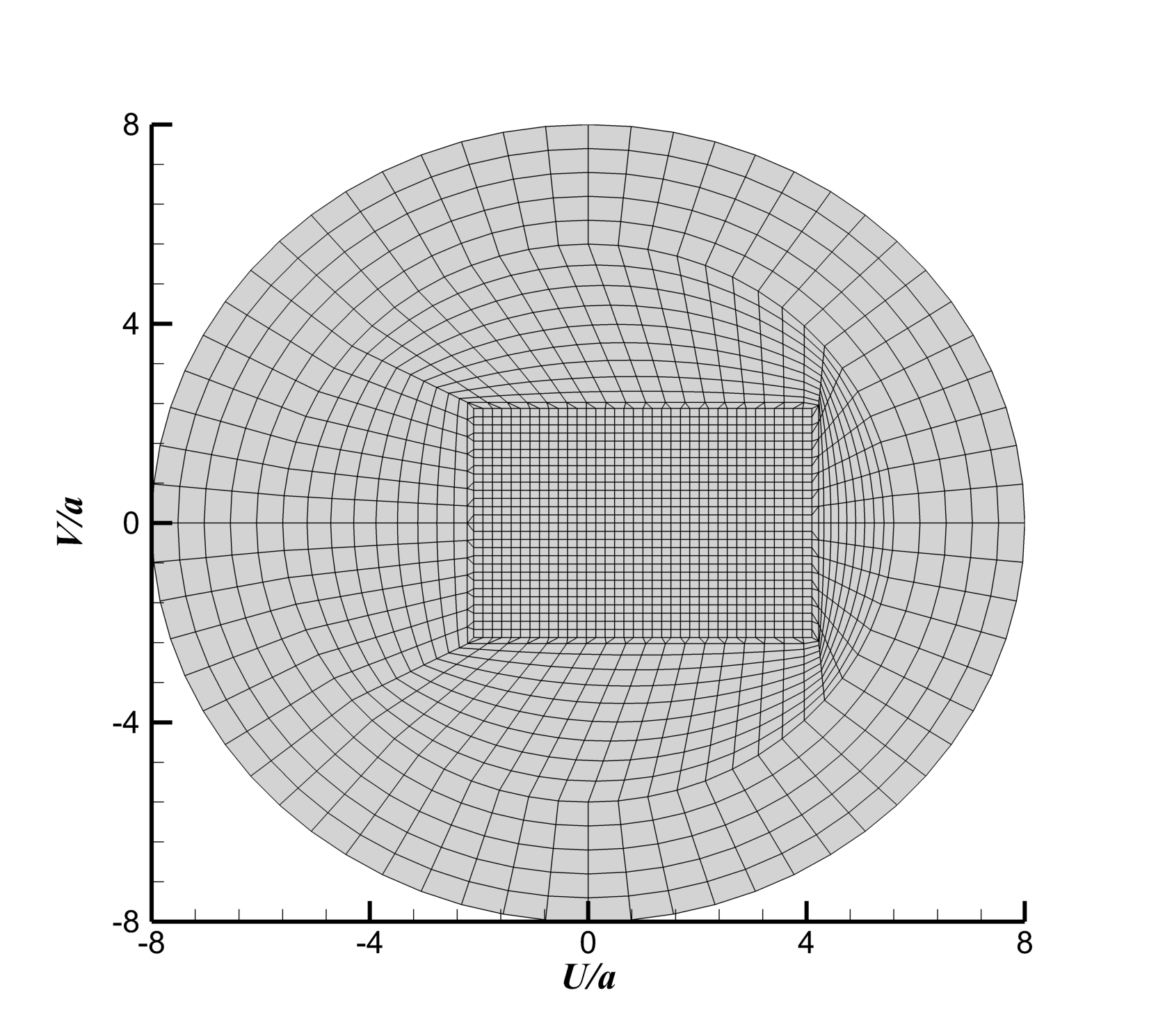}
    \caption{Nonuniform molecular-velocity mesh at $\Kn=0.5$ ($2032$ quadrature cells).}
    \label{fig:naca_velocity_mesh}
\end{figure}

The physical domain is discretized using a structured mesh of $224\times64$ quadrilateral cells with fixed connectivity (Fig.~\ref{fig:naca_mesh}). In velocity space, the $\Kn=0.01$ case employs a uniform $28\times28$ midpoint quadrature on $[-8a_\infty,8a_\infty]^2$, where $a_\infty$ is the free-stream sound speed, whereas the $\Kn=0.5$ case uses an unstructured $2032$-cell centroid quadrature refined around the wall and free-stream velocity states (Fig.~\ref{fig:naca_velocity_mesh}). Both discretizations remain fixed throughout the optimization.


Six symmetric Hicks--Henne bump functions~\cite{hicks1978wing} parameterize the upper and lower airfoil profiles:
\begin{equation}
    \begin{gathered}
        y_\pm(s,\vect a)=y_{\pm,0}(s)\pm c\sum_{i=1}^6 a_i b_i(s),\\
        b_i(s)=\sin^4\!\left(\pi s^{m_i}\right),\qquad
        m_i=\frac{\ln(1/2)}{\ln(s_i)},\quad s_i=\frac{i}{7},
        \quad s=\frac{x}{c}.
    \end{gathered}
    \label{eq:hh_parameterization}
\end{equation}
This parameterization preserves upper--lower symmetry while keeping the leading and trailing edges fixed. Because the surface perturbation acts strictly in the $y$-direction at fixed $x$, interior mesh nodes are displaced along the $y$-direction with weights proportional to the remaining reference radial-gridline arclength, decreasing linearly from unity on the airfoil surface to zero at the far-field boundary; the resulting linear mesh-deformation operator and its transpose are shared consistently by the tangent-linear and adjoint solvers.

The drag coefficient is defined as
\begin{equation}
    C_D=\frac{D}{\tfrac12\rho_\infty U_\infty^2c},
    \qquad
    D=\sum_{f\in\Gamma_w}
       \vect\Phi_f^{\,\mathrm{mom}}\cdot\vect e_x,
    \label{eq:drag_objective}
\end{equation}
where $\Gamma_w$ denotes the airfoil surface, $\vect e_x$ is the streamwise unit vector, and $\vect\Phi_f^{\,\mathrm{mom}}$ is the discrete wall momentum flux on face $f$, which incorporates both normal pressure and tangential shear stresses. The constrained drag-minimization problem is formulated as
\begin{equation}
    \begin{aligned}
        \underset{\vect a}{\operatorname{minimize}}\quad&C_D(\vect a),\\
        \text{subject to}\quad&A(\vect a)=A_0,\\
        &t(s_j,\vect a)\geq0.7t_0(s_j),\qquad j=1,\ldots,63,\\
        &-0.02\leq a_i\leq0.02,\qquad i=1,\ldots,6,
    \end{aligned}
    \label{eq:naca_optimization}
\end{equation}
where $A$ is the enclosed airfoil area and $t=y_+-y_-$ is the local thickness enforced at the interior paired wall stations. At each design iteration, the primal and adjoint systems are converged using the implicit point-relaxation scheme of Section~\ref{sec:implicit} and \ref{app:sweeps}, supplying the normalized objective $C_D/C_{D,0}$, discrete gradient, and geometric constraints to SLSQP.

\subsubsection{Airfoil shape-sensitivity verification}\label{sec:naca_sensitivity}

Discrete sensitivity accuracy is verified on the baseline NACA0012 airfoil with respect to the first Hicks--Henne coefficient $a_1$, using a central finite-difference step of $\epsilon=10^{-4}$.

\begin{figure}[!htbp]
    \centering
    \includegraphics[width=0.6\linewidth]{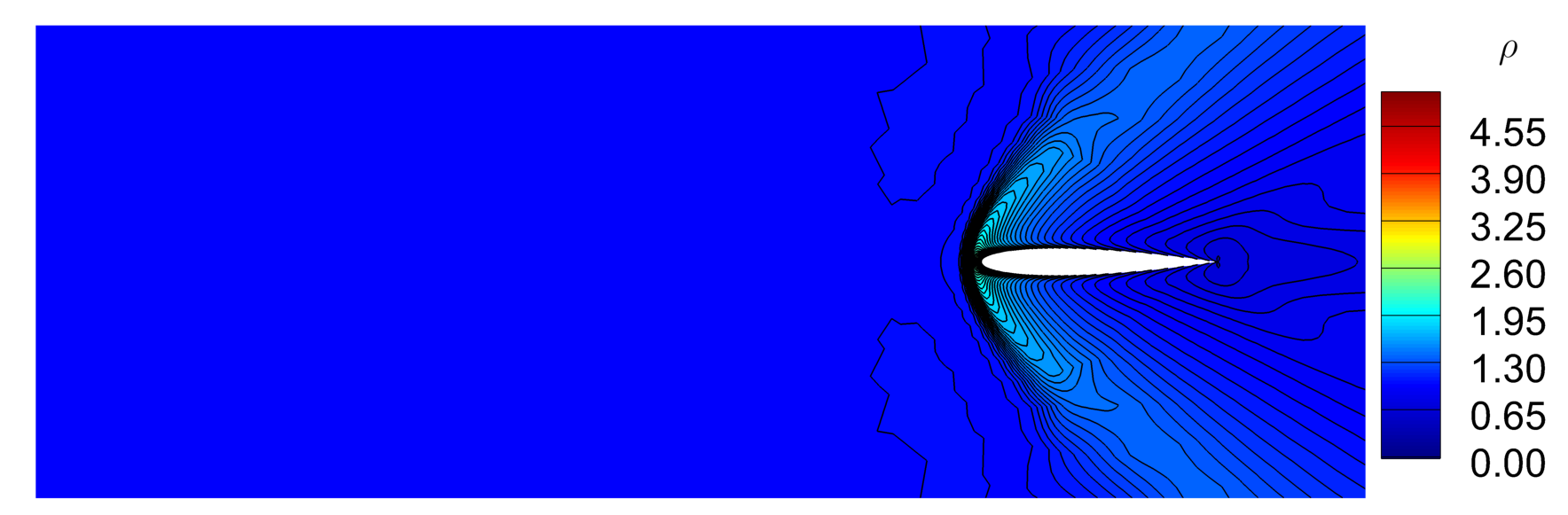}
    \par\vspace{0.15em}{\small (a) Primal density $\rho$}

    \vspace{0.35em}
    \includegraphics[width=0.6\linewidth]{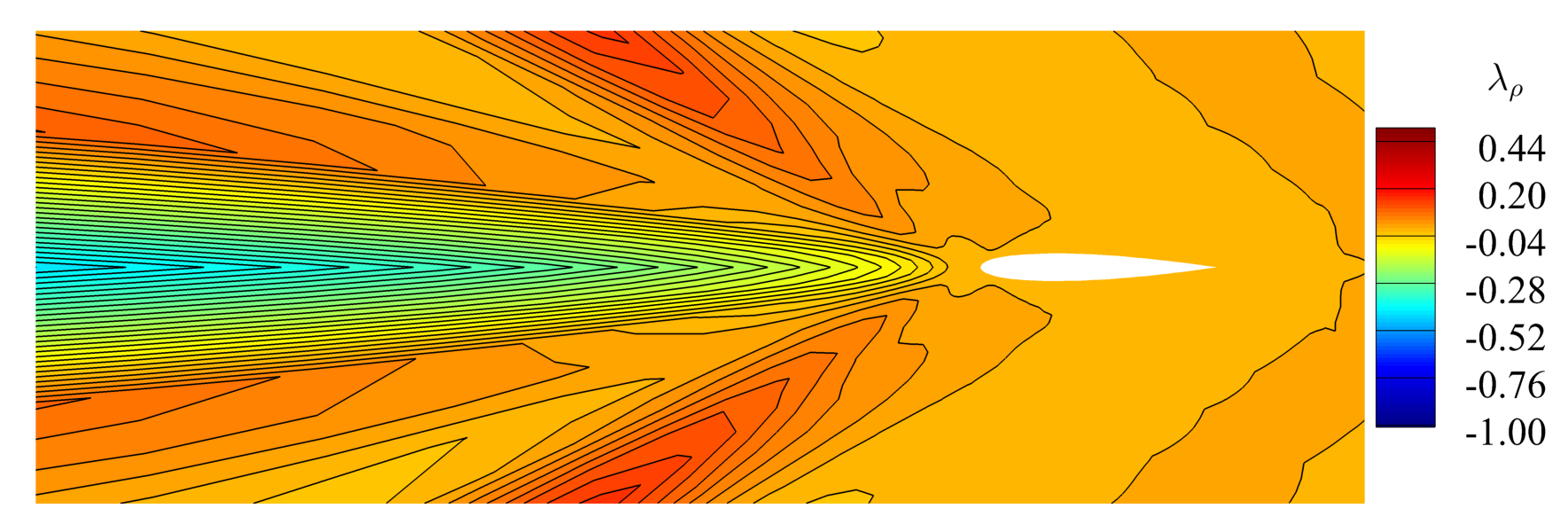}
    \par\vspace{0.15em}{\small (b) Adjoint $\lambda_\rho$}

    \vspace{0.35em}
    \includegraphics[width=0.6\linewidth]{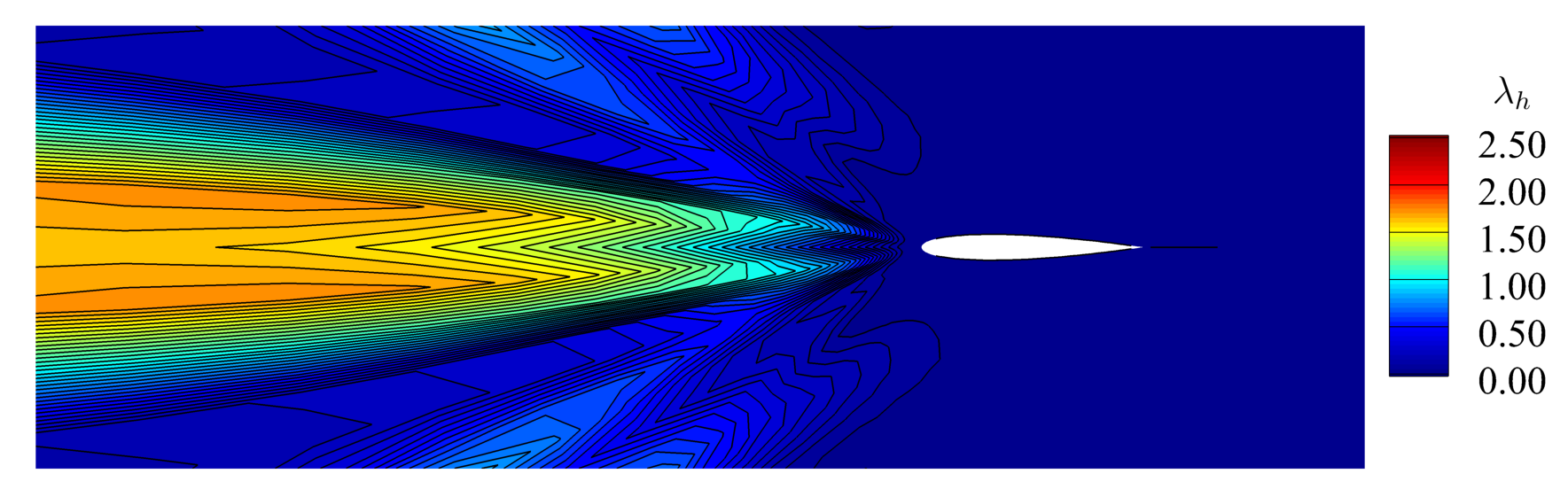}
    \par\vspace{0.15em}{\small (c) Adjoint $\lambda_h$}

    \caption{Representative converged primal and adjoint fields of the baseline NACA0012 airfoil at $\Kn=0.01$: (a) density $\rho$, (b) macroscopic density-adjoint component $\lambda_\rho$, and (c) microscopic adjoint component $\lambda_h$.}
    \label{fig:naca_adjoint_fields}
\end{figure}

Figure~\ref{fig:naca_adjoint_fields} shows the converged density and representative adjoint fields for the baseline NACA0012 airfoil at $\Kn=0.01$. 

The primal density field resolves the compression region ahead of the leading edge. Both adjoint components exhibit prominent structures extending upstream of the airfoil, with different spatial patterns in the microscopic and macroscopic variables. These fields provide a qualitative view of the discrete adjoint solution for the drag objective.
\FloatBarrier

\begin{figure}[!h]
    \centering
    \begin{minipage}[t]{0.48\linewidth}
        \centering
        \includegraphics[width=\linewidth]{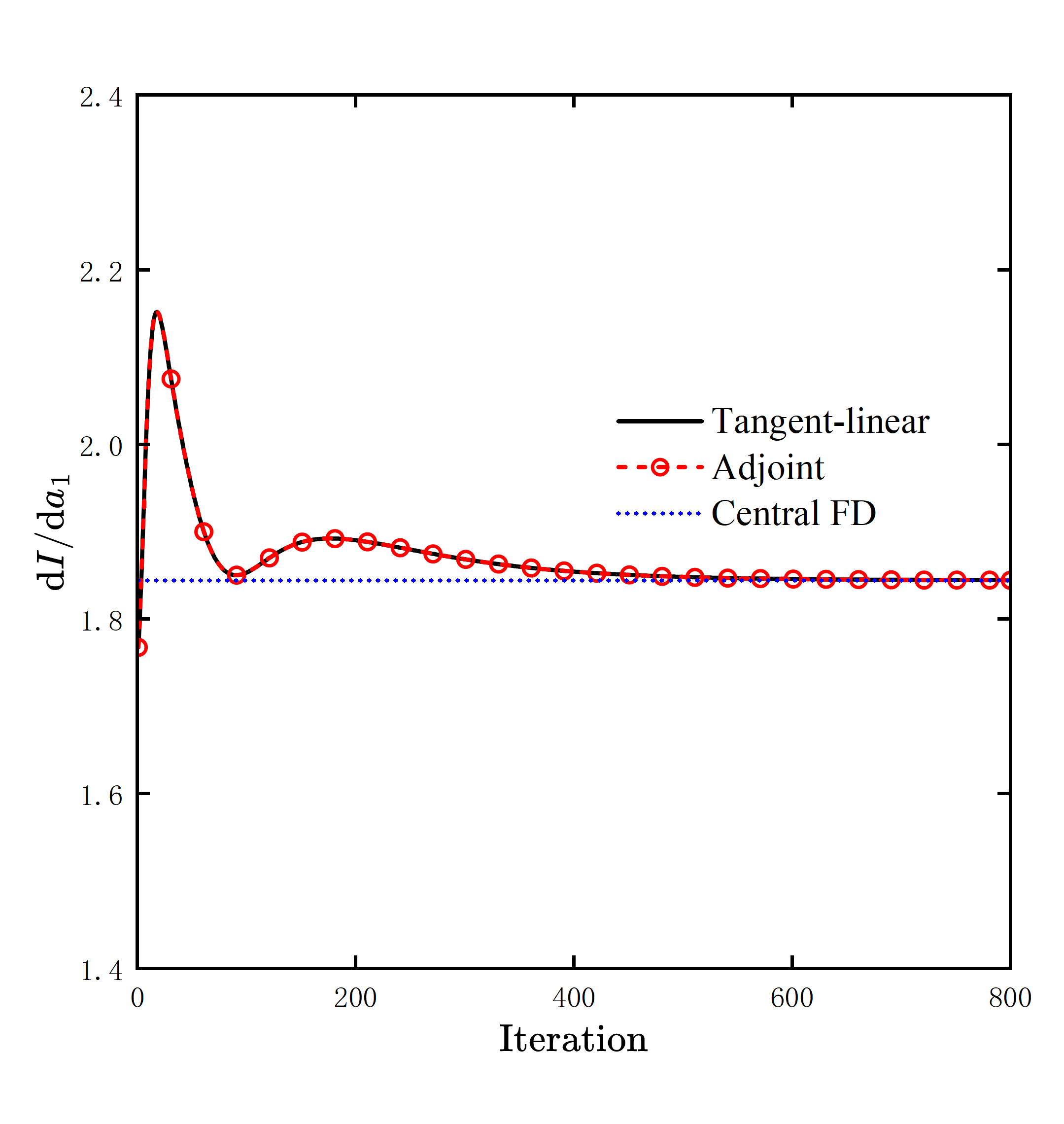}
        \par\smallskip{\small (a) $\Kn=0.01$}
    \end{minipage}\hfill
    \begin{minipage}[t]{0.48\linewidth}
        \centering
        \includegraphics[width=\linewidth]{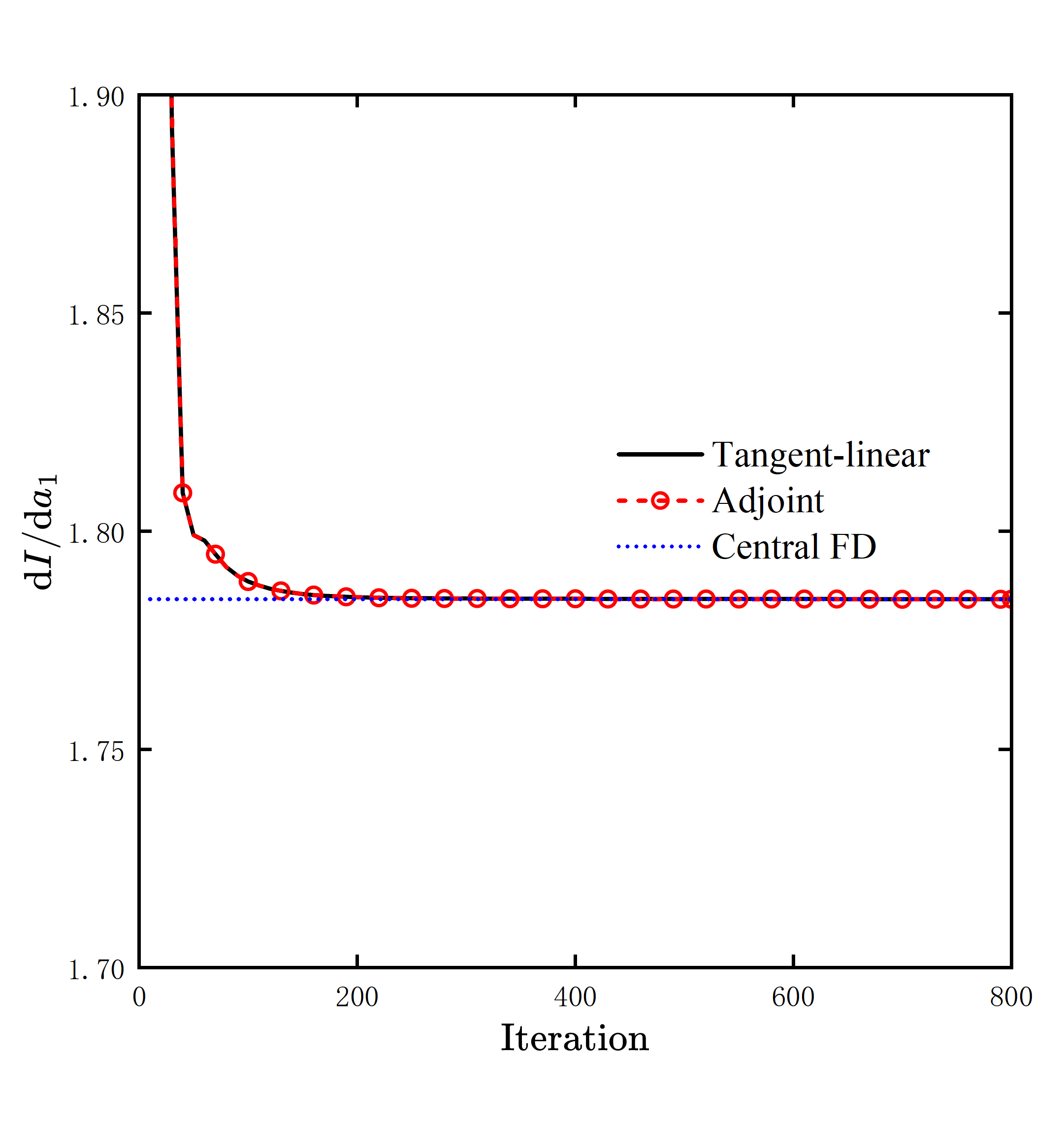}
        \par\smallskip{\small (b) $\Kn=0.5$}
    \end{minipage}
    \caption{Iteration histories of $\dd C_D/\dd a_1$ on the baseline NACA0012 airfoil, compared against central FD benchmarks ($\epsilon=10^{-4}$).}
    \label{fig:naca_sensitivity}
\end{figure}

As shown in Fig.~\ref{fig:naca_sensitivity}, the TL and adjoint histories overlap at every iteration, with maximum stepwise differences of $2.07\times10^{-14}$ at $\Kn=0.01$ and $1.58\times10^{-14}$ at $\Kn=0.5$.

The converged adjoint derivatives are $1.84411548$ at $\Kn=0.01$ and $1.78438564$ at $\Kn=0.5$, compared with central FD benchmarks of $1.84404064$ and $1.78446183$, yielding relative errors of $4.06\times10^{-5}$ and $4.27\times10^{-5}$, respectively. Across the full six-parameter design vector, the relative $\ell_2$ discrepancies between the adjoint and FD gradients are $8.23\times10^{-5}$ at $\Kn=0.01$ and $5.13\times10^{-4}$ at $\Kn=0.5$.
\FloatBarrier

\subsubsection{Drag reduction and active geometric constraints}\label{sec:naca_results}

\begin{figure}[h]
    \centering
    \begin{minipage}[t]{0.48\linewidth}
        \centering
        \includegraphics[width=\linewidth]{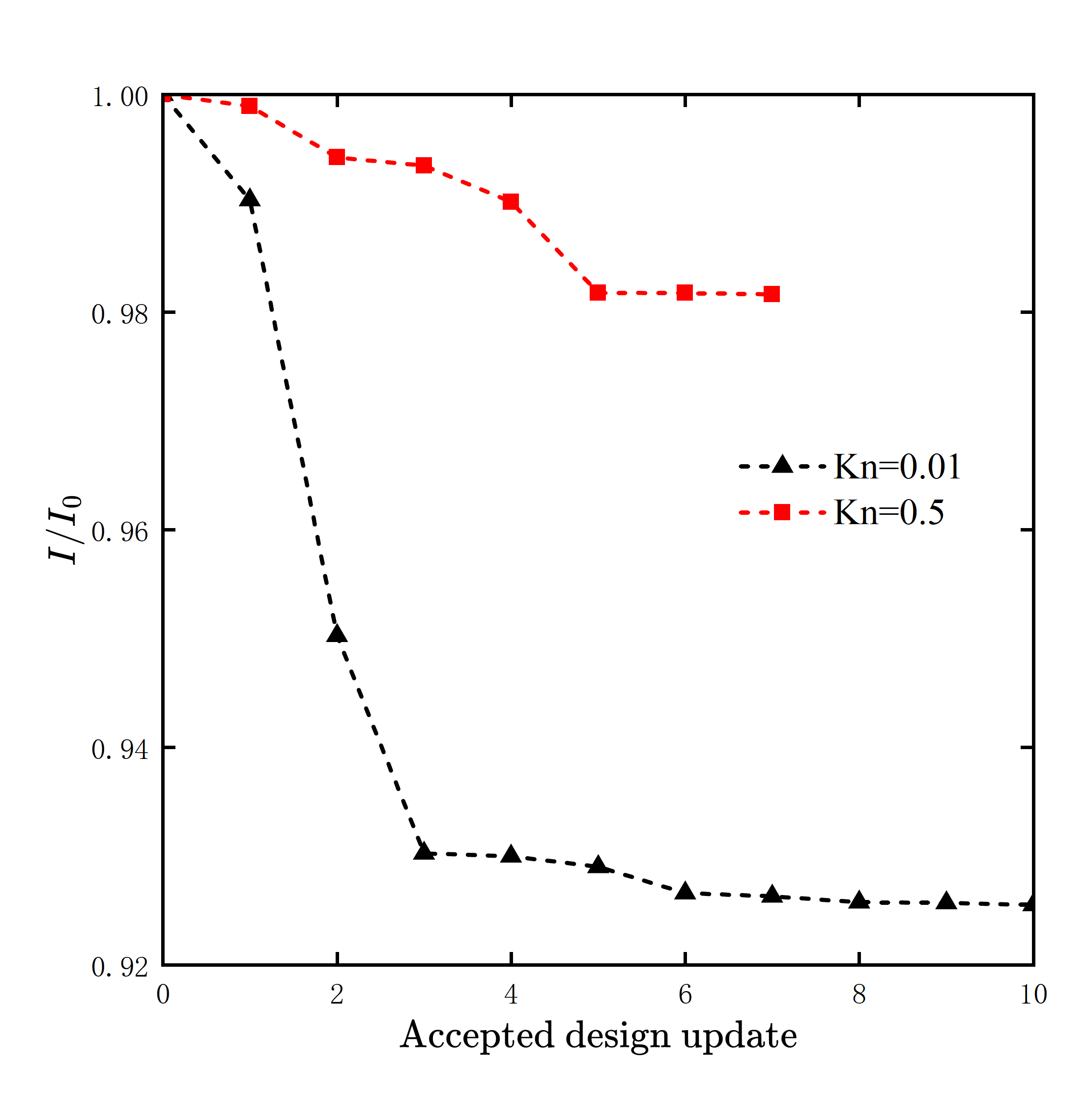}
        \par\smallskip{\small (a) Normalized drag}
    \end{minipage}\hfill
    \begin{minipage}[t]{0.48\linewidth}
        \centering
        \includegraphics[width=\linewidth]{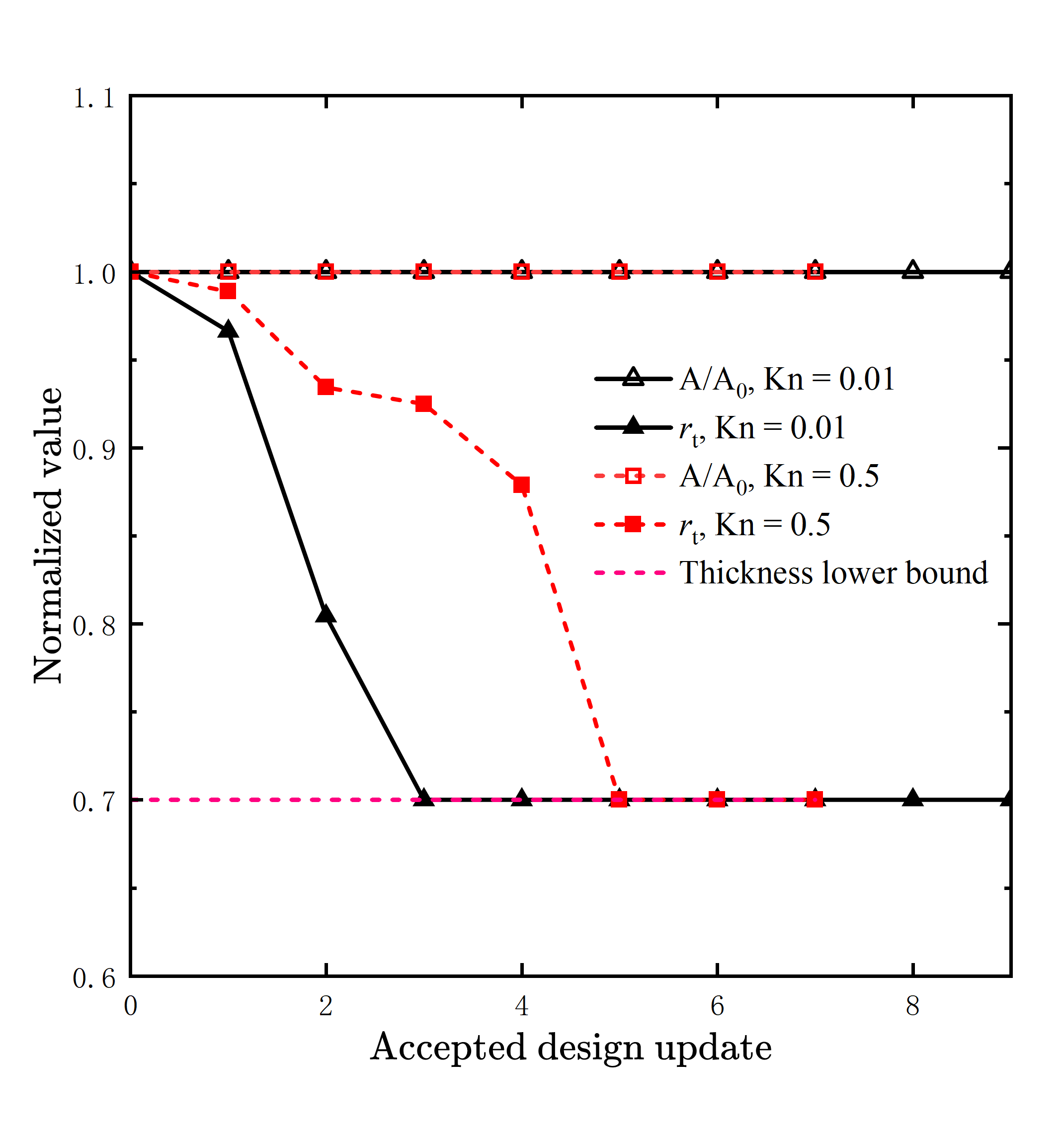}
        \par\smallskip{\small (b) Area and local thickness}
    \end{minipage}
    \caption{Airfoil optimization histories. The minimum thickness ratio is $r_t=\min_j[t(s_j)/t_0(s_j)]$, with lower bound $0.7$.}
    \label{fig:naca_optimization_history}
\end{figure}

Figure~\ref{fig:naca_optimization_history} shows the convergence histories of the normalized drag and geometric constraints. At $\Kn=0.01$, $C_D$ decreases from $0.26372433$ to $0.24405465$ ($7.46\%$ reduction) over ten design updates; at $\Kn=0.5$, $C_D$ decreases from $0.69286305$ to $0.68014212$ ($1.84\%$ reduction) over seven updates. In both cases, most of the drag reduction occurs during the early iterations before the objective plateaus.

As summarized in Table~\ref{tab:naca_results}, the enclosed airfoil area remains unchanged ($A/A_0=1.0000$) and the minimum local thickness constraint ($r_t=0.7000$) becomes active in both optimized designs, whereas the maximum shape-parameter magnitudes ($\max_i|a_i|=0.01293$ and $0.01417$) remain well within the $\pm0.02$ bounds.

\begin{table}[h]
    \centering
    \footnotesize
    \caption{Initial and optimized airfoil designs.}
    \label{tab:naca_results}
    \begin{tabular}{@{}lrr@{}}
        \toprule
        Quantity & $\Kn=0.01$ & $\Kn=0.5$\\
        \midrule
        Initial $C_D$ & 0.26372433 & 0.69286305\\
        Optimized $C_D$ & 0.24405465 & 0.68014212\\
        Drag reduction (\%) & 7.4584 & 1.8360\\
        $A/A_0$ & 1.0000 & 1.0000\\
        $\min_j(t_j/t_{j,0})$ & 0.7000 & 0.7000\\
        $\max_i|a_i|$ & 0.01293 & 0.01417\\
        \bottomrule
    \end{tabular}
\end{table}

\begin{figure}[h]
    \centering
    \begin{minipage}[t]{0.48\linewidth}
        \centering
        \includegraphics[width=\linewidth]{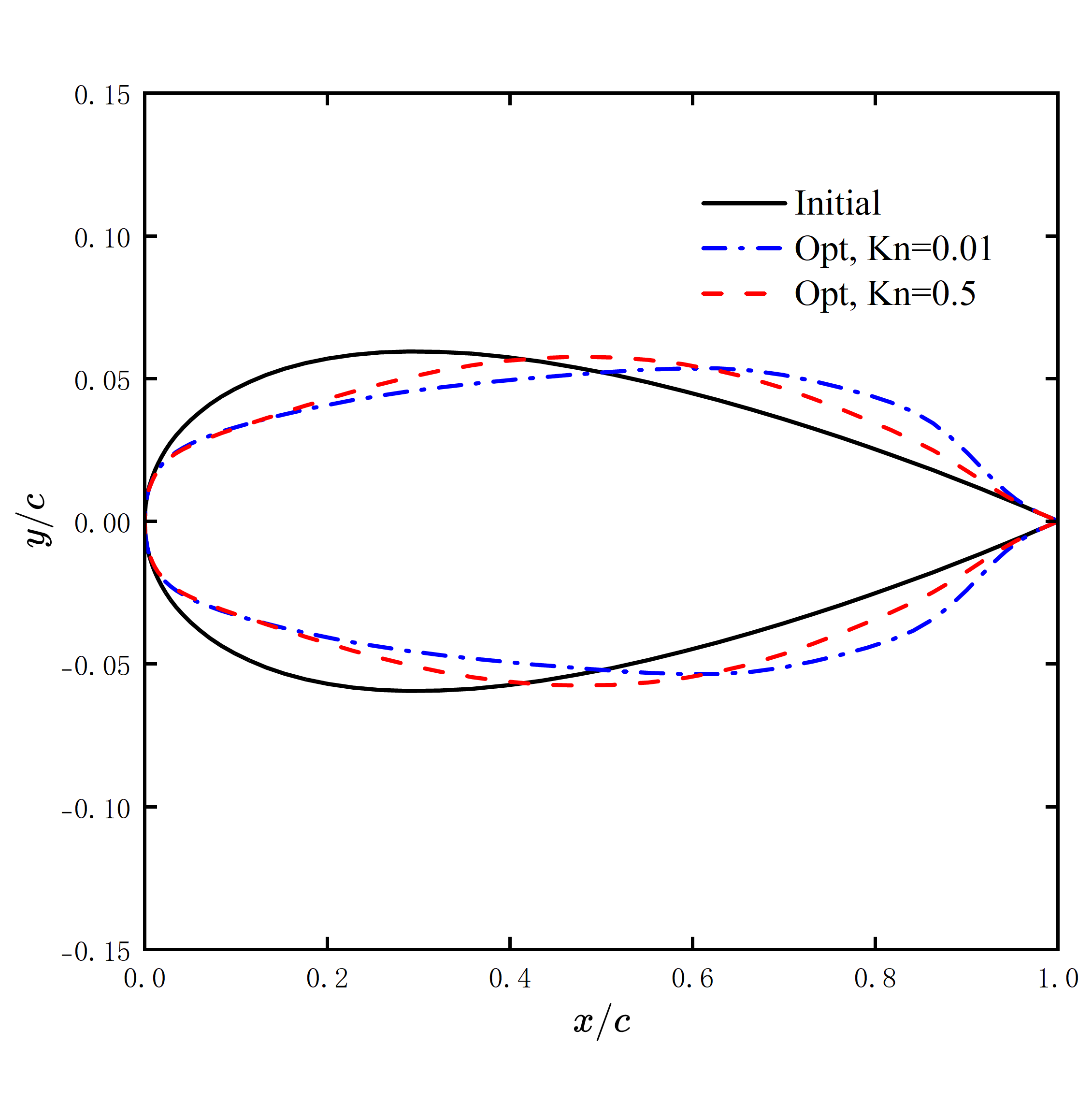}
        \par\smallskip{\small (a) Wall contours}
    \end{minipage}\hfill
    \begin{minipage}[t]{0.48\linewidth}
        \centering
        \includegraphics[width=\linewidth]{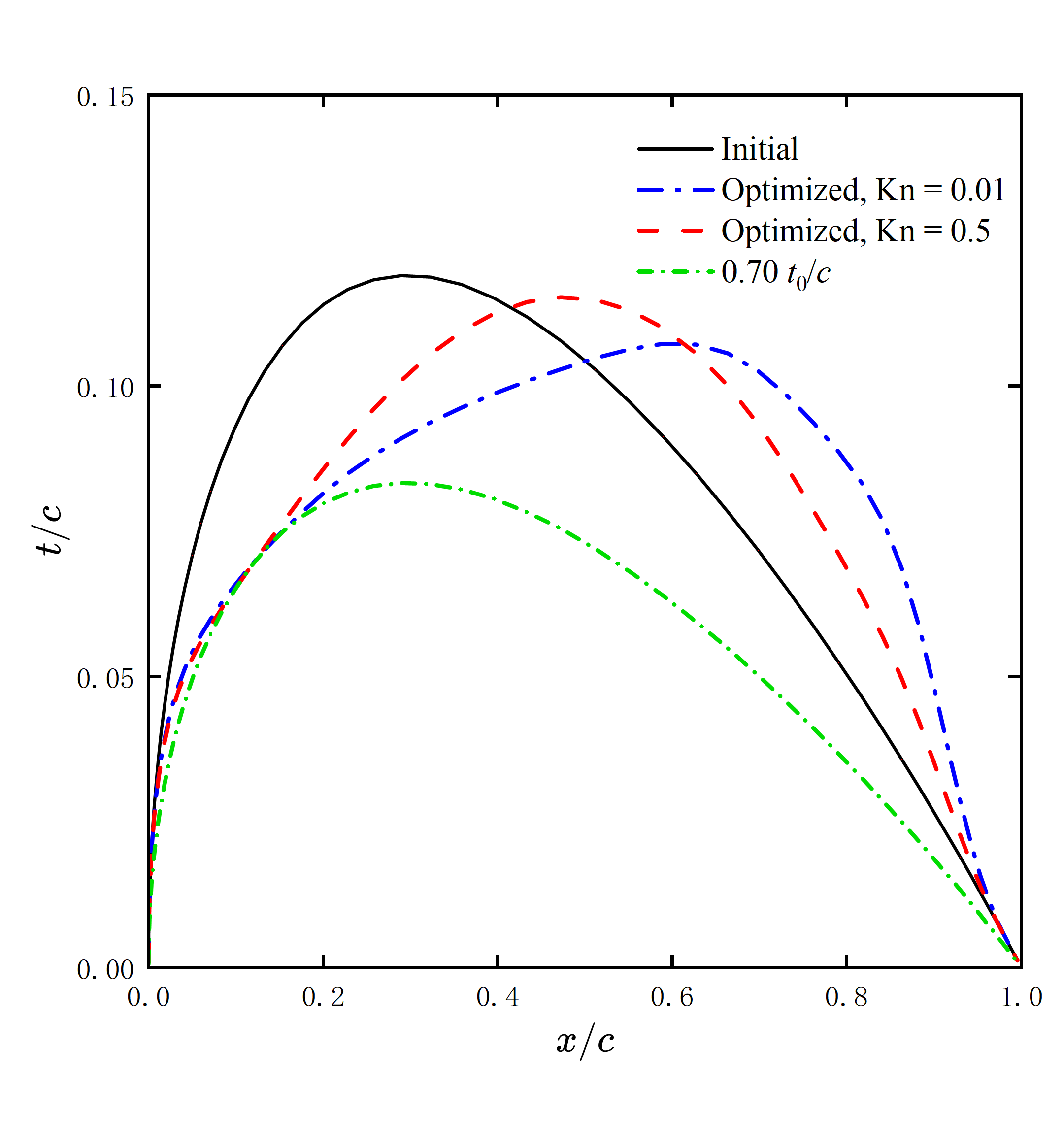}
        \par\smallskip{\small (b) Thickness distributions}
    \end{minipage}
    \caption{Comparison of initial and optimized airfoil shapes.}
    \label{fig:naca_shape_comparison}
\end{figure}

Figure~\ref{fig:naca_shape_comparison} compares the baseline and optimized airfoil contours and chordwise thickness profiles. To reduce drag at constant enclosed area, both designs reduce the thickness of the forward section down to the $0.7t_0$ lower bound and redistribute the conserved volume aftward, shifting the location of maximum thickness downstream. This geometric redistribution is substantially more pronounced at $\Kn=0.01$ than at $\Kn=0.5$: in the near-continuum supersonic regime, wave drag---which scales strongly with forward surface inclination---accounts for a larger share of the total drag, whereas at $\Kn=0.5$ thick diffused shock layers and dominant viscous shear stress limit the drag-reduction potential of forebody-slope modification.

\begin{figure}[h]
    \centering
    \begin{minipage}[t]{0.48\linewidth}
        \centering
        \includegraphics[width=\linewidth]{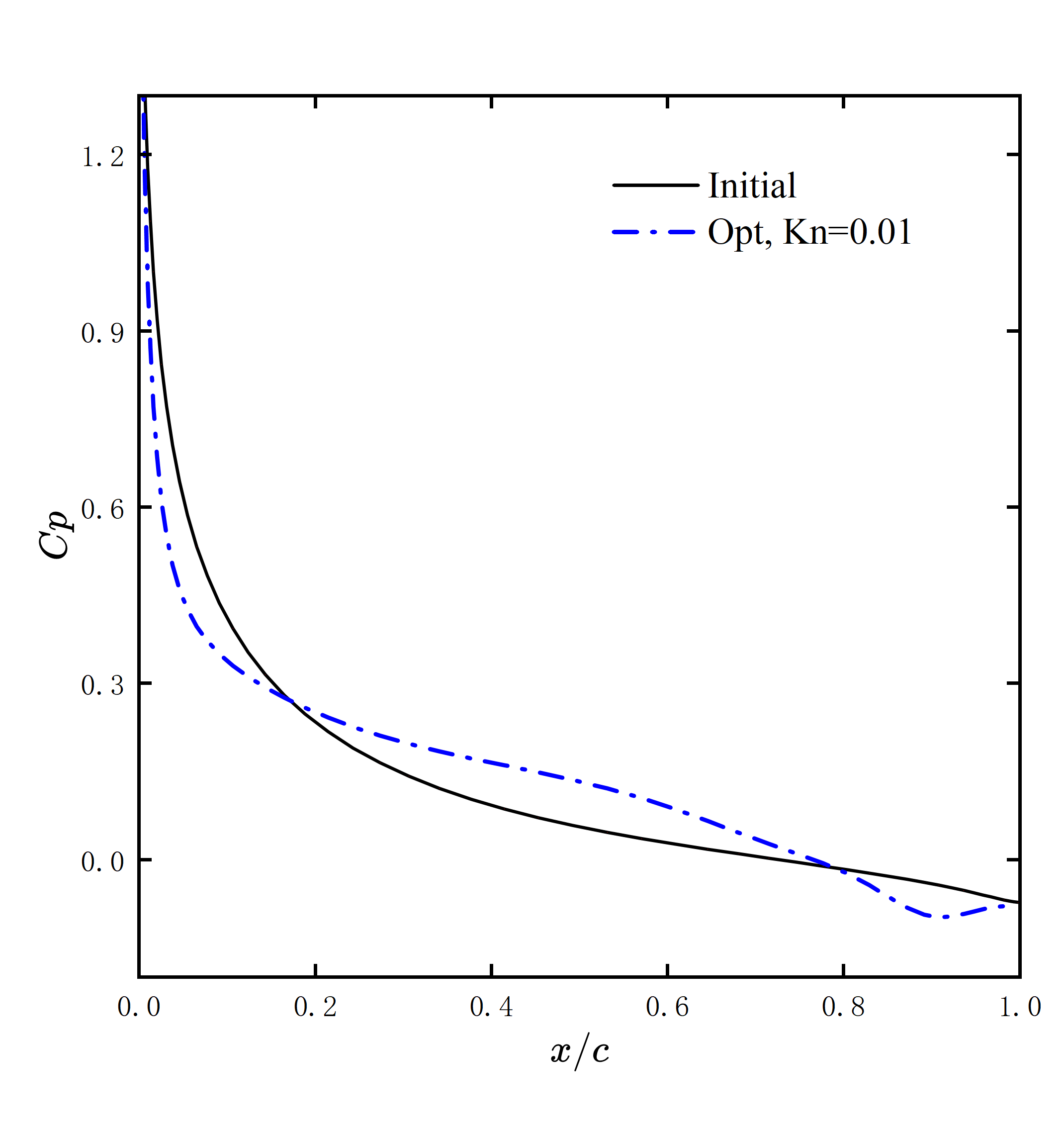}
        \par\smallskip{\small (a) $\Kn=0.01$}
    \end{minipage}\hfill
    \begin{minipage}[t]{0.48\linewidth}
        \centering
        \includegraphics[width=\linewidth]{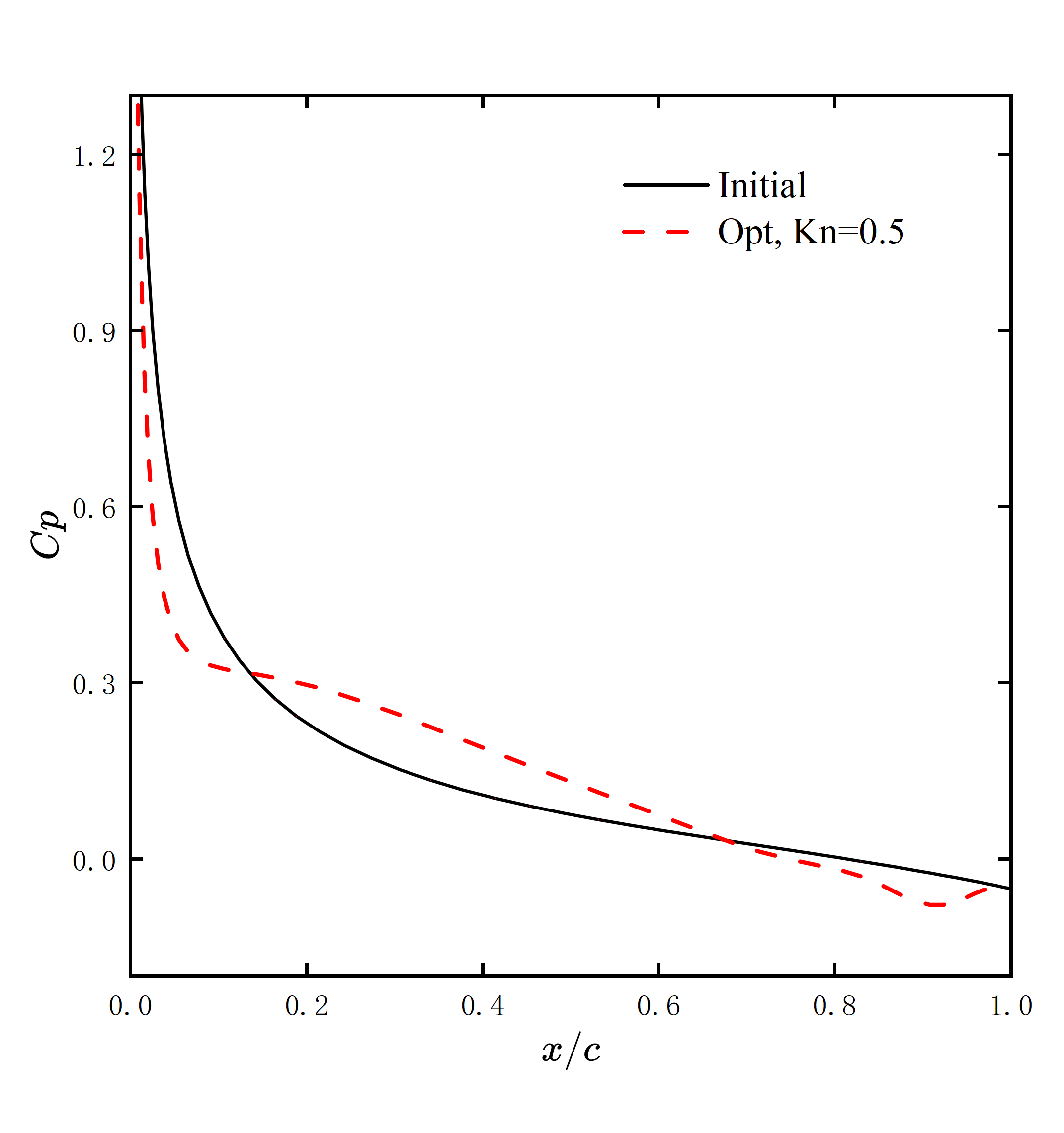}
        \par\smallskip{\small (b) $\Kn=0.5$}
    \end{minipage}
    \caption{Initial and optimized wall-pressure distributions.}
    \label{fig:naca_cp}
\end{figure}

The wall pressure coefficient is defined as
\begin{equation}
    C_p=\frac{\Pi_{nn,w}-p_\infty}
                   {\tfrac12\rho_\infty U_\infty^2},
    \label{eq:wall_cp}
\end{equation}
where $\Pi_{nn,w}$ denotes the normal stress exerted on the airfoil surface and $p_\infty$ is the free-stream static pressure, capturing both equilibrium pressure and nonequilibrium normal stress. As shown in Fig.~\ref{fig:naca_cp}, optimization redistributes the chordwise normal loading: although the leading-edge stagnation peak rises slightly (from $1.8081$ to $1.8419$ at $\Kn=0.01$, and from $2.6775$ to $2.7232$ at $\Kn=0.5$), $C_p$ decreases across the mid-forebody where the surface slope is reduced and increases slightly over the aft portion where area is recovered. Coupled with the reduced forward surface inclination, this chordwise pressure redistribution lowers the net streamwise pressure drag while accounting for tangential shear contributions in Eq.~\eqref{eq:drag_objective}.

\begin{figure}[h]
    \centering
    \begin{minipage}[t]{0.48\linewidth}
        \centering
        \includegraphics[width=\linewidth]{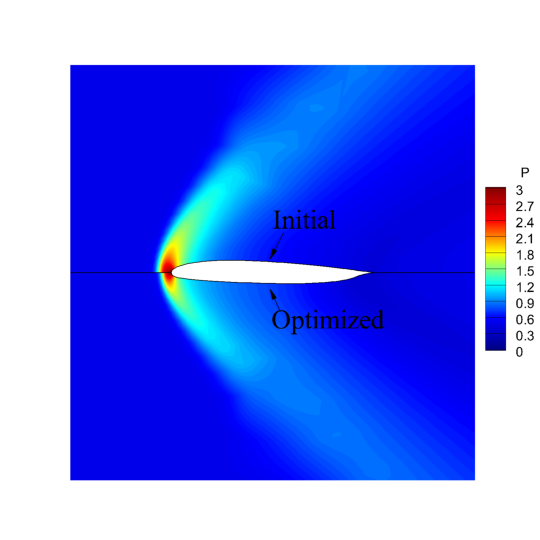}
        \par\smallskip{\small (a) $\Kn=0.01$}
    \end{minipage}\hfill
    \begin{minipage}[t]{0.48\linewidth}
        \centering
        \includegraphics[width=\linewidth]{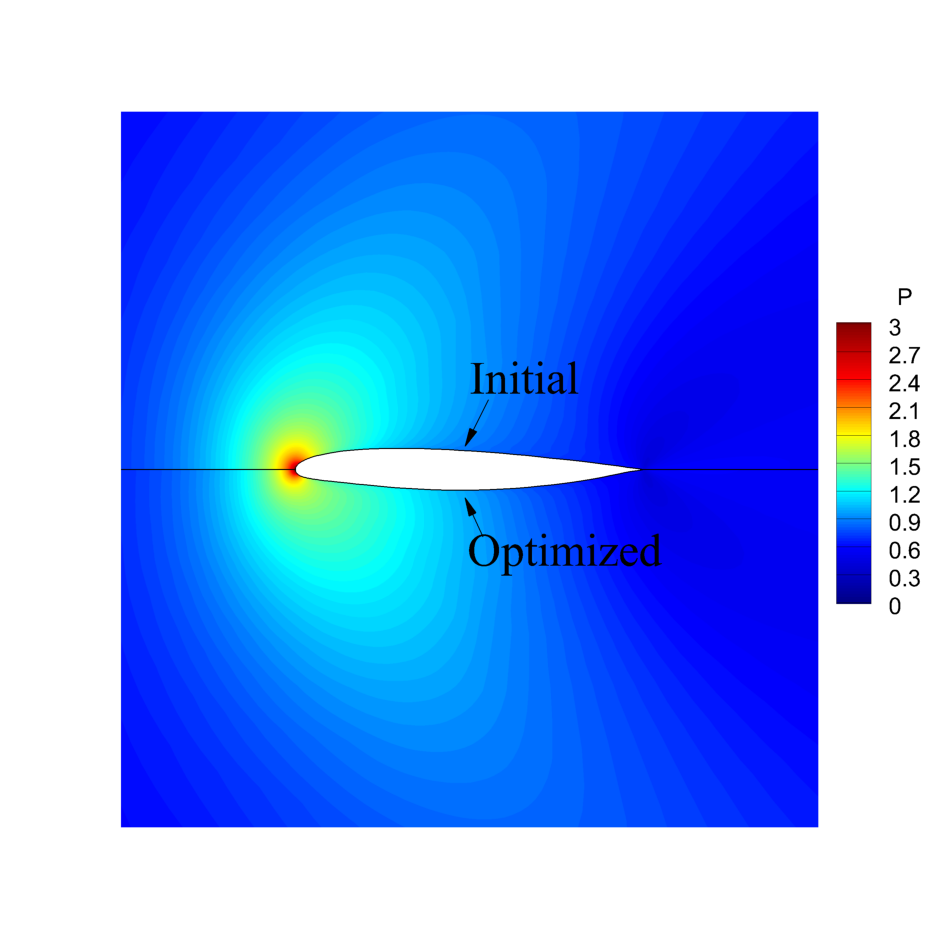}
        \par\smallskip{\small (b) $\Kn=0.5$}
    \end{minipage}
    \caption{Nondimensional pressure fields around the initial (upper half) and optimized (lower half) airfoils.}
    \label{fig:naca_pressure_fields}
\end{figure}

\begin{figure}[h]
    \centering
    \begin{minipage}[t]{0.48\linewidth}
        \centering
        \includegraphics[width=\linewidth]{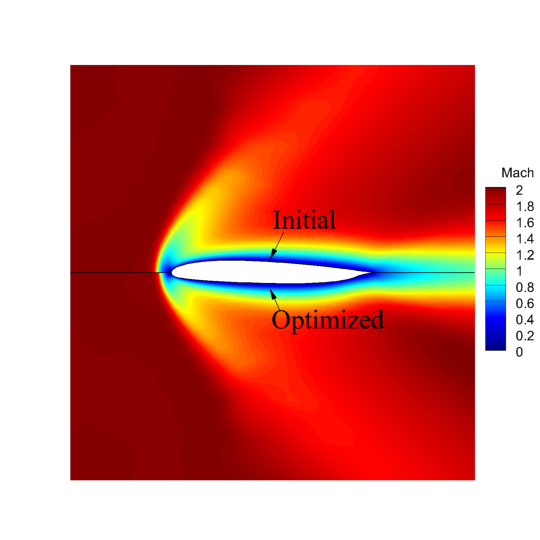}
        \par\smallskip{\small (a) $\Kn=0.01$}
    \end{minipage}\hfill
    \begin{minipage}[t]{0.48\linewidth}
        \centering
        \includegraphics[width=\linewidth]{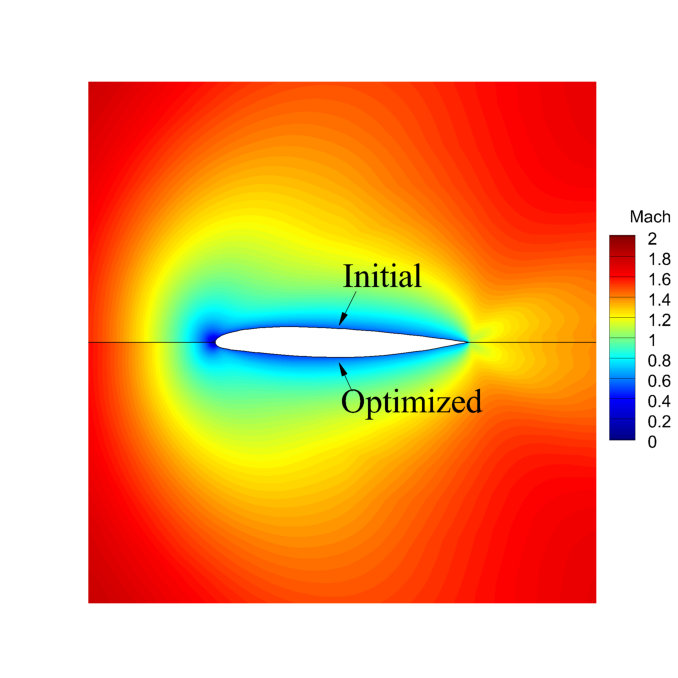}
        \par\smallskip{\small (b) $\Kn=0.5$}
    \end{minipage}
    \caption{Mach-number fields around the initial (upper half) and optimized (lower half) airfoils.}
    \label{fig:naca_mach_fields}
\end{figure}

Figures~\ref{fig:naca_pressure_fields} and~\ref{fig:naca_mach_fields} compare the pressure and Mach-number fields of the initial (upper half) and optimized (lower half) configurations. At $\Kn=0.01$, a well-defined detached bow shock forms ahead of the leading edge; thinning the forward section weakens the oblique compression over the forebody---visibly narrowing the high-pressure and post-shock deceleration layer along the forward shoulder---and yields a slightly narrower low-Mach wake downstream. At $\Kn=0.5$, rarefaction smears the upstream bow shock into a broad, radially diffused compression halo and envelops the airfoil in a thick subsonic viscous layer. Comparing the upper and lower halves shows that forebody thinning draws the elevated-pressure halo and inner subsonic deceleration zone slightly closer to the forward surface, whereas aft volume recovery modestly widens the low-Mach envelope and shifts the pressure contours downstream toward the trailing edge. Because strong molecular diffusion dominates momentum transport in this transitional regime, the geometric modification induces gradual adjustments across the diffuse flow envelope rather than sharp wave-structure changes. Furthermore, re-evaluating the baseline and optimized airfoils on a finer physical mesh ($447\times127$) confirms good spatial grid convergence, with relative differences in the drag-reduction rates of only $0.3\%$ and $0.9\%$ at $\Kn=0.01$ and $0.5$, respectively, compared with the results obtained on the $224\times64$ mesh.

\FloatBarrier

\section{Conclusions}\label{sec:conclusions}

A matrix-free discrete adjoint method has been developed for the steady implicit UGKS to enable sensitivity analysis and shape optimization. Through the exact linearization and transposition of the converged fixed-point operator, the method combines local AD of complex nonlinear kinetic kernels with analytically derived tangent and transpose operators for the coupled macroscopic--microscopic implicit updates and mass-conservation projection. This formulation preserves exact discrete consistency across both flow and geometric dependencies without assembling global Jacobians or incurring the memory overhead of end-to-end reverse-mode AD. For closed and periodic domains, the prescribed total gas inventory and its geometric derivatives are incorporated consistently, while mesh sensitivities across all problem configurations couple directly with arbitrary shape parameterizations.

Numerical benchmarks spanning continuum, transitional, and free-molecular regimes confirm the accuracy and computational efficiency of the framework. In lid-driven cavity tests, the tangent-linear and adjoint solvers exhibit matching asymptotic convergence rates. Across the cavity, Knudsen pump, and airfoil benchmarks, both formulations maintain high stepwise consistency with discrepancies approaching machine precision, and the converged gradients closely match independent central finite-difference estimates. In a six-parameter pump benchmark, obtaining the complete gradient from a single adjoint calculation achieves speedups of $7.54\times$ and $3.88\times$ over tangent-linear and finite-difference evaluations, respectively.

Constrained shape optimizations of a periodic S-shaped Knudsen pump and a Mach-2 NACA0012 airfoil demonstrate the capability of the framework for internal and external multiscale flows. Under constant-area and local width or thickness bounds, the optimizer achieves normalized mass-flow gains of $9.81\%$ ($\Kn=0.1$) and $29.68\%$ ($\Kn=10$) for the thermal micropump by suppressing parasitic recirculation, and aerodynamic drag reductions of $7.46\%$ ($\Kn=0.01$) and $1.84\%$ ($\Kn=0.5$) for the supersonic airfoil by reducing forward thickness and shifting the maximum thickness downstream. Since the adjoint computational cost is independent of the number of design variables, extending the present method to high-dimensional design spaces and three-dimensional multiscale flows with complex geometries will be pursued in future work.

\section*{Acknowledgements}

The current research is supported by the National Key R\&D Program of China (Grant No. 2022YFA1004500), the National Natural Science Foundation of China (92371107), and the Hong Kong Research Grants Council (16208324).

\appendix

\section{Point-relaxation, LU-SGS, and mass correction}\label{app:sweeps}

This appendix documents the detailed sweep coefficients, boundary preconditioning treatments, and bordered matrix formulation underlying the implicit iteration in Section~\ref{sec:implicit}.

For each cell $c$, the local spectral wave speed and characteristic spectral time entering Eq.~\eqref{eq:fluxtime} are evaluated as
\begin{equation}
    \begin{gathered}
        a_{cf}^{\mathrm{spec}} = a_{s,c}+\max\!\left(\left|\vect{U}_c\cdot\vect{n}_{cf}\right|,\max_k\left|\vect{u}_k\cdot\vect{n}_{cf}\right|\right),\\
        t_c^{\mathrm{spec}} = \frac{2A_c}{\displaystyle\sum_{f\in\partial\Omega_c}L_f a_{cf}^{\mathrm{spec}}},
        \qquad
        a_{s,c}=\sqrt{\frac{\gamma p_c}{\rho_c}}.
    \end{gathered}
    \label{eq:spectraltime}
\end{equation}

In the macroscopic predictor [Eqs.~\eqref{eq:app_prflux}--\eqref{eq:app_prvisit}], the Euler flux along unit normal $\vect{n}$ is given by
\begin{equation}
    \vect{F}_E(\vect{W};\vect{n})=
    \begin{pmatrix}
        \rho U_n\\
        \rho\vect{U}U_n+p\vect{n}\\
        (\rho E+p)U_n
    \end{pmatrix},
    \qquad
    U_n=\vect{U}\cdot\vect{n},
    \label{eq:app_eulerflux}
\end{equation}
and the interface dissipation coefficient reads $r_{cj}=\kappa_r\bigl(|\vect{U}_{cj}\cdot\vect{n}_{cj}|+a_{s,cj}\bigr)$, where $a_{s,cj}$ is the sound speed evaluated from the face-averaged conservative state and $\kappa_r$ is a numerical dissipation constant. During the point-relaxation passes, the newly computed increment in Eq.~\eqref{eq:app_prvisit} overwrites the previous value in cell $c$ in place, while $D_c^W$ and $\vect{b}_c^W$ remain frozen; two forward--backward sweep pairs are executed per implicit iteration. At a solid wall with prescribed fixed velocity, the macroscopic ghost-cell increment is approximated via the reflection relation
\begin{equation}
    \Delta\vect{W}_g=\operatorname{diag}(1,-1,-1,1)\,\Delta\vect{W}_c,
    \label{eq:app_prwall}
\end{equation}
evaluated using the adjacent interior cell value prior to its current-pass update. This treatment is exact for a stationary wall under equal-energy reflection and serves as a simplified diagonal preconditioning treatment for moving walls without affecting the converged steady state.

For the microscopic LU-SGS correction in Eq.~\eqref{eq:app_kinmatrix}, let $u_{n,cj,k}=\vect{u}_k\cdot\vect{n}_{cj}$. Using the predicted collision relaxation time $\widetilde\tau_c=\tau(\widetilde{\vect{W}}_c)$, the frozen upwind coefficients are
\begin{equation}
    \begin{gathered}
        D_{c,k}=\frac{1}{\Delta\widetilde t_c}+\frac{1}{\widetilde\tau_c}+\sum_{f\in\partial\Omega_c}\frac{L_f}{A_c}\max\!\left(\vect{u}_k\cdot\vect{n}_{cf},0\right),\\
        C_{cj,k}=\frac{L_{cj}}{A_c}\min\!\left(u_{n,cj,k},0\right).
    \end{gathered}
    \label{eq:app_kincoeff}
\end{equation}
For each reduced distribution $z\in\{h,b\}$, the cellwise forward and backward sweeps read
\begin{equation}
    \begin{gathered}
        y^z_{c,k}=\frac{r^z_{c,k}-\displaystyle\sum_{j\in\mathcal L(c)}C_{cj,k}y^z_{j,k}}{D_{c,k}},\\[5pt]
        \Delta z_{c,k}=y^z_{c,k}-\frac{\displaystyle\sum_{j\in\mathcal U(c)}C_{cj,k}\Delta z_{j,k}}{D_{c,k}},
    \end{gathered}
    \label{eq:app_kinsweeps}
\end{equation}
where $\mathcal L(c)$ and $\mathcal U(c)$ denote the lower- and upper-index neighbors in the sweep ordering, and the resulting increment $\Delta z_{c,k}$ is scaled by $\theta$ in forming $\mathcal P_{\mathrm{kin}}\vect{r}$. At a diffuse-reflection wall, the reflected increments are coupled to the incident increments across the discrete velocity space. Using the wall-to-gas unit normal $\vect{n}_w$ and Eqs.~\eqref{eq:wallmaxwell}--\eqref{eq:wallmassflux}, the wall-density increment at fixed wall temperature is
\begin{equation}
    \begin{gathered}
        \Delta\rho_w=-\frac{\displaystyle\sum_{u_{n,k}^{w}<0}w_k u_{n,k}^{w}y^h_{\mathrm{gas},k}}{J_{\mathrm{out}}^{(1)}},\\
        (\Delta h_{w,k},\Delta b_{w,k})=\Delta\rho_w\bigl(H_{w,k}^{(1)},B_{w,k}^{(1)}\bigr),
        \qquad u_{n,k}^{w}>0.
    \end{gathered}
    \label{eq:app_kinwall}
\end{equation}
A preliminary forward pass supplies the incident increments $y^h_{\mathrm{gas},k}$, from which $\Delta\rho_w$ determines the reflected ghost increments for the subsequent sweeps: ghost increments on lower-index boundaries ($\mathcal L$) are initialized to zero, whereas the reflected values are enforced on upper-index ghost boundaries ($\mathcal U$) during the backward sweep. The same diffuse-wall coupling is included in $\mathcal P_{\mathrm{kin}}$ when evaluating the mass-response vector $\vect{z}_M$.

For the global mass correction in Eqs.~\eqref{eq:massresponse}--\eqref{eq:massschur}, the normalized kinetic density source $\vect{s}$ is defined as
\begin{equation}
    \begin{gathered}
        s_{h,c,k}
        =\frac{\exp(-|\vect{u}_k|^2)}
                {\displaystyle\sum_{j=1}^{N_v}
                 w_j\exp(-|\vect{u}_j|^2)},\\
        s_{b,c,k}=0,
    \end{gathered}
    \label{eq:masssource}
\end{equation}
which satisfies $\sum_{k=1}^{N_v} w_k s_{h,c,k}=1$ and defines an algebraic mass-correction mode. Assuming $\mathcal P_{\mathrm{kin}}$ is invertible and setting $\mathsf K=\mathcal P_{\mathrm{kin}}^{-1}$, the mass-constrained update can be written equivalently as the bordered linear system
\begin{equation}
    \begin{bmatrix}
        \mathsf K&\vect{s}\\
        \vect{m}^T&0
    \end{bmatrix}
    \begin{bmatrix}
        \Delta\vect{f}\\
        \chi
    \end{bmatrix}
    =
    \begin{bmatrix}
        \vect{r}\\
        M_\star-\vect{m}^T\vect{f}^n
    \end{bmatrix}.
    \label{eq:app_massborder}
\end{equation}
Eliminating the first block row yields
\begin{equation}
    \Delta\vect{f}=\mathcal P_{\mathrm{kin}}\vect{r}-\mathcal P_{\mathrm{kin}}\vect{s}\,\chi.
    \label{eq:app_mass_df}
\end{equation}
Substituting Eq.~\eqref{eq:app_mass_df} together with Eqs.~\eqref{eq:microprovisional} and~\eqref{eq:massresponse} into the second block row reduces the mass constraint to
\begin{equation}
    \vect{m}^T\widehat{\vect{f}}-\chi\,\vect{m}^T\vect{z}_M=M_\star,
    \label{eq:app_masselimination}
\end{equation}
whose solution gives the Schur-complement mass multiplier $\chi$ in Eq.~\eqref{eq:massschur}.

\section{Tangent and transpose of the implicit update}\label{app:transpose}

The tangent and transpose operations are evaluated at the converged steady state, where both the physical residuals and implicit increments vanish identically. Because the point-relaxation and LU-SGS preconditioning coefficients are frozen at this base state, their derivatives multiply vanishing increments and drop out of the linearization, whereas the derivatives of the physical residuals and collision sources are retained.

For the macroscopic point-relaxation predictor, the neighboring-cell flux difference $\mathcal H_{cj}$ in Eq.~\eqref{eq:app_prflux} is linearized at zero increment. Because the two Euler-flux evaluations coincide at this state, their derivatives with respect to the base state and interface geometry cancel, while the derivative of $r_{cj}$ multiplies a vanishing increment. The directional derivative therefore reduces to
\begin{equation}
    \left[L_{cj}\mathcal H_{cj}(\Delta\vect{W}_j)\right]_d
    =\frac{L_{cj}}{2}\left[\mathsf A_{E,j}(\vect{n}_{cj})-r_{cj}\mathsf I\right]\Delta\vect{W}_{j,d}
    =\mathsf H_{cj}\Delta\vect{W}_{j,d},
    \label{eq:app_pr_tangent_flux}
\end{equation}
where $\mathsf A_{E,j}$ and $\mathsf H_{cj}$ are defined in Section~\ref{sec:implicit_derivatives}. Similarly, differentiating the area-weighted residual $\vect{b}_c^W$ in Eq.~\eqref{eq:app_prvisit} at zero residual and noting that the derivative of $D_c^W$ multiplies a vanishing increment yields the tangent of a single local point-relaxation update:
\begin{equation}
    \begin{gathered}
        (\vect{b}_c^W)_d=-A_c(\mathcal R_c^W)_d,\\
        \Delta\vect{W}_{c,d}^{\mathrm{new}}=\frac{1}{D_c^W}\left[(\vect{b}_c^W)_d-\sum_{j\in\mathcal N(c)}\mathsf H_{cj}\Delta\vect{W}_{j,d}^{\mathrm{avail}}\right],
    \end{gathered}
    \label{eq:app_pr_tangent_visit}
\end{equation}
where the superscript $\mathrm{avail}$ denotes the neighbor increment available at that stage of the sweep. Initialized from zero increments, the tangent sweep executes the two forward--backward point-relaxation pairs in the same sequence as the primal solver, enforcing the wall-ghost relation in Eq.~\eqref{eq:app_prwall} at the identical sweep stage.

For the transpose of this linearized local update, an adjoint seed $\overline{\Delta\vect{W}}_c^{\mathrm{new}}$ on the updated increment contributes to the source vector $\vect{b}_c^W$ and neighbor increments via
\begin{equation}
    \left(\overline{\vect{b}}_c^W\right)_{\mathrm{visit}}=\frac{1}{D_c^W}\overline{\Delta\vect{W}}_c^{\mathrm{new}},
    \qquad
    \left(\overline{\Delta\vect{W}}_j^{\mathrm{avail}}\right)_{\mathrm{visit}}=-\frac{1}{D_c^W}\mathsf H_{cj}^T\overline{\Delta\vect{W}}_c^{\mathrm{new}}.
    \label{eq:app_pr_reverse_visit}
\end{equation}
These adjoint contributions are accumulated by traversing the cell updates in strictly reverse sweep order. Because each local update overwrites the owner-cell increment, the prior value in cell $c$ receives adjoint contributions solely from operations that read it prior to overwriting, including the wall-ghost relation. Back-propagating through the macroscopic predictor in Eq.~\eqref{eq:macroprediction} passes its adjoint seed directly to $\vect{W}$ and, scaled by $\theta$, to the final increment $\Delta\vect{W}$. Finally, back-propagating through the area-weighted residual relation in Eq.~\eqref{eq:app_pr_tangent_visit} scales $\overline{\vect{b}}_c^W$ by $-A_c$ to form the macroscopic residual seed, recovering Eq.~\eqref{eq:pr_transpose_visit}.

For the microscopic correction, the collision sources evaluated at the current and predicted macroscopic states are denoted by $S^z$ and $\widetilde S^z$, respectively. Differentiating Eq.~\eqref{eq:microdefect} and the collision source defined in Section~\ref{sec:implicit} gives
\begin{equation}
    \begin{gathered}
        (r^z)_d=-(\mathcal R^z)_d-(S^z)_d+(\widetilde S^z)_d,
        \qquad z\in\{h,b\},\\
        (S^z)_d=\frac{(z^S)_d-z_d}{\tau}-\frac{z^S-z}{\tau^2}\tau_d,
    \end{gathered}
    \label{eq:app_micro_rhs_tangent}
\end{equation}
where $(S^z)_d$ and $(\widetilde S^z)_d$ are evaluated using $\vect{W}_d$ and $\widetilde{\vect{W}}_d$, respectively, while both retain the variation of the current reduced distributions $z_d$. In reverse mode, an adjoint seed $\overline{\vect{r}}$ contributes $-\overline{\vect{r}}$ to both the microscopic residual seed and the current collision-source seed, and $+\overline{\vect{r}}$ to the predicted collision-source seed, which are subsequently back-propagated through the Shakhov target distributions and relaxation times.

For interior and periodic couplings, the microscopic LU-SGS operator $\mathcal P_{\mathrm{kin}}$ in Eq.~\eqref{eq:app_kinmatrix} acts linearly on $\vect{r}$ in Eq.~\eqref{eq:microprovisional}. An adjoint seed $\overline{\widehat{\vect{f}}}$ therefore contributes directly to the seed on $\vect{f}$ and propagates to the defect seed $\overline{\vect{r}}$ via the two reverse triangular sweeps in Eq.~\eqref{eq:micro_transpose_sweeps}. The diffuse-wall increment coupling in Eq.~\eqref{eq:app_kinwall} is incorporated directly within these reverse sweeps. With $\overline{\Delta h}_{w,k}$ and $\overline{\Delta b}_{w,k}$ denoting the adjoint seeds on the reflected wall increments at the converged base state, their contribution to the wall-density-increment seed and the resulting contribution to the incident forward-sweep increment read
\begin{equation}
    \begin{gathered}
        \overline{\Delta\rho}_w=\sum_{u_{n,k}^w>0}\left[H_{w,k}^{(1)}\overline{\Delta h}_{w,k}+B_{w,k}^{(1)}\overline{\Delta b}_{w,k}\right],\\
        \left(\overline{y}_{\mathrm{gas},k}^h\right)_{\mathrm{wall}}=-\frac{w_ku_{n,k}^w}{J_{\mathrm{out}}^{(1)}}\overline{\Delta\rho}_w,
        \qquad u_{n,k}^w<0.
    \end{gathered}
    \label{eq:app_kin_wall_reverse}
\end{equation}
These adjoint contributions are accumulated in the reverse sequence of the wall-density reduction and microscopic sweeps with all boundary coupling coefficients frozen at the converged state.

For closed and periodic systems with a prescribed total gas mass $M_\star$, differentiating the mass constraint in Eq.~\eqref{eq:massconstraint} under state and mesh perturbations yields
\begin{equation}
    \begin{gathered}
        \sum_c\left(A_c\rho_{c,d}+\rho_c A_{c,d}\right)=0,\\
        \vect{m}^T\vect{f}_d+(\vect{m}_d)^T\vect{f}=0,
        \qquad
        (\vect{m}_d)^T\vect{f}=\sum_c\rho_c A_{c,d},
    \end{gathered}
    \label{eq:massshape}
\end{equation}
where $A_{c,d}$ denotes the cell-area variation. Differentiating the scalar mass correction in Eq.~\eqref{eq:massschur} prior to imposing steady-state convergence gives
\begin{equation}
    \begin{gathered}
        \sigma_d=\vect{m}_d^T\vect{z}_M+\vect{m}^T\vect{z}_{M,d},\\
        \chi_d=\frac{\vect{m}^T\widehat{\vect{f}}_d+\vect{m}_d^T\widehat{\vect{f}}-M_{\star,d}-\chi\sigma_d}{\sigma},\\
        \vect{f}_d^+=\widehat{\vect{f}}_d-\vect{z}_M\chi_d-\chi\vect{z}_{M,d}.
    \end{gathered}
    \label{eq:app_mass_full_tangent}
\end{equation}
For a fixed gas mass evaluated at the converged fixed point, where the mass multiplier $\chi$ and mass variation $M_{\star,d}$ vanish and the provisional distribution coincides with $\vect{f}$, all terms multiplying $\chi$ drop out identically. Introducing the linear mass projection operator
\begin{equation}
    \Pi=\mathsf I-\frac{\vect{z}_M\vect{m}^T}{\sigma},
    \qquad
    \vect{m}^T\Pi=\vect{0}^T,
    \label{eq:massprojector}
\end{equation}
the tangent variations of the accepted microscopic and macroscopic states reduce to
\begin{equation}
    \vect{f}_d^+=\Pi\widehat{\vect{f}}_d-\frac{\vect{z}_M}{\sigma}(\vect{m}_d)^T\vect{f},
    \qquad
    \vect{W}_d^+=\mathcal M\vect{f}_d^+.
    \label{eq:massbordertangent}
\end{equation}
The geometric contribution $(\vect{m}_d)^T\vect{f}$ generally remains nonzero even for area-preserving shape perturbations because local cell areas deform nonuniformly, whereas it vanishes on a fixed mesh. By the orthogonality property in Eq.~\eqref{eq:massprojector}, the projected tangent update strictly satisfies Eq.~\eqref{eq:massshape}.

For the corresponding transpose operation, combining intermediate adjoint seeds on the accepted state through moment recovery yields
\begin{equation}
    \vect{p}=\overline{\vect{f}}^{+}+\mathcal M^T\overline{\vect{W}}^{+},
    \qquad
    \nu=\frac{\vect{z}_M^T\vect{p}}{\sigma}.
    \label{eq:momenttranspose}
\end{equation}
From Eq.~\eqref{eq:massbordertangent}, the discrete inner-product identity
\begin{equation}
    (\overline{\vect{f}}^+)^T\vect{f}_d^++(\overline{\vect{W}}^+)^T\vect{W}_d^+
    =\vect{p}^T\Pi\widehat{\vect{f}}_d-\nu(\vect{m}_d)^T\vect{f}
    =(\Pi^T\vect{p})^T\widehat{\vect{f}}_d-\sum_c\rho_c\nu A_{c,d}
    \label{eq:app_mass_duality}
\end{equation}
directly recovers the transposed provisional-state seed $\overline{\widehat{\vect{f}}}$ and the cell-area adjoint contribution $\overline{A}_c^{\mathrm{mass}}$ stated in Eq.~\eqref{eq:massborderadjoint}. Retaining both $(\vect{m}_d)^T\vect{f}$ and $\overline{A}_c^{\mathrm{mass}}$ preserves exact discrete duality under arbitrary mesh deformation. In reverse mode, this mass-projection transpose is executed first, prior to back-propagating through the microscopic correction and macroscopic predictor.

\section{Geometry and boundary derivatives}\label{app:geometry}

This appendix summarizes the directional derivative relations evaluated by the local AD kernels in Section~\ref{sec:local_ad} for mesh metrics, spatial reconstruction, flux-evolution intervals, kinetic boundaries, and objective normalizations. Here, the subscript $d$ denotes a directional derivative as in Section~\ref{sec:discretefixedpoint}, while the discrete molecular velocity nodes $\vect{u}_k$ and quadrature weights $w_k$ remain fixed under mesh deformation.

For an interface $f$ bounded by vertices $\vect{x}_{f,1}$ and $\vect{x}_{f,2}$, with fixed outward quarter-turn rotation matrix $\mathsf R_{cf}$, the face metrics and their directional derivatives read
\begin{equation}
    \begin{gathered}
        \vect{e}_f=\vect{x}_{f,2}-\vect{x}_{f,1},
        \qquad
        L_f=|\vect{e}_f|,
        \qquad
        \vect{n}_{cf}=\frac{\mathsf R_{cf}\vect{e}_f}{L_f},\\
        (\vect{x}_f)_d=\frac12\left((\vect{x}_{f,1})_d+(\vect{x}_{f,2})_d\right),
        \qquad
        (L_f)_d=\frac{\vect{e}_f\cdot(\vect{e}_f)_d}{L_f},\\
        (\vect{n}_{cf})_d=\mathsf R_{cf}\left[\frac{(\vect{e}_f)_d}{L_f}-\frac{\vect{e}_f(L_f)_d}{L_f^2}\right].
    \end{gathered}
    \label{eq:app_face_metric_derivatives}
\end{equation}
For a counterclockwise polygonal cell with vertices $\vect{x}_{c,i}$, the signed cell area and its variation are given by
\begin{equation}
    \begin{gathered}
        A_c=\frac12\sum_i\vect{x}_{c,i}\times\vect{x}_{c,i+1},\\
        (A_c)_d=\frac12\sum_i\left[(\vect{x}_{c,i})_d\times\vect{x}_{c,i+1}+\vect{x}_{c,i}\times(\vect{x}_{c,i+1})_d\right],
    \end{gathered}
    \label{eq:app_area_derivative}
\end{equation}
where $\times$ denotes the two-dimensional scalar cross product under cyclic vertex indexing. Cell-centroid derivatives follow analogously from the discrete primal centroid formula, with both areas and centroids evaluated in cell-local shifted coordinates consistently across the primal, tangent, and transpose calculations.

For the weighted least-squares gradient reconstruction in Section~\ref{sec:ugks}, with geometric weights $\varpi_{cj}$, the normal equations and their directional derivative read
\begin{equation}
    \begin{gathered}
        \mathsf L_c^{\mathrm{LS}}=\sum_{j\in\mathcal N(c)}\varpi_{cj}\vect{r}_{cj}\vect{r}_{cj}^T,
        \qquad
        \vect{b}_c^{\mathrm{LS}}=\sum_{j\in\mathcal N(c)}\varpi_{cj}\vect{r}_{cj}(z_j-z_c),\\
        \vect{r}_{cj}=\vect{x}_j-\vect{x}_c,
        \qquad
        \mathsf L_c^{\mathrm{LS}}(\nabla z_c)_d=(\vect{b}_c^{\mathrm{LS}})_d-(\mathsf L_c^{\mathrm{LS}})_d\nabla z_c.
    \end{gathered}
    \label{eq:app_ls_tangent}
\end{equation}
Both $\vect{r}_{cj}$ and $\varpi_{cj}$ vary with the mesh coordinates, incorporating the periodic translation vector where applicable. The limited face reconstruction and its directional derivative are
\begin{equation}
    \begin{gathered}
        z_{c,f}=z_c+\varphi_c\nabla z_c\cdot(\vect{x}_f-\vect{x}_c),\\
        (z_{c,f})_d=z_{c,d}+(\varphi_c)_d\nabla z_c\cdot(\vect{x}_f-\vect{x}_c)\\
        +\varphi_c(\nabla z_c)_d\cdot(\vect{x}_f-\vect{x}_c)
        +\varphi_c\nabla z_c\cdot\bigl((\vect{x}_f)_d-(\vect{x}_c)_d\bigr).
    \end{gathered}
    \label{eq:app_reconstruction_tangent}
\end{equation}
The Venkatakrishnan limiter $\varphi_c$~\cite{venkatakrishnan1995convergence} is differentiated along its active local branch, accounting for variations in the stencil extrema and cell-to-face offsets; its area-dependent regularization threshold satisfies
\begin{equation}
    \epsilon_c^2\propto A_c^{3/2},
    \qquad
    (\epsilon_c^2)_d=\frac32\frac{\epsilon_c^2}{A_c}(A_c)_d.
    \label{eq:app_limiter_scale_derivative}
\end{equation}
For the UGKS interface equilibrium expansion~\cite{xu2010unified,huang2012unified}, differentiating the local moment systems for the normal and tangential slopes yields
\begin{equation}
    \begin{gathered}
        \mathsf M_0\vect{\alpha}_\ell=\vect{W}_{0,\ell},
        \qquad \ell\in\{n,t\},\\
        \mathsf M_0(\vect{\alpha}_\ell)_d=(\vect{W}_{0,\ell})_d-(\mathsf M_0)_d\vect{\alpha}_\ell,
    \end{gathered}
    \label{eq:app_equilibrium_slope_derivative}
\end{equation}
where $\mathsf M_0$ denotes the local Maxwellian moment matrix. Differentiating the temporal compatibility relation analogously yields the temporal slope variation $(\vect{\alpha}_0)_d$, accounting for variations in both $\mathsf{M}_0$ and the moment-integral vector. Variations in the local normal and tangential molecular velocity components induced by interface rotation $(\vect{n}_{cf})_d$ are propagated consistently.

For the local interface evolution interval in Eqs.~\eqref{eq:fluxtime} and~\eqref{eq:spectraltime}, the weighted spectral sum and characteristic time are written as
\begin{equation}
    \mathcal S_c^{\mathrm{spec}}=\sum_{f\in\partial\Omega_c}L_fa_{cf}^{\mathrm{spec}},
    \qquad
    t_c^{\mathrm{spec}}=\frac{2A_c}{\mathcal S_c^{\mathrm{spec}}}.
    \label{eq:app_spectral_sum}
\end{equation}
With $s(f)$ denoting the cell attaining the minimum spectral time at interface $f$, differentiating along the active branch gives
\begin{equation}
    \begin{gathered}
        (\vartheta_f)_d=\vartheta_f\left[\frac{(A_{s(f)})_d}{A_{s(f)}}-\frac{(\mathcal S_{s(f)}^{\mathrm{spec}})_d}{\mathcal S_{s(f)}^{\mathrm{spec}}}\right],\\
        (\mathcal S_c^{\mathrm{spec}})_d=\sum_{f\in\partial\Omega_c}\left[(L_f)_d a_{cf}^{\mathrm{spec}}+L_f(a_{cf}^{\mathrm{spec}})_d\right].
    \end{gathered}
    \label{eq:app_face_time_derivative}
\end{equation}
Decomposing the local spectral wave speed into acoustic and convective parts leads to
\begin{equation}
    \begin{gathered}
        a_{cf}^{\mathrm{spec}}=a_{s,c}+q_{cf},
        \qquad
        q_{cf}=\max\!\left(|\vect{U}_c\cdot\vect{n}_{cf}|,\max_k|\vect{u}_k\cdot\vect{n}_{cf}|\right),\\
        (a_{s,c})_d=\frac{a_{s,c}}2\left(\frac{p_{c,d}}{p_c}-\frac{\rho_{c,d}}{\rho_c}\right),\\
        (q_{cf})_d=\begin{cases}
            \operatorname{sgn}(\vect{U}_c\cdot\vect{n}_{cf})\bigl(\vect{U}_{c,d}\cdot\vect{n}_{cf}+\vect{U}_c\cdot(\vect{n}_{cf})_d\bigr),&\text{macroscopic branch},\\
            \operatorname{sgn}(\vect{u}_{k_*}\cdot\vect{n}_{cf})\vect{u}_{k_*}\cdot(\vect{n}_{cf})_d,&\text{velocity-node branch},
        \end{cases}
    \end{gathered}
    \label{eq:app_spectral_speed_derivative}
\end{equation}
where $k_*$ denotes the maximizing discrete velocity node. All derivatives are evaluated along the active branch selected at the converged state, excluding branch switching at ties.

Because $\vartheta_f$ enters both the time-averaging prefactor and the upper integration limit in Eq.~\eqref{eq:microflux}, its explicit partial derivative at fixed interface distribution reads
\begin{equation}
    \left.\frac{\partial\Phi^z_{cf,k}}{\partial\vartheta_f}\right|_{\mathrm{interface}}=-\frac{\Phi^z_{cf,k}}{\vartheta_f}+\frac{L_f}{\vartheta_f}(\vect{u}_k\cdot\vect{n}_{cf})z_f(\vartheta_f,\vect{u}_k),
    \qquad z\in\{h,b\}.
    \label{eq:app_flux_time_partial}
\end{equation}
This term is combined with the derivatives of the interface distribution, edge length, and unit normal in differentiating the full interface flux. In reverse mode, adjoint seeds accumulated on $\vartheta_f$ are propagated via Eq.~\eqref{eq:app_face_time_derivative} back to the controlling cell state and mesh metrics.

For diffuse reflection at a solid wall with fixed velocity $\vect{U}_w$, differentiating the wall-relative normal velocity in Eq.~\eqref{eq:wall_normal_velocity} and the normal mass fluxes in Section~\ref{sec:boundaries} yields
\begin{equation}
    \begin{gathered}
        \vect{c}_{w,k}=\vect{u}_k-\vect{U}_w,
        \qquad
        (u_{n,k}^w)_d=\vect{c}_{w,k}\cdot(\vect{n}_w)_d,\\
        (J_{\mathrm{in}})_d=\sum_{u_{n,k}^w<0}w_k\left[(u_{n,k}^w)_d h_{\mathrm{in},k}+u_{n,k}^w(h_{\mathrm{in},k})_d\right],\\
        (J_{\mathrm{out}}^{(1)})_d=\sum_{u_{n,k}^w>0}w_k\left[(u_{n,k}^w)_d H_{w,k}^{(1)}+u_{n,k}^w(H_{w,k}^{(1)})_d\right],
    \end{gathered}
    \label{eq:app_wall_flux_derivatives}
\end{equation}
where $(h_{\mathrm{in},k})_d$ follows from differentiating the interior upwind reconstruction at the wall face. For a prescribed wall-temperature perturbation $(T_w)_d$ extracted from $\vect{\mathcal{B}}_d$, the variations of the unit-density wall Maxwellians are
\begin{equation}
    \begin{gathered}
        (H_{w,k}^{(1)})_d=H_{w,k}^{(1)}\left(-\frac1{T_w}+\frac{|\vect{c}_{w,k}|^2}{T_w^2}\right)(T_w)_d,\\
        (B_{w,k}^{(1)})_d=B_{w,k}^{(1)}\frac{|\vect{c}_{w,k}|^2}{T_w^2}(T_w)_d.
    \end{gathered}
    \label{eq:app_wall_maxwell_derivatives}
\end{equation}
Differentiating the zero-net-mass-flux condition in Eq.~\eqref{eq:wallrho} and the reflected distributions in Eq.~\eqref{eq:wall} then yields
\begin{equation}
    \begin{gathered}
        (\rho_w)_d=-\frac{(J_{\mathrm{in}})_d+\rho_w(J_{\mathrm{out}}^{(1)})_d}{J_{\mathrm{out}}^{(1)}},\\
        (h_{w,k})_d=H_{w,k}^{(1)}(\rho_w)_d+\rho_w(H_{w,k}^{(1)})_d,
        \qquad
        (b_{w,k})_d=B_{w,k}^{(1)}(\rho_w)_d+\rho_w(B_{w,k}^{(1)})_d,
        \qquad u_{n,k}^w>0.
    \end{gathered}
    \label{eq:wall_distribution_tangent}
\end{equation}

At far-field boundaries, the prescribed external Maxwellians depend on the free-stream state variations $(\rho_{\infty,d},\vect{U}_{\infty,d},T_{\infty,d})$ supplied by $\vect{\mathcal{B}}_d$, giving
\begin{equation}
    \begin{gathered}
        \vect{c}_{\infty,k}=\vect{u}_k-\vect{U}_{\infty},\\
        (H_{\infty,k})_d=H_{\infty,k}\left[\frac{\rho_{\infty,d}}{\rho_{\infty}}+\left(-\frac1{T_{\infty}}+\frac{|\vect{c}_{\infty,k}|^2}{T_{\infty}^2}\right)T_{\infty,d}+\frac{2\vect{c}_{\infty,k}\cdot\vect{U}_{\infty,d}}{T_{\infty}}\right],\\
        (B_{\infty,k})_d=\frac{T_{\infty}}{2}(H_{\infty,k})_d+\frac{H_{\infty,k}}2T_{\infty,d},
    \end{gathered}
    \label{eq:app_inlet_outlet_derivatives}
\end{equation}
while variations of the outgoing distributions are supplied by the interior cell reconstruction. For periodic face pairs linked by the translation vector $\vect{d}$,
\begin{equation}
    \vect{x}_{f_R}=\vect{x}_{f_L}+\vect{d},
    \qquad
    \vect{d}_d=(\vect{x}_{f_R})_d-(\vect{x}_{f_L})_d,
    \label{eq:app_periodic_translation_derivative}
\end{equation}
which enters the periodic stencil displacements, with transpose sensitivities accumulated symmetrically across both paired boundaries while preserving opposite outward normal orientations.

For the density-normalized mass-flow objective of the periodic micropump in Eq.~\eqref{eq:objective}, differentiating the section-averaged flux $\mathcal F$ and domain-averaged density $\bar\rho$ at fixed reference width $D_a$ yields
\begin{equation}
    \begin{gathered}
        I_d=\frac{\mathcal F_d}{\bar\rho D_a}-\frac{I}{\bar\rho}\bar\rho_d,
        \qquad
        \bar\rho_d=\frac1{A}\sum_c\left[A_c\rho_{c,d}+(\rho_c-\bar\rho)A_{c,d}\right],\\
        \left.\frac{\partial I}{\partial A_c}\right|_{\mathcal F,\rho}=\frac{I}{A}\left(1-\frac{\rho_c}{\bar\rho}\right).
    \end{gathered}
    \label{eq:app_normalized_objective_derivative}
\end{equation}
This explicit cell-area sensitivity is combined with the state and mass-projection terms during nodal gradient assembly. When both the total gas mass and total domain area are held fixed, the total variation $\bar\rho_d$ vanishes identically; nevertheless, both the state and explicit cell-area partial derivatives in Eq.~\eqref{eq:app_normalized_objective_derivative} must be retained individually so that their contributions cancel consistently under nonuniform local cell deformation and remain exact for unconstrained area perturbations.

For the periodic micropump, interior nodal displacements $\vect{d}_I^{\mathrm{mesh}}$ are computed from prescribed boundary displacements $\vect{d}_B^{\mathrm{mesh}}$ via a linear-elasticity mesh-deformation analogy assembled on the reference grid:
\begin{equation}
    \mathsf K_{II}\vect{d}_I^{\mathrm{mesh}}=-\mathsf K_{IB}\vect{d}_B^{\mathrm{mesh}},
    \qquad
    \mathsf K_{II}(\vect{d}_I^{\mathrm{mesh}})_d=-\mathsf K_{IB}(\vect{d}_B^{\mathrm{mesh}})_d.
    \label{eq:app_elastic_mesh_derivative}
\end{equation}
Because the elasticity system is formed on the fixed reference mesh, the stiffness submatrices $\mathsf K_{II}$ and $\mathsf K_{IB}$ remain constant. For the airfoil cases, the radial grid-line blending defined in Section~\ref{sec:naca_setup} specifies the linear nodal displacement operator directly. In both configurations, geometric adjoint contributions from cell metrics, fluxes, periodic couplings, and the mass projection in Eq.~\eqref{eq:massborderadjoint} are accumulated at the mesh nodes before projection onto the design space via $P_a^T$ in Eq.~\eqref{eq:controlgradients}.

Finally, for the symmetric channel-width parameterization of the periodic micropump, with fixed centerline coordinates $\vect{x}_i^{\mathrm{mid}}$ and equal-and-opposite wall displacements $\vect{\delta}_i$, the net signed-area contribution from the opposing walls satisfies
\begin{equation}
    \begin{gathered}
        (\vect{x}_i^{\mathrm{mid}}+\vect{\delta}_i)\times(\vect{x}_{i+1}^{\mathrm{mid}}+\vect{\delta}_{i+1})
        -(\vect{x}_i^{\mathrm{mid}}-\vect{\delta}_i)\times(\vect{x}_{i+1}^{\mathrm{mid}}-\vect{\delta}_{i+1})\\
        =2\left(\vect{x}_i^{\mathrm{mid}}\times\vect{\delta}_{i+1}+\vect{\delta}_i\times\vect{x}_{i+1}^{\mathrm{mid}}\right).
    \end{gathered}
    \label{eq:app_symmetric_area}
\end{equation}
Because the quadratic displacement cross-terms cancel identically, the total domain area is strictly affine in the shape parameters $\vect{a}$ whenever the end boundaries are fixed and $\vect{\delta}_i$ depends affinely on $\vect{a}$, whereas individual cell-area variations are still evaluated locally via Eq.~\eqref{eq:app_area_derivative}.

\bibliographystyle{elsarticle-num}
\bibliography{main}
\end{document}